\documentclass[twocolumn]{aastex631}
\hypersetup{linkcolor=red,citecolor=blue,filecolor=cyan,urlcolor=magenta}

\usepackage{acronym}  
\usepackage{amssymb} 
\usepackage{amsbsy}
\usepackage{amsmath}
\usepackage{longtable}
\usepackage{multirow}
\usepackage{units}

\acrodef{PMPS}[PMPS]{Parkes Multibeam Pulsar Survey}
\acrodef{SMPS}[SMPS]{Swinburne Intermediate-latitude Pulsar Survey}
\acrodef{HTRU}[HTRU]{High Time Resolution Universe}
\acrodef{ISM}[ISM]{interstellar medium}
\acrodef{CNN}[CNN]{convolutional neural network}
\acrodef{MDN}[MDN]{mixture density network}
\acrodef{ReLU}[ReLU]{rectified linear unit}
\acrodef{SBI}[SBI]{simulation-based inference}
\acrodef{MCMC}[MCMC]{Markov Chain Monte Carlo}
\acrodef{DM}[DM]{dispersion measure}
\acrodef{SNR}[SNR]{signal-to-noise ratio}
\acrodef{NPE}[NPE]{neural posterior estimation}
\acrodef{NLE}[NLE]{neural likelihood estimation}
\acrodef{NRE}[NRE]{neural ratio estimation}
\acrodef{CI}[CI]{credible interval}
\acrodef{HDR}[HDR]{highest-density region}
\acrodef{KDE}[KDE]{kernel density estimation}

\newcommand{\bt}{\boldsymbol{\theta}}
\newcommand{\bp}{\boldsymbol{\psi}}
\newcommand{\bx}{\boldsymbol{x}}
\newcommand{\pr}{\mathcal{P}}

\submitjournal{ApJ}

\shorttitle{Isolated pulsar population synthesis with simulation-based inference}
\shortauthors{Graber et al.}

\begin{document}

\title{Isolated pulsar population synthesis with simulation-based inference}


\correspondingauthor{Vanessa Graber}
\email{v.graber@herts.ac.uk}

\author[0000-0002-6558-1681]{Vanessa Graber}
\affiliation{Institute of Space Sciences (CSIC-ICE), Campus UAB, Carrer de Can Magrans s/n, 08193, Barcelona, Spain}
\affiliation{Institut d'Estudis Espacials de Catalunya (IEEC), Carrer Gran Capit\`a 2--4, 08034 Barcelona, Spain}
\affiliation{Centre for Astrophysics Research, Department of Physics, Astronomy and Mathematics, University of Hertfordshire, College Lane, Hatfield AL10 9AB, UK}

\author[0000-0003-2781-9107]{Michele Ronchi}
\affiliation{Institute of Space Sciences (CSIC-ICE), Campus UAB, Carrer de Can Magrans s/n, 08193, Barcelona, Spain}
\affiliation{Institut d'Estudis Espacials de Catalunya (IEEC), Carrer Gran Capit\`a 2--4, 08034 Barcelona, Spain}

\author[0000-0002-8118-255X]{Celsa Pardo-Araujo}
\affiliation{Institute of Space Sciences (CSIC-ICE), Campus UAB, Carrer de Can Magrans s/n, 08193, Barcelona, Spain}
\affiliation{Institut d'Estudis Espacials de Catalunya (IEEC), Carrer Gran Capit\`a 2--4, 08034 Barcelona, Spain}

\author[0000-0003-2177-6388]{Nanda Rea} 
\affiliation{Institute of Space Sciences (CSIC-ICE), Campus UAB, Carrer de Can Magrans s/n, 08193, Barcelona, Spain}
\affiliation{Institut d'Estudis Espacials de Catalunya (IEEC), Carrer Gran Capit\`a 2--4, 08034 Barcelona, Spain}


\begin{abstract}

We combine pulsar population synthesis with simulation-based inference (SBI) to constrain the magnetorotational properties of isolated Galactic radio pulsars. We first develop a framework to model neutron star birth properties and their dynamical and magnetorotational evolution. We specifically sample initial magnetic field strengths, $B$, and spin periods, $P$, from lognormal distributions and capture the late-time magnetic field decay with a power law. Each lognormal is described by a mean, $\mu_{\log B}, \mu_{\log P}$, and standard deviation, $\sigma_{\log B}, \sigma_{\log P}$, while the power law is characterized by the index, $a_{\rm late}$. We subsequently model the stars' radio emission and observational biases to mimic detections with three radio surveys, and we produce a large database of synthetic $P$--$\dot{P}$ diagrams by varying our five magnetorotational input parameters. We then follow an SBI approach that focuses on neural posterior estimation and train deep neural networks to infer the parameters' posterior distributions. After successfully validating these individual neural density estimators on simulated data, we use an ensemble of networks to infer the posterior distributions for the observed pulsar population. We obtain $\mu_{\log B} = 13.10^{+0.08}_{-0.10}$, $\sigma_{\log B} = 0.45^{+0.05}_{-0.05}$ and $\mu_{\log P} = -1.00^{+0.26}_{-0.21}$, $\sigma_{\log P} = 0.38^{+0.33}_{-0.18}$ for the lognormal distributions and $a_{\rm late} = -1.80^{+0.65}_{-0.61}$ for the power law at the $95\%$ credible interval. We contrast our results with previous studies and highlight uncertainties of the inferred $a_{\rm late}$ value. Our approach represents a crucial step toward robust statistical inference for complex population synthesis frameworks and forms the basis for future multiwavelength analyses of Galactic pulsars.

\end{abstract}




\section{Introduction}
\label{sec:intro}

As one of the end points of stellar evolution of massive stars, neutron stars are influenced by many extremes of physics, including strong gravity, large densities, fast rotation, and extreme magnetic fields. Consequently, these compact objects have been connected with several of the most energetic transient phenomena in our Universe, such as fast radio bursts, superluminous supernovae, ultraluminous X-ray sources, long- and short-duration gamma-ray bursts, and gravitational-wave emission \citep[e.g.,][]{Bachetti2014, Berger2014, Metzger2014,  Abbott2017a, Margalit2018, Petroff2022}. Accurately modeling these processes requires a detailed understanding of neutron star properties, which also set constraints on massive stellar evolution. Inferring the birth properties of neutron stars and the physics that govern their subsequent evolution is, thus, crucial for other fields of astrophysics.

Detecting and accurately characterizing individual objects within the entire neutron star population is, hence, critical. As a result, the number of known pulsars (those neutron stars that emit regular electromagnetic pulses) has steadily increased since the first detection in 1967 \citep{Hewish1968}, and we currently know of around 3500 of these objects \citep{Manchester2005}.\footnote{\url{https://www.atnf.csiro.au/research/pulsar/psrcat/}; v2.1.0} These are visible across the full electromagnetic spectrum, and their emission is predominantly driven by their enormous rotational energy reservoirs. Roughly 400 of these sources are confirmed to be in binaries, of which the majority were strongly influenced by accretion from their companions and spun up to short spin periods earlier in their lives. The remaining $\sim 3100$ sources are primarily isolated neutron stars. Due to observational limitations and diverse emission properties, we cannot detect these with a single telescope, but instead have to focus on certain subpopulations. With around 1100 members, a subset of isolated radio pulsars constitutes the largest fraction of neutron stars detected in a single survey \citep{Posselt2023}. However, these numbers only cover a tiny portion of the overall neutron star population. We can provide a rough estimate of the neutron stars in the Milky Way by multiplying their birth rate \citep[a core-collapse supernova rate of $\sim 2$ per century; see][]{Keane2008, Rozwadowska2021} by the age of the Milky Way \citep[$\sim 13$ billion years; see, e.g.,][]{Conroy2022, Xiang2022} to arrive at a total of 260 million Galactic neutron stars.

To bridge the gap between expected and observed neutron stars, we take advantage of population synthesis. This approach relies on producing a large catalog of synthetic pulsar populations that are passed through a set of filters to mimic observational constraints. The resulting populations are then contrasted with the true observed sample to find those parameter regions that best explain the data. Although different versions of this methodology have been applied to pulsar data for several decades \citep[e.g.,][]{Narayan1990, Lorimer2004, Faucher2006, Gonthier2007, Bates2014, Gullon2014, Gullon2015, Cieslar2020}, the complexity of models that capture the properties of observed Galactic neutron stars significantly complicates the comparison between the simulated populations and the observed one. This is especially true if we are interested in quantifying uncertainties for our neutron star parameters, because Bayesian \ac{MCMC} or nested sampling methods (the standard tools for this kind of question; see, e.g., \citet{Feroz2009, Foreman-Mackey2013, Sharma2017, Ashton2019, Speagle2020}) become infeasible for pulsar population synthesis unless significant simplifications for simulation models and the likelihood function are made \citep{Cieslar2020}. The main reason for this is that we can no longer write down an explicit likelihood for realistic neutron star simulation frameworks. In this paper, we thus focus on \acf{SBI} (also known as likelihood-free inference; for a recent review see \citet{Cranmer2020}) in the context of pulsar population synthesis for the first time.

In the past few years, \ac{SBI} has successfully challenged traditional approaches such as approximate Bayesian computation \citep[e.g.,][]{Rubin1984, Beaumont2002, Dean2011, Frazier2017} in those areas of science that rely on complex simulators, which lead to intractable likelihoods. The existence of such a simulator, essentially acting as a forward model, is the only requirement for \ac{SBI}. As such, the approach is ideal for astrophysics and has been recently applied to parameter estimation in, e.g., cosmology \citep{Alsing2019, Hahn2023, Lemos2023, Lin2023}, high-energy astrophysics \citep{Huppenkothen2022, Mishra-Sharma2022}, gravitational-wave astronomy \citep{Dax2021, Cheung2022, Bhardwaj2023} and exoplanet research \citep{Vasist2023}. \ac{SBI} is particularly powerful in combination with neural networks, whose benefits for pulsar population synthesis studies were outlined in \citet{Ronchi2021} by inferring point estimates for the dynamical properties of radio pulsars in the Milky Way.

\begin{figure*}
\centering
\includegraphics[width=0.85\textwidth]{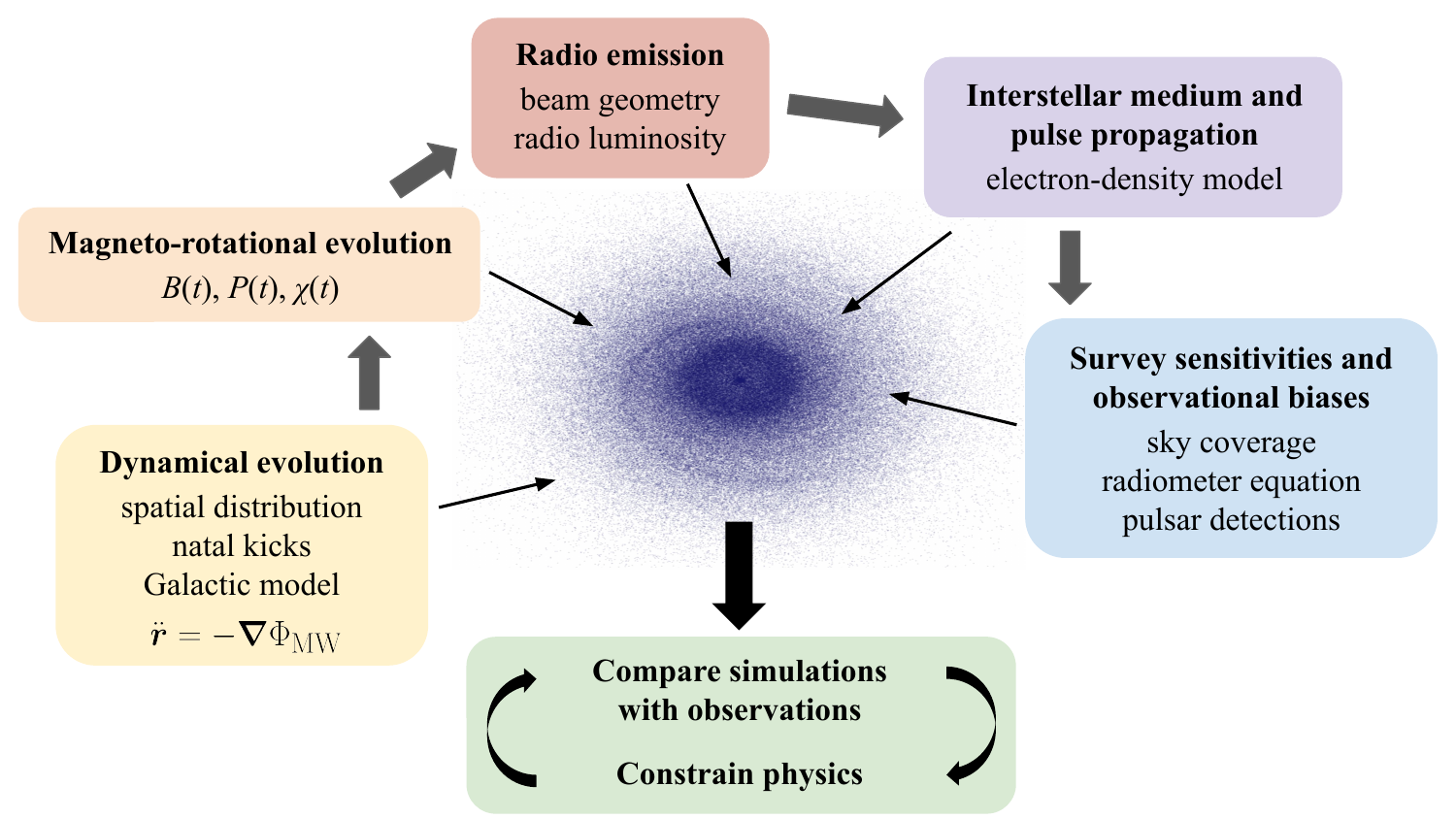}
\caption{The key ingredients for pulsar population synthesis. Starting from the bottom left, this approach relies on modeling the neutron stars' dynamical evolution, as well as their magnetorotational properties. For a given beaming geometry and luminosity model, we then determine the pulsars' radio emission and its propagation across the Galaxy toward Earth. For the neutron stars pointing toward us, we subsequently invoke survey limitations and sensitivity thresholds to determine those objects that are detectable. The resulting synthetic populations are compared to the observed ones to constrain input physics.}
\label{fig:Pop_flowchart}
\end{figure*}

In this study, we take a Bayesian perspective to infer posteriors of neutron star parameters using \ac{SBI}. For this purpose, we model the Galactic neutron star dynamics, the magnetorotational evolution and the radio emission properties. We then run snapshots of the total pulsar population at the current time through a set of filters to mimic observational limitations. The resulting simulation output are synthetic $P$--$\dot{P}$ diagrams (where $P$ and $\dot{P}$ denote the pulsar spin period and its time derivative, respectively) of the observed pulsar population. We then construct an \ac{SBI} pipeline, which we train, validate, and test on a large database of these synthetic $P$--$\dot{P}$ diagrams to infer posterior distributions of our input parameters. We specifically focus on five parameters related to the initial period distribution of pulsars and their magnetic field properties that crucially affect the positions of stars in the $P$--$\dot{P}$ plane. We then apply our optimized deep-learning framework, for the first time, to the radio pulsars detected in the \ac{PMPS} \citep{Manchester2001, Lorimer2006}, the \ac{SMPS} \citep{Edwards2001, Jacoby2009} and the low- and mid-latitude \ac{HTRU} survey \citep{Keith2010} (all recorded with Murriyang, the Parkes radio telescope).

The paper is structured as follows: Section~\ref{sec:popsyn} summarizes our population synthesis framework. We then provide a general overview of \ac{SBI} and our choice of setup in Sections~\ref{sec:sbi_overview} and \ref{sec:architecture}, respectively, whereas Section~\ref{sec:experiments} summarizes the machine-learning experiments conducted for this study. We next address network training and inference results plus corresponding validation approaches in Section~\ref{sec:results}, specifically benchmarking our pipeline on test simulations before applying it to the observed pulsar population. Finally, we provide a detailed discussion of our approach and results, as well as an outlook into the future, in Section~\ref{sec:conclusions}.


\section{Pulsar population synthesis}
\label{sec:popsyn}

\subsection{Overview}
\label{sec:pop_overview}

The key ingredients for our pulsar-population synthesis model are summarized in Figure~\ref{fig:Pop_flowchart}. We first require a prescription for the star's dynamical properties to populate our synthetic Galaxy with neutron stars. To this end, we model their birth positions and velocities plus their subsequent dynamical evolution in the Milky Way. We further capture the stars' initial magnetic and rotational characteristics in addition to their evolution. For both these aspects, our framework broadly follows earlier works (see, e.g., \citet{Faucher2006, Gullon2014, Cieslar2020, Ronchi2021}) and our simulator employs a Monte-Carlo approach to sample relevant parameters at birth from corresponding probability density functions. We note that we save computation time by not evolving the dynamical properties for each single simulation. As the dynamical and magnetorotational properties are independent, we instead simulate a single dynamical database for a large number of current pulsar positions and velocities, and we subsequently sample from these distributions before determining the magnetorotational evolution. Next, we characterize the stars' radio emission by implementing a realistic beaming geometry. We then simulate detections by propagating the corresponding radio pulses across the Galaxy for a specific electron density model. The resulting emission for those pulsars pointing toward Earth is then contrasted to observational biases and sensitivity thresholds for a given radio survey to determine which synthetic pulsars would be detected. The resulting mock populations are then compared to the observed populations to constrain relevant model parameters. We explore \ac{SBI} for this purpose as outlined in detail in Section~\ref{sec:sbi}.


\subsection{Dynamical Evolution}
\label{sec:dyn_evol}

To create our dynamical database from which we sample neutron star positions and velocities, we simulate $10^7$ neutron stars from birth to today. For each object, we randomly assign an age sampled from a uniform distribution up to a maximum age of $10^{8} \,$yr, which ensures that our synthetic Milky Way is populated with a sufficient number of neutron stars within a reasonable computation time. As sources older than $10^{8} \,$yr are no longer detectable as radio pulsars (see below), this approach provides a realistic description of the current positions and velocities of these objects.

We then define a cylindrical reference frame, $(r, \phi, z)$, whose origin is located at the Galactic center. Here $r$, $\phi$ and $z$ denote the distance from the origin in kpc, the azimuthal angle in radians, and the distance from the Galactic plane in kpc, respectively. In particular, we position our Sun at $r=\unit[8.3]{kpc}$, $\phi= \pi/2$, and $z=\unit[0.02]{kpc}$ \citep[see][and references therein]{Pichardo2012}.

To determine the birth locations of individual neutron stars, we first focus on the distributions of their massive progenitors in the $(r, \phi)$-plane and along $z$ separately. Considering the distribution of free electrons as a tracer of star formation in the Milky Way that correlates with the massive OB stars that evolve into neutron stars, we sample the initial positions in $r, \phi$ according to the Galactic electron density distribution of \cite{Yao2017}. This will also allow consistency when relating pulsar distances with their dispersion measures in Section~\ref{sec:obs_lims}. In addition, as the Galactic matter distribution is not static, we assume that the Milky Way rotates rigidly in clockwise direction with an angular velocity $\Omega = 2 \pi /T$, where $T \approx \unit[250]{Myr}$ \citep{Vallee2017, Skowron2019}. For a given stellar age, we can thus retrace the angular coordinate, $\phi$, at birth.

Moreover, we assume that pulsar birth positions along the $z$-direction follow an exponential disk model \citep{Wainscoat1992} and sample from a probability density function of the form
\begin{equation}
	\mathcal{P}(z) = \frac{1}{h_c} \exp\left(-\frac{ \lvert z \rvert}{ h_c } \right).
\end{equation}
We follow the pulsar population studies of \citet{Gullon2014} and \citet{Ronchi2021} and set the characteristic scale height, $h_c$, to a fiducial value of $\unit[0.18]{kpc}$. Note that this is consistent with the distribution of young, massive stars in our Galaxy \citep{Li2019}. We then randomly assign each star's $z$-coordinate a positive or negative sign to distribute our population above and below the Galactic plane.

Next, we focus on the pulsars' birth velocities, which are a combination of the kick velocity, $\boldsymbol{v}_k$, imparted during the supernova owing to explosion asymmetries \citep[see][and references therein]{Coleman2022, Janka2022} and the velocity, $\boldsymbol{v}_{\rm pr}$, inherited from the progenitors' orbital Galactic motion. Specifically, we sample the magnitude of the kick velocities, $v_k \equiv |\boldsymbol{v}_k|$, from a Maxwell distribution,
\begin{equation}
	\mathcal{P}(v_k) = \sqrt{ \frac{2}{\pi} }  \frac{v_k^2}{\sigma_k^3}
		\exp\left(-\frac{v_k^2}{ \sigma_k^2 } \right),
		\label{eq:pdf_maxwell_kick}
\end{equation}
and then assign a random direction to determine the kick along the $r$-, $\phi$- and $z$-directions. For the dispersion parameter, $\sigma_k$, we take a fiducial value of $\sigma_k \approx \unit[260]{km \, s^{-1}}$ \citep{Hobbs2005}, which is broadly consistent with observed proper motions of radio pulsars \citep{Hobbs2005, Faucher2006}. See, however, \citet{Verbunt2017} and \citet{Igoshev2020} who find that a double Maxwellian characterizes the data better.

The second velocity component due to the progenitors' motion depends on the Galactic gravitational potential, $\Phi_{\rm MW}$, and points along the azimuthal direction:
\begin{equation}
	\boldsymbol{v}_{pr} = \sqrt{ r \, \frac{\partial \Phi_{\rm MW} \left( r,z \right)}{\partial r} } \, \hat{\boldsymbol{\phi}},
\end{equation}
where $\hat{\boldsymbol{\phi}}$ is a unit vector in $\phi$-direction. For this study, we consider a Galactic potential that is given as the sum of four components, i.e., the nucleus, $\Phi_n$, the bulge, $\Phi_b$, the disk, $\Phi_d$, and the halo, $\Phi_h$, \citep{Marchetti2019}. The nucleus and bulge contributions are described by a spherical Hernquist potential \citep{Hernquist1990}:
\begin{equation}
	\Phi_{n, b} = -\frac{ G M_{n, b}}{ R_{n, b} +  R},
\end{equation}
where $R = \sqrt{r^2 + z^2}$ is the spherical radial coordinate and $G$ the gravitational constant. The disk has a cylindrical Miyamoto--Nagai potential of the form \citep{Miyamoto1975}
\begin{equation}
	\Phi_d = -\frac{ G M_d}{ \sqrt{ \left(a_d + \sqrt{z^2+b_d^2}\right)^2 + r^2} },
\end{equation}
where $a_d$ and $b_d$ represent the scale length and scale height of the disk, respectively. Finally, the halo is characterized by a spherical Navarro--Frenk--White potential \citep{Navarro1996}:
\begin{equation}
	\Phi_h = -\frac{ G M_h}{ R } \ln{ \left( 1 + \frac{R}{R_h}\right) }.
\end{equation}
The free parameters, $M_{n, b, d, h}$, $R_{n,b,h}$, $a_d$ and $b_d$, can be obtained through fits of the Milky Way's mass profile and are given in Table~2 of \citet{Ronchi2021} (see also \citet{Bovy2015} and Table~1 of \citet{Marchetti2019}).

After determining the initial positions and velocities for each of our $10^7$ neutron stars, we perform the dynamical evolution by solving the Newtonian equation of motion in cylindrical coordinates, $\ddot{\boldsymbol{r}} = -\boldsymbol{\nabla} \Phi_{\rm MW}$, according to the stars' respective ages. This way, we obtain a database of current pulsar positions and velocities in the Milky Way.


\subsection{Magnetorotational Evolution}
\label{sec:mr_evol}

The primary diagnostic for the pulsar population is the $P$--$\dot{P}$ diagram. For our study, we focus on rotation-powered radio pulsars, which are the easiest to detect and constitute the largest class of neutron stars. Corresponding period and period derivative measurements for this population are enabled via radio timing. To first order, radio pulsars can be approximated as rotating magnetic dipoles, implying that their spin-down is driven by electromagnetic dipole radiation. The locations of individual neutron stars and the shape of the population's distribution in the $P$--$\dot{P}$ plane are, hence, determined by their dipolar magnetic fields and rotation periods at birth and the subsequent magnetorotational evolution. The latter couples the evolution of the pulsar period, $P$, the dipolar magnetic field strength, $B$, at the pole, and the inclination angle, $\chi$, between the magnetic and the rotation axis.

To capture these physics, we first sample the misalignment angle at birth, $\chi_0$, randomly in the range $[0, \pi/2]$ according to the probability density \citep{Gullon2014}
\begin{equation}
	\mathcal{P}(\chi_0) = \sin \chi_0.
\end{equation}
We then sample the logarithm of the initial magnetic field, $B_0$ (measured in G), and the initial period, $P_0$ (measured in s), for each pulsar from normal distributions of the form \citep{Popov2010, Gullon2014, Igoshev2020, Igoshev2022, Xu2023}
\begin{align}
	\mathcal{P}(\log B_0) &= \frac{1}{\sqrt{2 \pi} \sigma_{\log B}}
		\, \exp\left(-\frac{\log B_0 - \mu_{\log B}}{2 \sigma_{\log B}^2} \right),
			\label{eqn:B_pdf} \\[1.8ex]
  	\mathcal{P}(\log P_0) &= \frac{1}{\sqrt{2 \pi} \sigma_{\log P}}
      		\, \exp\left(-\frac{\log P_0 - \mu_{\log P}}{2 \sigma_{\log P}^2} \right).
			\label{eqn:P_pdf}
\end{align}
The means, $\mu_{\log B}, \mu_{\log P}$, and the standard deviations, $\sigma_{\log B}, \sigma_{\log P}$, are free parameters of our model and four of those parameters, whose posteriors we set out to infer with our \ac{SBI} approach in Section~\ref{sec:sbi}. We will specifically explore the ranges $\mu_{\log B} \in [12, 14]$, $\mu_{\log P} \in [-1.5, -0.3]$, $\sigma_{\log B} \in [0.1, 1.0]$ and $\sigma_{\log P} \in [0.1, 1.0]$ to encompass results of earlier analyses \citep[e.g.,][]{Gullon2014}.

Assuming that pulsars spin down owing to dipolar emission, we follow \citet{Spitkovsky2006} and \citet{Philippov2014} and solve the following coupled differential equations:
\begin{align}
	\dot{P} &= \frac{\pi^2}{c^3}\frac{B^2 R_{\rm NS}^6}{I_{\rm NS} P} \left( \kappa_0 + \kappa_1 \sin^2 \chi \right),
		\label{eqn:P_ode} \\[1.8ex]
	\dot{\chi} &= -\frac{\pi^2}{c^3}\frac{B^2 R_{\rm NS}^6}{I_{\rm NS} P^2} \, \kappa_2 \sin\chi \cos\chi,
		\label{eqn:chi_ode}
\end{align}
where $c$ is the speed of light, $R_{\rm NS} \approx \unit[11]{km}$ is the neutron star radius and $I_{\rm NS} \simeq 2 M_{\rm NS} R_{\rm NS}^2 / 5 \approx \unit[1.36 \times 10^{45}]{g \, cm^{2}}$ is the stellar moment of inertia (for a fiducial mass $M_{\rm NS} \approx \unit[1.4]{M_{\odot}}$). For realistic pulsars surrounded by plasma-filled magnetospheres, we choose $\kappa_0 \simeq \kappa_1 \simeq \kappa_2 \simeq 1$, and we note that Equation~\eqref{eqn:chi_ode} implies that $\chi$ decreases with time, i.e., our pulsars move toward alignment.

\begin{figure}
\includegraphics[width=0.95\columnwidth]{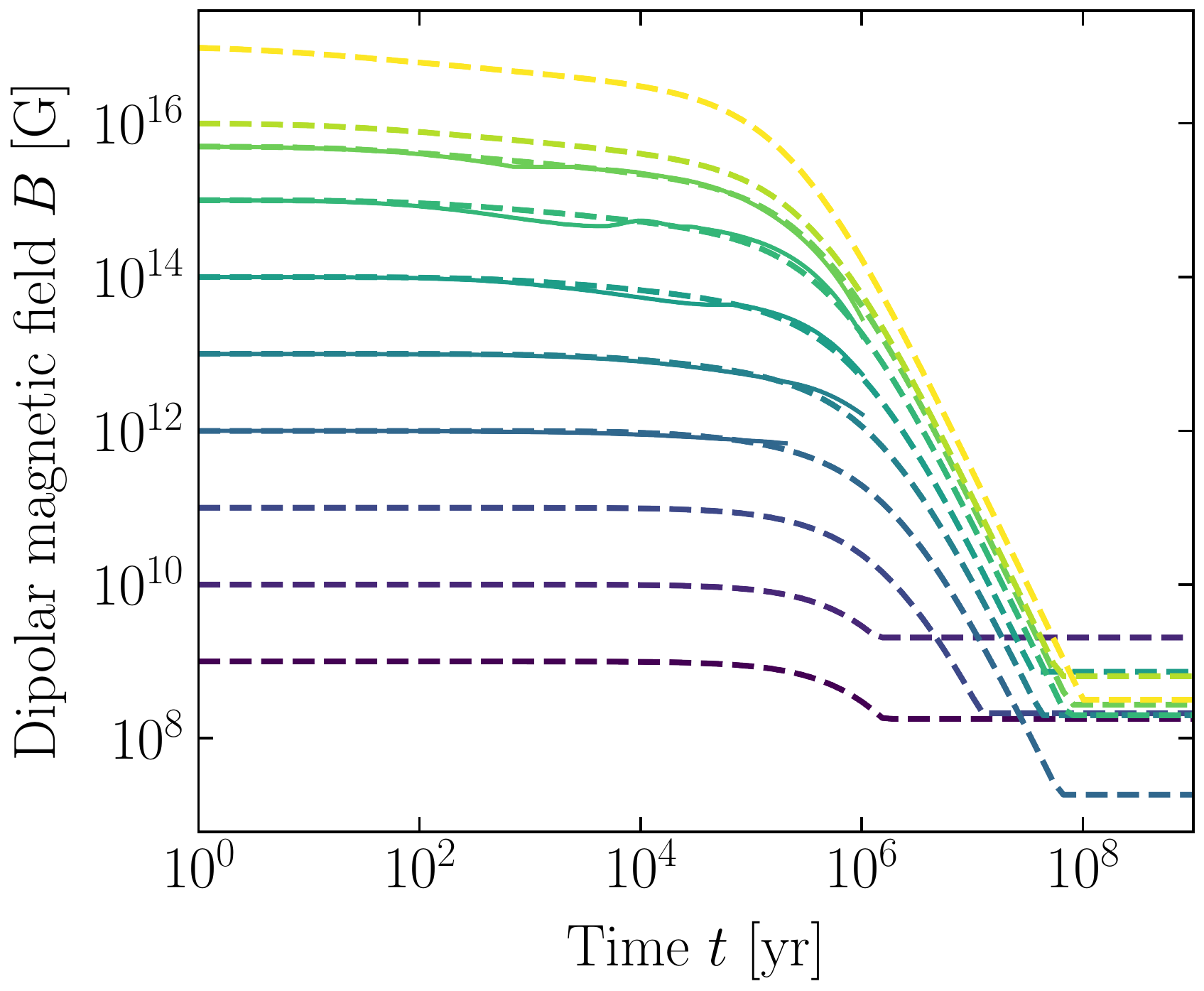}
\caption{Illustration of the $B$-field parameterization used for this study. The five solid curves represent realistic two-dimensional simulations of magnetothermal evolution in the neutron star crust \citep{Vigano2021}. We fit these together with the late-time power-law evolution of the magnetic field with several broken power laws. The dashed curves shown here are determined for $a_{\rm late} = -3.0$. The colors represent the initial magnetic field strength, $B_0$. To avoid the field decaying to unrealistically small numbers at very late times, we sample the final fields from a Gaussian distribution. The procedure, which allows us to easily extract the dipolar field strength, $B$, at different times, $t$, to study the magnetorotational evolution of our synthetic pulsars, is described in detail in Appendix~\ref{app:B-field}.}
\label{fig:B_fields}
\end{figure}

The final ingredient is a suitable prescription for the evolution of the dipolar magnetic field strength. While the $B$-field decay in the neutron star crust is typically assumed to be driven by the combined action of the Hall effect and ohmic dissipation \citep[e.g.,][]{Aguilera2008}, changes in the magnetic field are strongly coupled to the thermal properties of the neutron star interior \citep[e.g.,][]{Pons2019}. This is particularly important for strongly magnetized neutron stars with fields above $\sim \unit[10^{13}]{G}$ and, hence, relevant for a significant fraction of our simulated pulsar population. In the past decade, several theoretical and numerical efforts have begun to unveil the complex processes of magnetothermal evolution \citep[e.g.,][]{Vigano2013, Vigano2021, DeGrandis2021, Igoshev2021, Dehman2023}. As corresponding simulations are highly time-consuming, we instead develop a new approach, outlined in detail in Appendix \ref{app:B-field} and summarized in Figure~\ref{fig:B_fields}, that parameterizes a range of magnetothermal simulations for different magnetic field strengths \citep{Vigano2021}. This prescription allows us to extract magnetic fields up to pulsar ages of around $\unit[10^6]{yr}$. Above this value, current numerical simulations become unreliable because they rely on implementations of complex microphysics that are unsuitable for cold, old stars. Moreover, they do not capture the highly uncertain physics of neutron star cores, which become relevant at large ages. We instead incorporate the cores' field evolution at late times by means of a power law of the form
\begin{equation}
	B(t) \propto \left(1+ \frac{t}{\tau_{\rm late}} \right)^{a_{\rm late}},
		\label{eqn:B_late}
\end{equation}
where $\tau_{\rm late} \approx \unit[2 \times 10^6]{yr}$, $t$ is the time, and the power-law index, $a_{\rm late}$, is the fifth free parameter of our model. We note that although the details of core field evolution are not known, Eq.~\eqref{eqn:B_late} is physically motivated because several known mechanisms exhibit similar power-law behavior (see Appendix \ref{app:B-field}). Hence, we will explore the parameter range $a_{\rm late} \in [-3.0, -0.5]$. Finally, to prevent the dipolar magnetic field from decaying to arbitrarily small values (in disagreement with observations of old, recycled millisecond pulsars; see, e.g., \citet{Lorimer2008}), we assume that the field eventually reaches a constant value. Therefore, we sample the logarithm of the field, $B_{\rm final}$, from a normal distribution with a mean $\mu_{\log B, {\rm final}} = 8.5$ and a standard deviation $\sigma_{\log B, {\rm final}} = 0.5$ in line with observations of old pulsars.

Following this prescription allows us to determine the spin periods, dipolar field strengths, and misalignment angles for our simulated pulsars at the current time.


\subsection{Emission Characteristics}
\label{sec:emission_phys}

We next implement a prescription for the radio emission geometry to determine those pulsars whose beams sweep over Earth and are, in principle, detectable. In the canonical model of radio pulsars, their emission is produced close to the stellar surface in the cone-shaped, open field-line region \citep{Lorimer2012, Johnston2020}. Assuming that this entire region is involved in the emission, geometric considerations allow us to estimate the half opening angle of the emission beam, $\rho_b$ (in rad), via \citep{Gangadhara2001}
\begin{equation}
	\rho_b \simeq \sqrt{\frac{9 \pi r_{\rm em}}{2 c P}},
		\label{eqn:rho_b}
\end{equation}
where $r_{\rm em}$ is the emission height. The latter is thought to be period independent, and we set it to $\unit[300]{km}$ following \citet{Johnston2020} (see also references therein). Note that several studies of pulsars with stable emission profiles have recovered this $\rho_b \propto P^{-1/2}$ behavior \citep[e.g.,][]{Kramer1994, Maciesiak2011b, Skrzypczak2018}. Knowledge of $\rho_b$, then, allows us to obtain the solid angle, $\Omega_b$, covered by a pulsar's two radio beams. More specifically,
\begin{equation}
	\Omega_b = 4 \pi (1 - \cos \rho_b).
		\label{eqn:omega_b}
\end{equation}
As we do not expect biases in how we observe this conal emission for any given pulsar, we draw a random line-of-sight angle, $\alpha$, with respect to the rotation axis in the range $[0, \pi/2]$ using the probability density $\sin \alpha $. Combined with the half opening angle, $\rho_b$, and the evolved inclination angle, $\chi$, we can then determine those pulsars whose radio beams are visible from Earth. We note that as a result of this purely geometric argument, between $\sim 60-95\%$ of our generated pulsars (depending on the specific choice of magnetorotational parameters) are typically not detectable.

We proceed with determining the emission characteristics of those neutron stars that point toward Earth. In particular, we follow \citet{Maciesiak2011a} and express the intrinsic pulse width (measured in s) of our simulated pulsars as follows:
\begin{equation}
	w_{\rm int} = \frac{2}{\pi} \arcsin{ \sqrt{ \frac{\sin^2\left( \frac{\rho_b}{2} \right) 
				- \sin^2\left( \frac{\alpha - \chi}{2} \right)}
		{\sin \left( \alpha \right) \sin \left(\chi \right)}}} \, P.
			\label{eqn:instr_pulsewidth} 
\end{equation}
Finally, as the radio emission is ultimately driven by the stars' rotational energy reservoir, we assume that the intrinsic radio luminosity, $L_{\rm int}$ (in ${\rm erg} \, {\rm s}^{-1}$), for each star depends on the spin-down power, $|\dot{E}_{\rm rot}| = 4 \pi^2 I_{\rm NS} \dot{P} / P^3$. In particular, we consider
\begin{equation}
	L_{\rm int} = L_0 \sqrt{ \frac{\dot{P}}{P^3}},
        \label{eqn:luminosity}
\end{equation}
where $L_0$ is a normalization factor whose logarithm we sample from a normal distribution with mean $\mu_{\log L} = 35.5$ and standard deviation $\sigma_{\log L} = 0.8$ \citep[see also][]{Faucher2006, Gullon2014} to eventually recover observed luminosities. 

\begin{deluxetable*}{c|cccc}
\tablecaption{Survey Parameters for \ac{PMPS}, \ac{SMPS}, and the Low- and Mid-latitude \ac{HTRU} Survey Taken from \citet{Manchester2001, Lorimer2006}, \citet{Edwards2001, Jacoby2009}, and \citet{Keith2010}, Respectively. \label{tab:SurveyParam}}
\tabletypesize{\small}
\tablecolumns{5}
\tablenum{1}
\tablewidth{0pt}
\tablehead{
\colhead{Survey} &
\colhead{PMPS} &
\colhead{SMPS} &
\colhead{HTRU mid} &
\colhead{HTRU low}
}
\startdata
Sky region & $-100^{\circ} < l < 50^{\circ}$ & $-100^{\circ} < l < 50^{\circ}$ & 
	$-120^{\circ} < l < 30^{\circ}$ & $-80^{\circ} < l < 30^{\circ}$  \\
 & $|b| < 5^{\circ}$ & $5^{\circ}<|b| < 30^{\circ}$ & $|b| < 15^{\circ}$ &  $|b| < 3.5^{\circ}$  \\
$f$ (GHz) & 1.374 & 1.374 & 1.352 & 1.352 \\
$\Delta f_{\rm ch}$ (kHz) & 3000 & 3000 & 390.625 & 390.625 \\
$\tau_{\rm samp}$ ($\mu$s) & 250 & 125 & 64 & 64 \\
$G$ (K$\, {\rm Jy}^{-1}$) &  0.735 & 0.735 & 0.735 & 0.735 \\
$n_{\rm pol}$ & 2 & 2 & 2 & 2 \\
$\Delta f_{\rm bw}$ (MHz) & 288 & 288 & 340 & 340 \\
$t_{\rm obs}$ (s) & 2100 & 265 & 540 & 4300 \\
$\beta$ & 1.5 & 1.5 & 1.5 & 1.5 \\
$T_{\rm sys}$ (K) & 21 & 21 & 23 & 23 \\
S/N threshold & 9 & 9 & 9 & 9
\enddata
\tablecomments{We provide the survey region where completeness is above $90\%$ in Galactic longitude ($l$) and latitude ($b$), the central observing frequency ($f$), the channel width ($\Delta f_{\rm ch}$), the sampling time ($\tau_{\rm samp}$), the telescope gain ($G$), the number of observed polarizations ($n_{\rm pol}$), the observing bandwidth ($\Delta f_{\rm bw}$), the integration time ($t_{\rm obs}$), the degradation factor ($\beta$), the system temperature ($T_{\rm sys}$), and the S/N threshold for each of the surveys. Corresponding units are given in brackets in the first column.}
\end{deluxetable*}


\subsection{Simulating Detections}
\label{sec:obs_lims}

Armed with the knowledge of intrinsic pulsar properties, we now turn to the possibility of detecting those objects whose emission beams cross our line of sight. First, the bolometric radio flux, $S$, that reaches us from any given simulated pulsar is equal to
\begin{equation}
	S = \frac{L_{\rm int}}{\Omega_b d^2}, 
\end{equation}
where $d$ is the distance known from the dynamical evolution outlined in Section~\ref{sec:dyn_evol}. To determine the corresponding radio flux density, $S_{f}$ (measured in Jy), at a specific observing frequency, $f$, we follow \citet{Lorimer2012} and assume that the radio emission spectrum follows a power law in $f$. In particular, we set the spectral index to $-1.6$ \citep{Jankowski2018}. We can, hence, approximate the total fluence of a pulse with width $w_{\rm int}$ as $S_{f} w_{\rm int}$. Assuming that this fluence stays constant as the radio signal propagates from the pulsar toward us, we estimate the flux density, $S_{f , {\rm obs}}$, that reaches Earth as
\begin{equation}
	S_{f , {\rm obs}} \simeq S_{f} \frac{ w_{\rm int}}{w_{\rm obs}}
	\label{eqn:obs_flux} 
\end{equation}
where $w_{\rm obs}$ is the observed pulse width.

\begin{figure*}
\centering
\includegraphics[height=0.78\columnwidth]{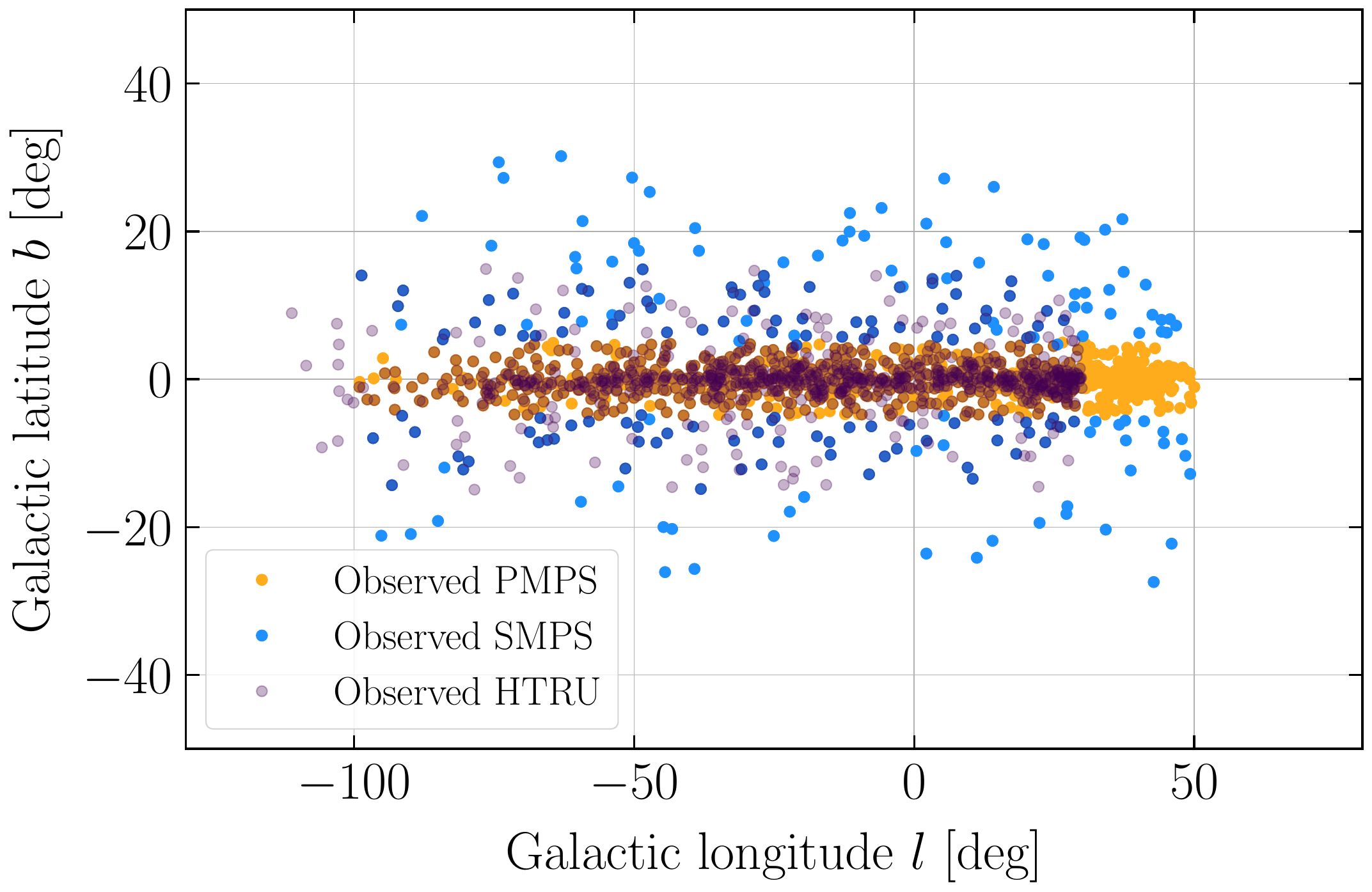}
\hspace{0.2cm}
\includegraphics[height=0.78\columnwidth]{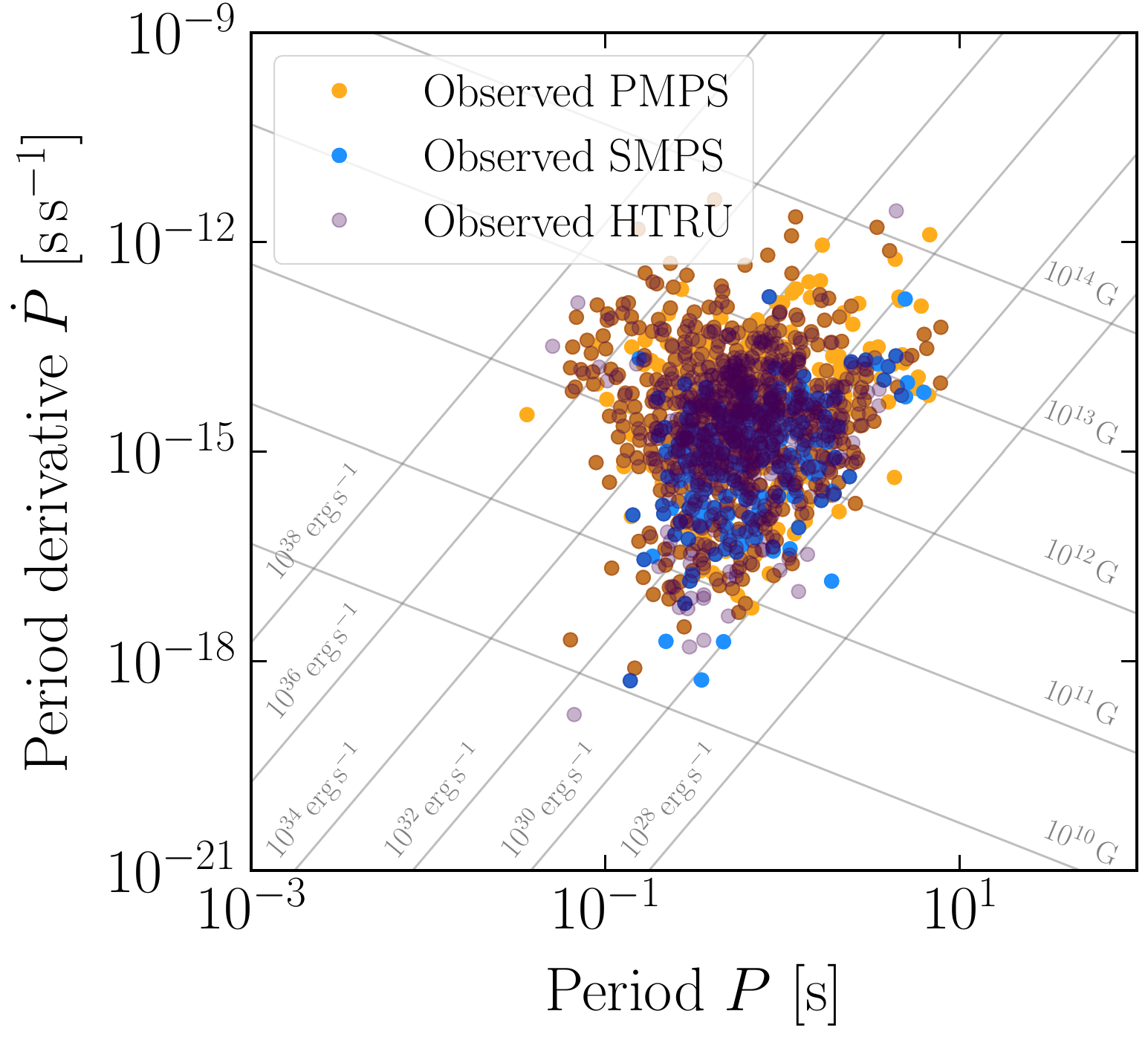}
\caption{Observed populations of isolated Galactic radio pulsars detected with \ac{PMPS}, \ac{SMPS} and the low- and mid-latitude \ac{HTRU} survey (highlighted in yellow, light blue, and purple, respectively). The left panel shows the distribution of these three populations in Galactic latitude, $b$, and longitude, $l$, while the right panel depicts the detected pulsars in the period, $P$, and period derivative, $\dot{P}$, plane. In the latter, we also give lines of constant spin-down power, $|\dot{E}_{\rm rot}|$, and constant dipolar surface magnetic field, $B$ (estimated via Equation~\eqref{eqn:P_ode} for an aligned rotator). Data taken from the ATNF Pulsar Catalogue \citep[][\url{https://www.atnf.csiro.au/research/pulsar/psrcat/}, v1.69]{Manchester2005}. Observational filters are described in detail in the text.}
\label{fig:pop_observed}
\end{figure*}

\begin{figure}[b]
\centering
\includegraphics[width=0.95\columnwidth]{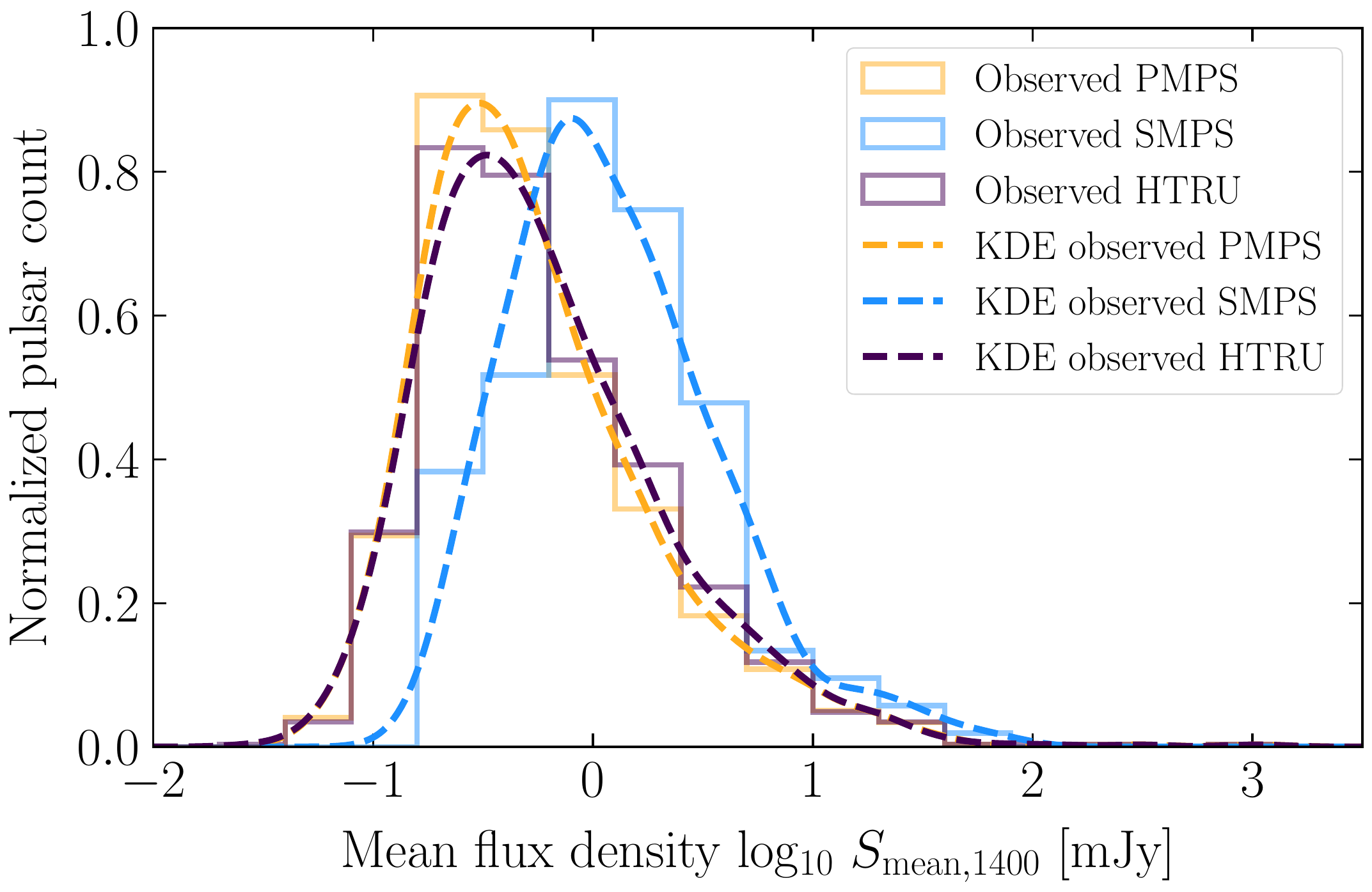}
\caption{Distributions of mean radio flux density measurements, $S_{{\rm mean}, 1400}$, at \unit[1400]{MHz} for the populations of isolated Galactic radio pulsars detected with \ac{PMPS}, \ac{SMPS} and the low- and mid-latitude \ac{HTRU} survey (in yellow, light blue, and purple, respectively). We show the normalized number of stars as a function of radio flux density as solid lines. Dashed lines represent the corresponding probability density functions obtained via \ac{KDE} using a Gaussian kernel. Data taken from the ATNF Pulsar Catalogue \citep[][\url{https://www.atnf.csiro.au/research/pulsar/psrcat/}, v1.69]{Manchester2005}.}
\label{fig:flux_observed}
\end{figure}

Specifically, as a radio pulse propagates, it experiences dispersion and scattering caused by interactions with the free electrons and density fluctuations in the \ac{ISM}, respectively. Both mechanisms result in a broader pulse when compared with the intrinsic width, $w_{\rm int}$. Further broadening is caused by instrumental effects, which are dominated by the sampling time, $\tau_{\rm samp}$, of the hardware used to record radio observations. Accounting for these processes, we can write the observed pulse width as \citep{Cordes2003}
\begin{equation}
	w_{\rm obs} \simeq \sqrt{w_{\rm int}^2 + \tau_{\rm samp}^2 
		+ \tau_{\rm DM}^2 + \tau_{\rm scat}^2}.
			\label{eqn:obs_pulsewidth} 
\end{equation}
We follow \citet{Bates2014} to determine $\tau_{\rm DM}$, encoding the pulse smearing due to dispersion for a single frequency channel of the telescope's receiver. Specifically,
\begin{equation}
	\tau_{\rm DM} = \frac{e^2}{\pi m_e c} \, \frac{\Delta f_{\rm ch}}{f^3} \, {\rm DM},
		\label{eqn:tau_DM} 
\end{equation}
where $e$ is the electronic charge, $m_{e}$ the corresponding mass, $\Delta f_{\rm ch}$ the hardware-specific width of a frequency channel at observing frequency $f$ and DM is the dispersion measure. We further use the empirical fit relationship from \citet{Krishnakumar2015} for $\tau_{\rm scat}$, the pulse smearing due to scattering of radio waves by an inhomogeneous and turbulent \ac{ISM}:
\begin{equation}
	\tau_{\rm scat} = 3.6 \times 10^{-9} {\rm DM}^{2.2} \left(1 + 1.94 \times 10^{-3} {\rm DM}^2 \right),
		\label{eqn:tau_scat} 
\end{equation}
where $\tau_{\rm scat}$ is measured in s. We moreover account for a significant scatter in the underlying data (see Figure~3 in \citet{Krishnakumar2015}) by drawing $\log \tau_{\rm scat}$ values from a Gaussian distribution around the fit in Equation~\eqref{eqn:tau_scat} with a standard deviation of 0.5. We also incorporate the fact that \citet{Krishnakumar2015} analyzed observations at $\unit[327]{MHz}$ by rescaling to a given observing frequency $f$, assuming a Kolmogorov spectrum, i.e., $\tau_{\rm scat} \propto f^{-4.4}$ \citep[see][for details]{Lorimer2012}. As $\tau_{\rm DM}$ and $\tau_{\rm scat}$ both depend on the pulsars' respective dispersion measure, we again employ the Galactic electron density distribution of \cite{Yao2017} to convert our simulated neutron star positions from Section~\ref{sec:dyn_evol} into DM values. 

At this stage, we require information for the radio surveys we want to emulate. We specifically focus on three surveys recorded with Murriyang, the Parkes radio telescope: the \acf{PMPS} \citep{Manchester2001, Lorimer2006}, the \acf{SMPS} \citep{Edwards2001, Jacoby2009}, and the low- and mid-latitude \acf{HTRU} survey \citep{Keith2010}. All relevant survey parameters (including the sampling time, $\tau_{\rm samp}$, the observing frequency, $f$, and the channel width, $\Delta f_{\rm ch}$, needed to calculate $w_{\rm obs}$) are summarized in Table~\ref{tab:SurveyParam}. 

To assess whether those simulated stars that cross our line of sight are detectable with a given survey, we first determine whether they are located in the surveys' fields of view. While \ac{PMPS} and \ac{HTRU} have a similar sky coverage, we highlight that \ac{SMPS} detected pulsars at higher Galactic latitude (see left panel of Figure~\ref{fig:pop_observed}). This survey is thus sensitive to older neutron stars that have had sufficient time to move away from their birth positions closer to the Galactic plane, providing complementary information on the pulsar population. For those objects that fall within our survey coverage, we subsequently establish whether they are sufficiently bright to be detected. To do so, we calculate the pulsars' signal-to-noise ratio using the radiometer equation \citep{Lorimer2012}:
\begin{equation}
        {\rm S/N} = \frac{ S_{\rm mean} G \sqrt{n_{\rm pol} \Delta f_{\rm bw} t_{\rm obs}} }
        		{ \beta \left[ T_{\rm sys} + T_{\rm sky }(l, b) \right] } 
		\sqrt{\frac{P- w_{\rm obs}}{w_{\rm obs}}}.
		\label{eqn:radiometer}
\end{equation}
Here $S_{\rm mean} \simeq S_{f, {\rm obs}} w_{\rm obs} / P$ denotes the mean flux density averaged over a single rotation period $P$, $G$ is the receiver gain \citep[see][for details]{Lorimer1993, Bates2014}, $n_{\rm pol}$ is the number of detected polarizations, $\Delta f_{\rm bw}$ is the observing bandwidth, $t_{\rm obs}$ is the integration time, and $\beta > 1 $ is a degradation factor that accounts for imperfections during the digitization of the signal. Moreover, $T_{\rm sys}$ denotes the system temperature, and $T_{\rm sky}(l, b)$ is the sky background temperature dominated by synchrotron emission of Galactic electrons, which varies strongly with latitude, $l$, and longitude, $b$. To model the latter, we use results from \citet{Remazeilles2015}, who provided a refined version of the temperature map of \citet{Haslam1981, Haslam1982}. As the underlying data were obtained at 408 MHz, we rescale to the relevant observing frequencies by assuming a power-law dependence of the form $T_{\rm sky} \propto f^{-2.6}$ \citep{Lawson1987, Johnston1992}. 

A synthetic pulsar counts as detected if the value obtained from Equation~\eqref{eqn:radiometer} exceeds the surveys' sensitivity thresholds. We aim to recover the numbers of detected isolated Galactic radio pulsars for each survey, i.e., 
\begin{align}
&\text{\ac{PMPS}: $1009$ observed pulsars}, \nonumber \\
&\text{\ac{SMPS}: $218$ observed pulsars}, \label{eqn:detected_objects} \\
&\text{\ac{HTRU}: $1023$ observed pulsars}. \nonumber
\end{align}
To obtain these values, we used the data from the ATNF Pulsar Catalogue \citep{Manchester2005}\footnote{\url{https://www.atnf.csiro.au/research/pulsar/psrcat/}; v1.69} and removed extragalactic sources and those in globular clusters. We further applied a cutoff in period ($P > \unit[0.01]{s}$) and period derivative ($\dot{P} > \unit[10^{-19}]{s \, s^{-1}}$; for those objects with measured $\dot{P}$ values because the above counts also include a small number of pulsars without $\dot{P}$ measurements) to remove those objects that have (likely) been spun up by accretion from a companion star and cannot be modeled with the framework discussed so far. The locations of those objects with known period and period derivatives are shown in the $P$--$\dot{P}$ plane in the right panel of Figure~\ref{fig:pop_observed}. The distribution of mean flux densities, $S_{\rm mean}$, measured at $\unit[1400]{MHz}$ for isolated Galactic pulsars in our three surveys as recorded in the ATNF catalog is shown in Figure~\ref{fig:flux_observed}. Note that this database does not contain flux measurements for all sources and that uncertainties on reported $S_{\rm mean}$ values can be large. We also note that $S_{\rm mean}$ values in the catalog do not form a homogeneous sample, as there is no standardized way for $S_{\rm mean}$ measurements to be reported in the literature. For example, in some cases $S_{\rm mean}$ is measured by observing a flux calibration source, while other values are estimated using the radiometer equation given by Equation~\eqref{eqn:radiometer}, introducing additional systematics due to different prescriptions for the S/N or pulse width. For ease of comparison with our simulated pulsar populations, we also show \acf{KDE} fits for the corresponding probability density functions obtained with a Gaussian kernel in Figure~\ref{fig:flux_observed}.


\begin{figure*}
\centering
\includegraphics[height=0.61\columnwidth]{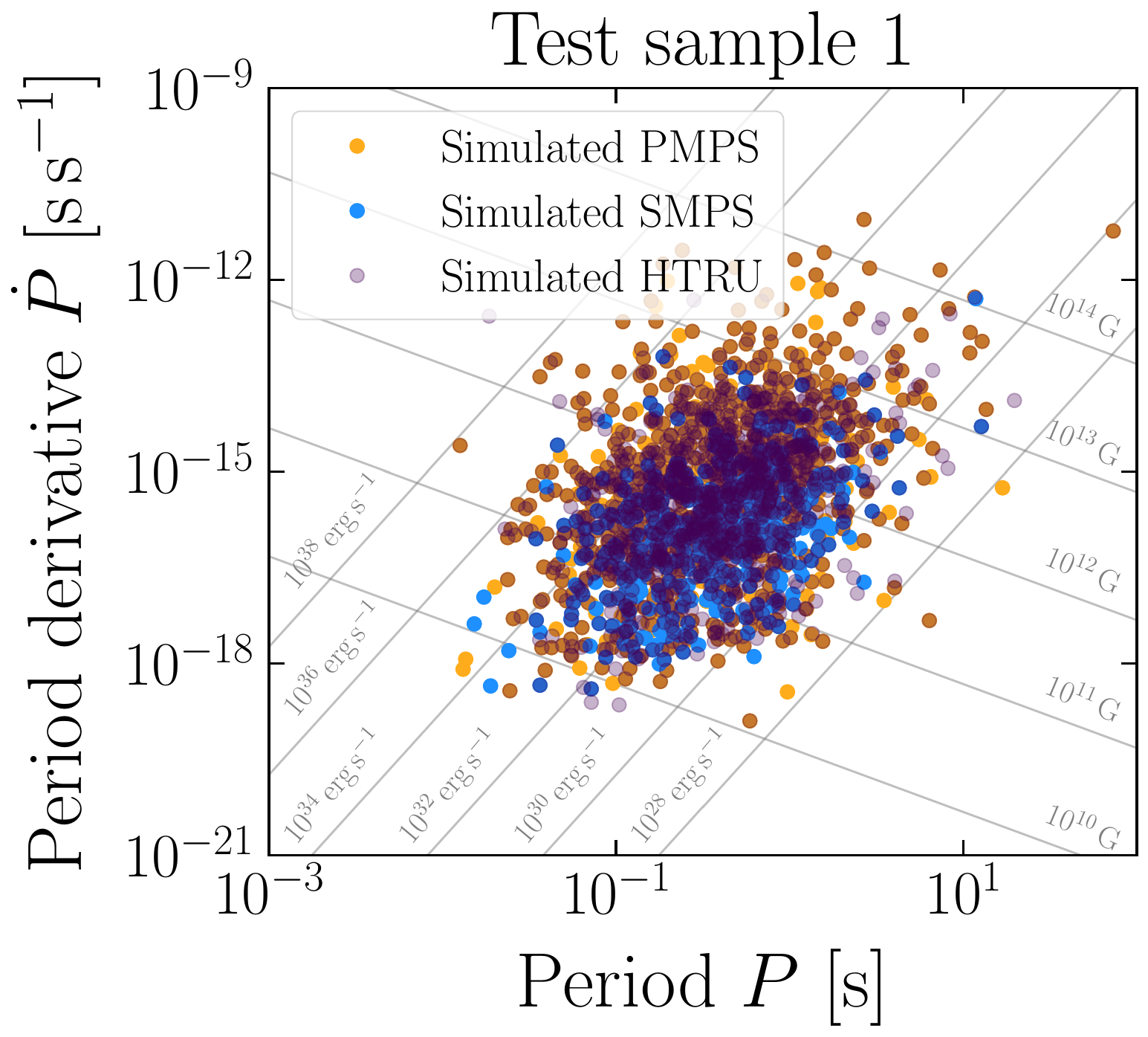}
\hspace{0.1cm}
\includegraphics[height=0.61\columnwidth]{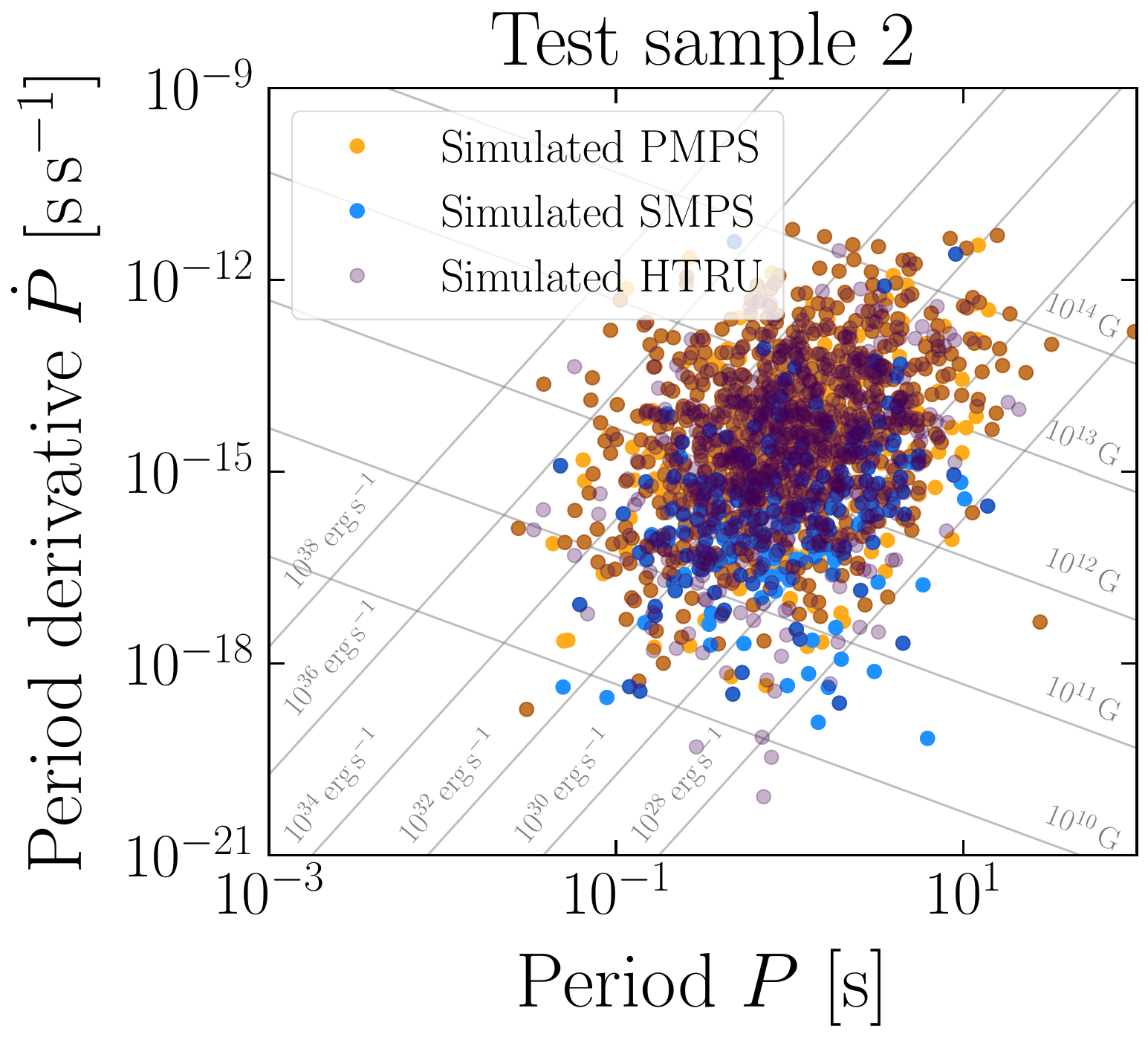}
\hspace{0.1cm}
\includegraphics[height=0.61\columnwidth]{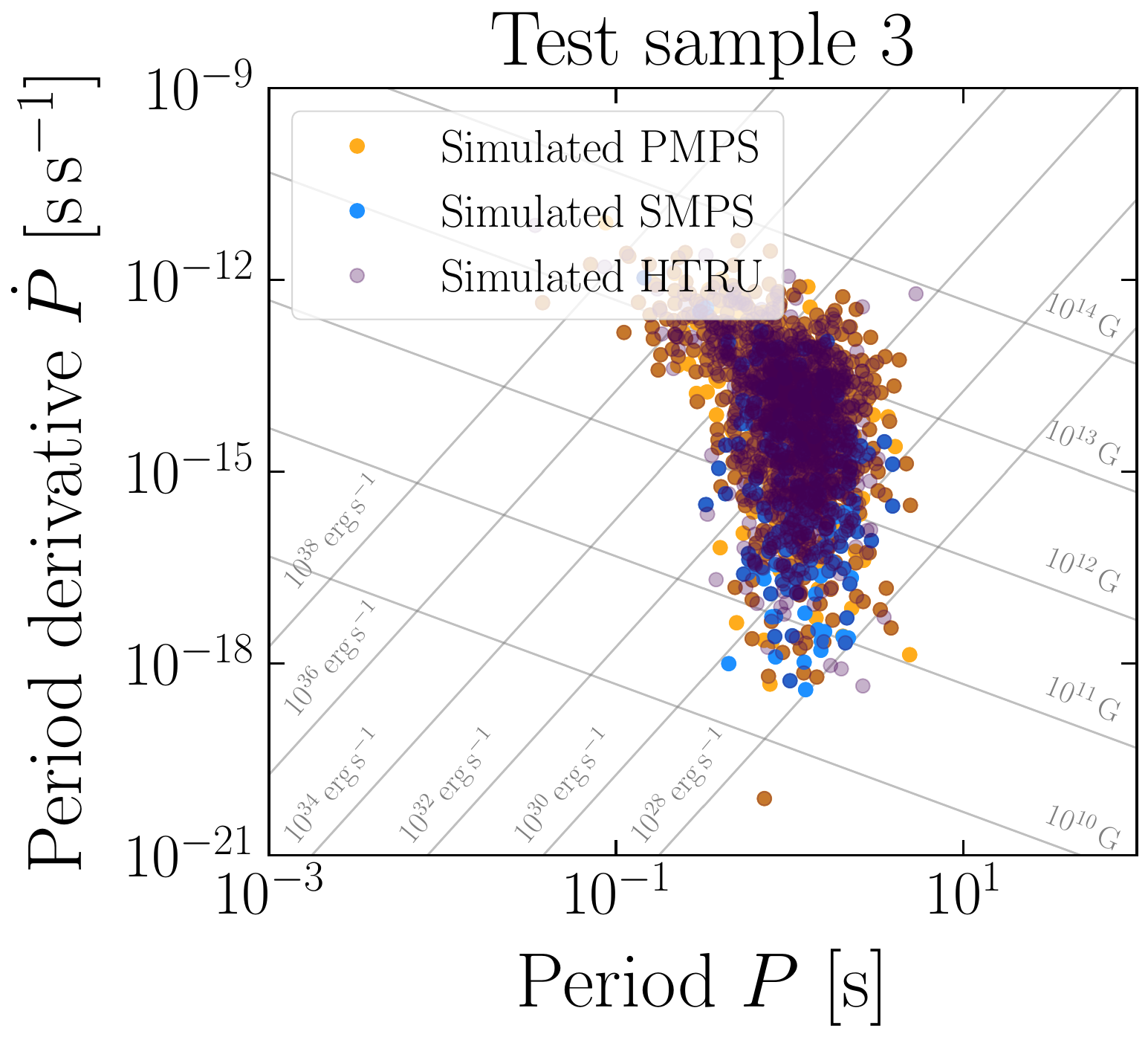} \vskip 0.2cm
\includegraphics[height=0.61\columnwidth]{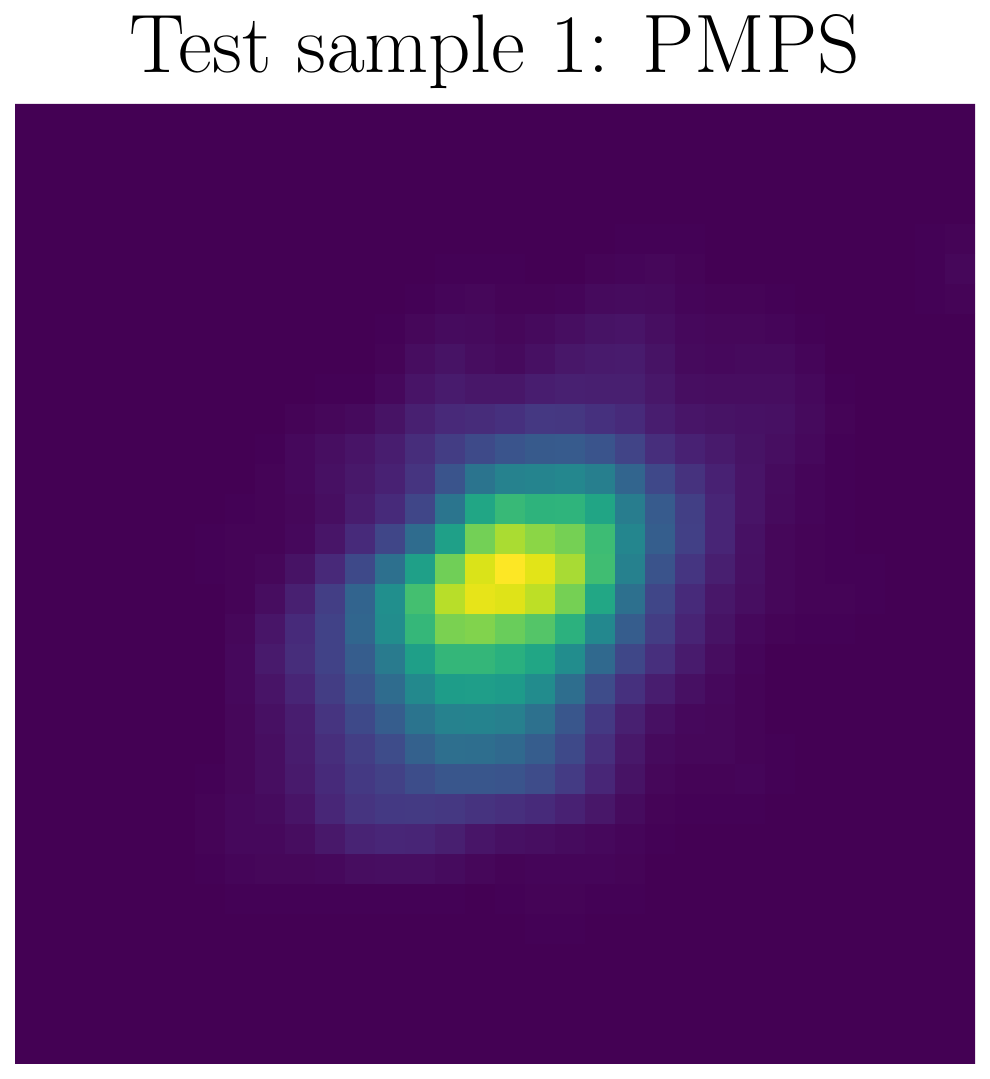}
\hspace{1cm}
\includegraphics[height=0.61\columnwidth]{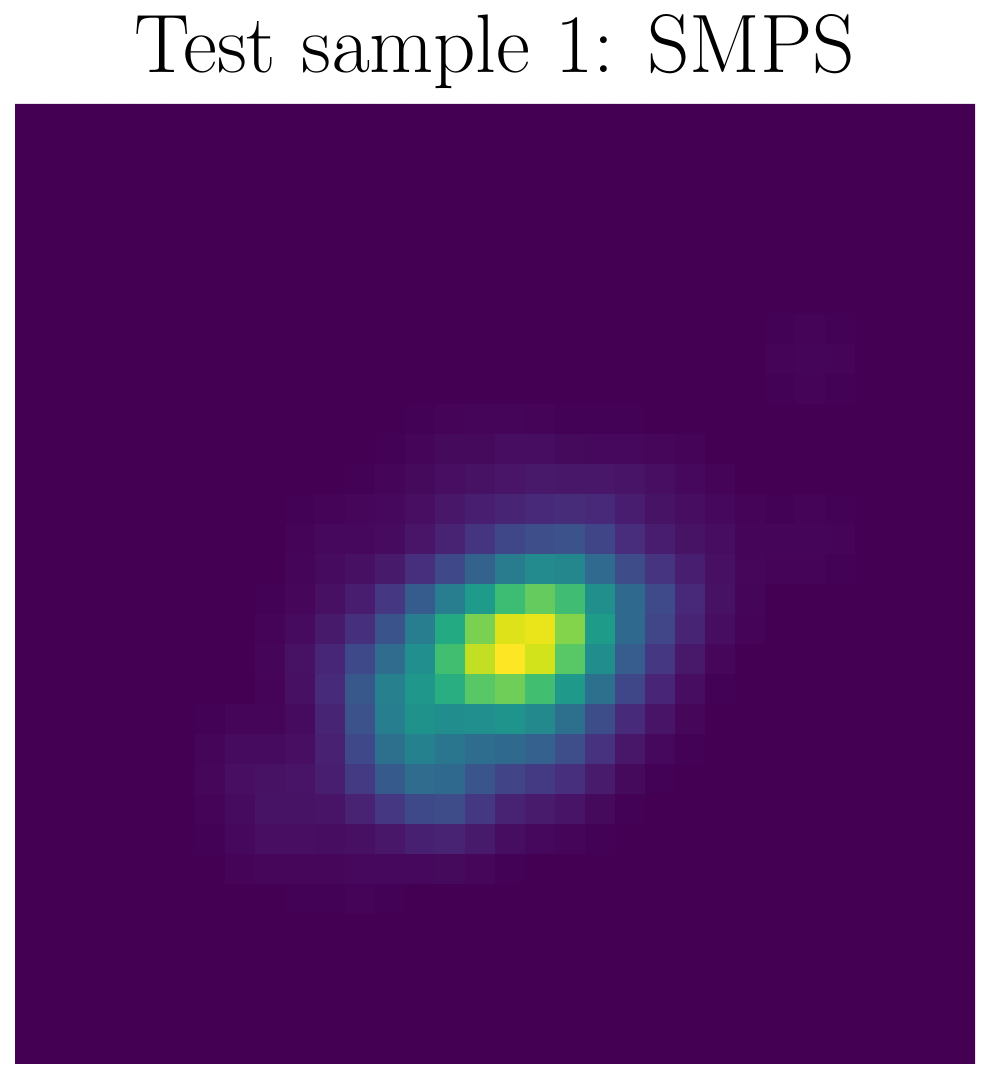}
\hspace{1cm}
\includegraphics[height=0.61\columnwidth]{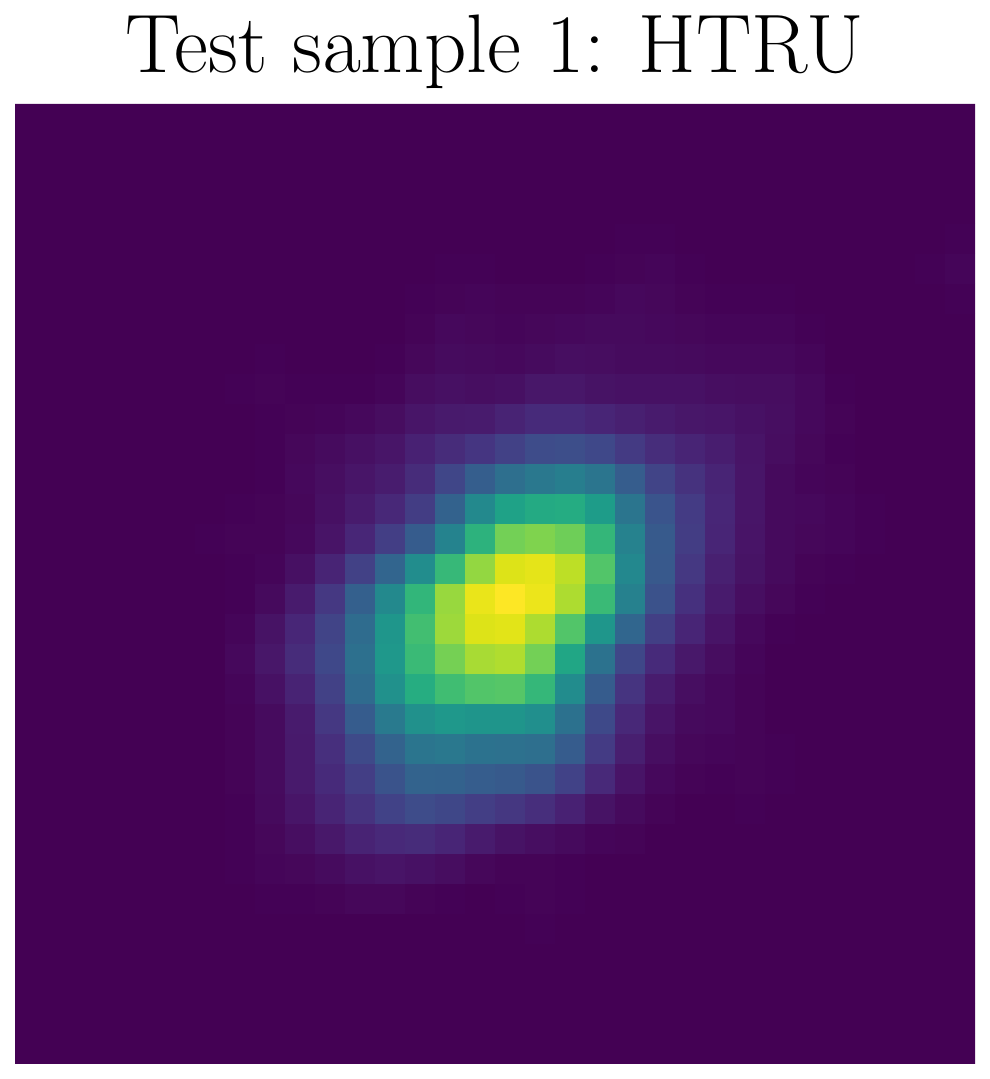}
\caption{Examples of simulated pulsar populations and the corresponding density maps, which are fed into the \ac{SBI} pipeline. The top row shows synthetic $P$--$\dot{P}$ diagrams for the three surveys considered in this study generated from three random sets of magnetorotational parameters. In particular, test sample 1 (top left) is the result of a simulation with $\mu_{\log B} \approx 13.19, \sigma_{\log B} \approx 0.96, \mu_{\log P} \approx -0.85 , \sigma_{\log P} \approx 0.51$ and $a_{\rm late} \approx -0.86$, while test sample 2 (top middle) was generated with $\mu_{\log B} \approx 13.86, \sigma_{\log B} \approx 0.88, \mu_{\log P} \approx -0.42 , \sigma_{\log P} \approx 0.61$, and $a_{\rm late} \approx -1.71$. Finally, test sample 3 (top right) corresponds to $\mu_{\log B} \approx 13.35, \sigma_{\log B} \approx 0.24, \mu_{\log P} \approx -1.25 , \sigma_{\log P} \approx 0.60$, and $a_{\rm late} \approx -2.38$. The bottom row shows the three density maps (one for each survey) generated with a resolution of 32 from the $P$--$\dot{P}$ diagram for test sample 1. Here dark blue encodes regions where no neutron stars are present, while yellow bins represent the largest density for the binned pulsar distribution.}
\label{fig:pop_simulated}
\end{figure*}

\subsection{Simulation Output}
\label{sec:sim_output}

To simulate our \textit{mock} observed pulsar populations, we do not make any assumptions on the neutron star birth rate. Instead, we randomly sample a subset of $10^5$ neutron stars from our dynamical database (see Section~\ref{sec:dyn_evol}). We subsequently evolve these stars magnetorotationally as outlined in Section~\ref{sec:mr_evol} and assess how many of them are detected by each of the three surveys (see Sections~\ref{sec:emission_phys} and \ref{sec:obs_lims}), saving their respective properties. We iterate this process until the number of detected stars matches the number of observed objects in all surveys. Note that we adaptively reduce the number of stars we draw from our dynamical database to $10^4$ and $5 \times 10^3$, once we have recovered 90\% and 95\% of the target values, respectively. The output of a single simulator run, which has a typical computation time of around $\unit[1]{hr}$, is a data frame containing the properties of those pulsars we can detect with \acs{PMPS}, \acs{SMPS}, and \acs{HTRU}, respectively. 

The location of the resulting synthetic population and the shape of the stars' distribution in the $P$--$\dot{P}$ plane are directly controlled by the magnetorotational parameters, $\mu_{\log B}, \sigma_{\log B}, \mu_{\log P}, \sigma_{\log P}$, and $a_{\rm late}$, the five parameters we want to infer. Three examples of synthetic $P$--$\dot{P}$ diagrams are shown in the top row of Figure~\ref{fig:pop_simulated}. 

We note that our prescription does not rely on a \textit{by-hand} implementation of a \textit{pulsar death line} \citep[e.g.,][]{Bhattacharya1992, Chen1993, Rudak1994, Zhang2000}, beyond which pulsar emission ceases, as done in most previous population synthesis studies \citep[e.g.,][]{Faucher2006, Bates2014, Cieslar2020}. We opt for this approach due to significant uncertainties around the radio emission process generally associated with the production of electron--positron pairs in pulsar magnetospheres above the polar caps \citep{Ruderman1975}. In particular, different assumptions on magnetic field strengths and geometries, pair production, and stellar properties (like mass and radius) lead to different death lines, effectively expanding into a \textit{death valley}. We thus avoid adopting a somewhat arbitrary choice for a single death line. In our simulations, pulsars instead become undetectable naturally if they approach the bottom right of the $P$--$\dot{P}$ plane. This is due to the evolution toward (i) smaller misalignment angles, $\chi$, resulting in smaller beaming fractions, and (ii) smaller $\dot{P}$ (and thus lower $|\dot{E}_{\rm rot}|$), ultimately leading to sources that are too faint to be detected.

At this point, we also highlight that our approach provides information on the number of total stars generated over a timescale of $\unit[10^8]{yr}$ (the oldest possible age for stars in our dynamical database), implying that we can directly determine the birth rate required to reproduce observations for a given survey. Although not the primary focus of this work, we note two things here: first, the number of detectable neutron stars per iteration step described above and, thus, the birth rate (as well as the distribution of stars in the $P$--$\dot{P}$ plane) depend strongly on the five magnetorotational parameters. For some parameter combinations, reaching the counts in Equation~\eqref{eqn:detected_objects} requires unrealistically large birth rates, and thus extensive computation time. To mitigate this issue, we stop our iterative simulation approach once the birth rate exceeds a conservative limit of five neutron stars per century \citep{Keane2008, Rozwadowska2021} even though this implies that we do not reach the numbers of observed objects in these simulations. We, however, still use these simulations in the following to assess whether our inference approach can identify those parameter combinations that require birth rates $\gtrsim 5$ as unreasonable from the distribution of stars in the $P$--$\dot{P}$ plane alone. Second, for a single simulation run, we generally do not obtain the same birth rate for all three surveys, and estimates can differ by a factor of $\sim 1-3$ neutron stars per century. In principle, we only expect the \textit{correct} physical simulator to produce the observed distributions of pulsars across different surveys. The correct simulation framework is, however, not known, and constraining the relevant physics is the main goal of our analysis. To explore this behavior, we thus produce neutron stars until the target values in all three surveys are reached (or exceeded). While this implies that the number of detected objects in some simulations can be larger than the observed number of stars for a given survey (by up to a factor of $\sim 3$), our focus on the location and shape of the distribution of pulsars in $P$ and $\dot{P}$ and not their total number (see below) circumvents this issue. We will, however, return to the issue of the birth rate in the discussion in Section~\ref{sec:NS_birthrate}, once we have explained our inference approach and provided results for our best estimates.

To provide a broad range of synthetic $P$--$\dot{P}$ diagrams for our inference pipeline, we explore the ranges outlined in Section~\ref{sec:mr_evol} and uniformly sample random combinations of the five parameters as follows:
\begin{align}
    \mu_{\log B} &\in \mathcal{U}(12, 14), \nonumber \\
    \sigma_{\log B} &\in \mathcal{U}(0.1, 1), \nonumber \\
        \mu_{\log P} & \in \mathcal{U}(-1.5, -0.3), \label{eqn:priors} \\
    \sigma_{\log P} &\in \mathcal{U}(0.1, 1), \nonumber  \\
    a_{\rm late} &\in \mathcal{U}(-3, -0.5) \nonumber.
\end{align}
We generate a total of 360,000 parameter combinations (which we refer to as our input parameters, labels, or ground truths below) and simulate the corresponding synthetic populations in parallel over the course of 6 weeks.

To represent the discrete output of our simulator in a way that can be processed by a neural network, we convert a single $P$--$\dot{P}$ diagram for three surveys as seen in the top row of Figure~\ref{fig:pop_simulated} into three two-dimensional density maps (one for each survey) by counting the number of stars within a given bin. In particular, we set the limits $P \in [0.001, 100]\, {\rm s}$ and $\dot{P} \in [10^{-21}, 10^{-9}] \, {\rm s \, s^{-1}}$ and test our inference procedure for a resolution of 32 and 64 bins. To avoid sharp edges in our binned distributions, we apply a smoothing Gaussian filter (with radius $4 \sigma$ and $\sigma=1$), which will also improve the stability during the training of our machine-learning pipeline. An example of the resulting density maps is shown in the bottom row of Figure~\ref{fig:pop_simulated} for one of our test simulations. 

The final preprocessing stage for our simulated data is either a normalization or a standardization step (depending on the choice of setup discussed below) to provide the neural network with signals and labels of similar magnitude. In the former case, the bins in each individual density map are rescaled such that they contain continuous values between $0$ and $1$. The same holds for the corresponding labels, which are normalized over the entire parameter ranges given in Equation~\eqref{eqn:priors}. On the other hand, standardization is achieved by using $z$-scores, so that the resulting information in each map has a mean of $0$ and standard deviation of $1$. The same method is applied to the labels across our entire set of simulations.


\begin{figure*}
\centering
\includegraphics[width = \textwidth]{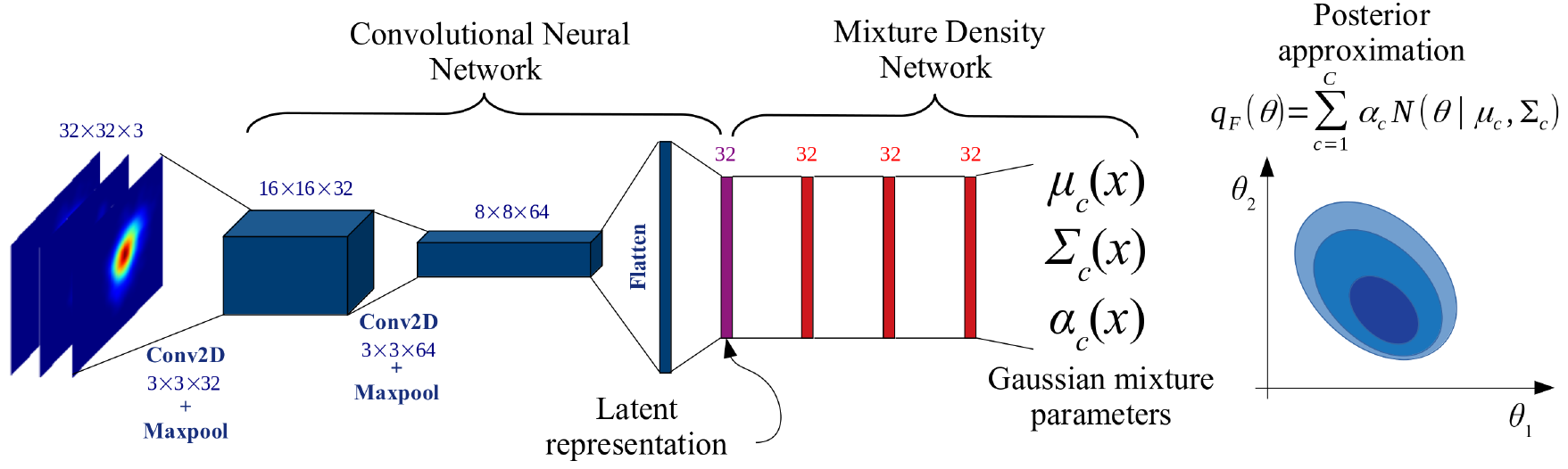}
\caption{Schematic representation of our inference pipeline for three input $P$--$\dot{P}$ maps (one for each survey) with resolution $32 \times 32$. A \ac{CNN} is first used to extract features from our images and produce a compressed representation of our simulation output, $\bx$. We then train a Gaussian \ac{MDN}, a flexible neural density estimator, on this latent representation to approximate the posterior distribution of the simulation input parameters, $\bt$.}
\label{fig:sbi_architecture}
\end{figure*}

\section{Simulation-based inference}
\label{sec:sbi}

\subsection{Overview}
\label{sec:sbi_overview}

The pulsar population synthesis pipeline summarized in Section~\ref{sec:popsyn} is a typical example of a stochastic forward model that aims to emulate real-world observations. We specifically introduced stochasticity by sampling relevant variables from underlying probability distributions using Monte Carlo techniques. In particular, given the input parameter, $\bt = \{ \theta_1, \theta_2, \dots\}$, our simulator generates a synthetic realization of the observed data, $\bx$. The key challenge is then to constrain our model parameters in such a way that they are consistent with true observations, $\bx_0$, and our prior knowledge, encoded in the prior distribution, $\pr(\bt)$. To this end, we want to compute the posterior distribution, $\pr(\bt | \bx)$, using Bayes's theorem
\begin{equation}
	\pr(\bt | \bx) = \frac{\pr(\bt) \pr(\bx |\bt)}{\pr(\bx)},
		\label{eqn:Bayes_theorem}
\end{equation}
where $\pr(\bx |\bt)$ is the likelihood of our data, $\bx$, given the parameter, $\bt$, and
\begin{equation}
	\pr(\bx) \equiv \int \pr(\bx |\bt') \pr(\bt') \, d \bt' ,
		\label{eqn:evidence}
\end{equation}
denotes the evidence obtained by marginalizing over all $\bt$. However, for complex simulators like ours, we typically cannot write down an explicit form of the likelihood function, so $\pr(\bx |\bt)$ is essentially intractable. In addition, even if the likelihood were tractable, Equation~\eqref{eqn:evidence} involves an integral over $\bt$, which becomes challenging for simulators with high-dimensional parameter spaces. 

\ac{SBI} circumvents these issues by taking advantage of the fact that our simulator encodes the likelihood function implicitly \citep[see][for a recent review]{Cranmer2020}. These approaches have been particularly successful in combination with deep-learning techniques because neural networks can be used to learn a probabilistic association between a given simulation outcome, $\bx$, and the input parameters, $\bt$. This allows an approximation of the posterior distribution, $\pr(\bt |\bx)$, without the need to explicitly compute the likelihood. Three approaches exist to achieve this goal:
\begin{itemize}
	\item \Ac{NPE}: The network learns to directly map the simulator output, $\bx$, onto the posterior distribution, $\pr(\bt |\bx)$, for the underlying parameters, $\bt$. This requires the use of a flexible neural density estimator such as a \textit{normalizing flow} or a \acf{MDN} \citep[e.g.][]{Papamakarios2016, Lueckmann2017, Greenberg2019, Dax2021, Mishra-Sharma2022, Hahn2023, Vasist2023}.
	\item \Ac{NLE}: The network emulates the simulator by learning an association between $\bt$ and $\bx$, thus providing direct access to an approximation of the likelihood, $\pr(\bx |\bt)$. Because the prior is known, the posterior can then be obtained by an additional \ac{MCMC} sampling step \citep[e.g.,][]{Papamakarios2018, Alsing2019}.
	\item \Ac{NRE}: Here the network learns the likelihood-to-evidence ratio, $r(\bt, \bx) \equiv \pr(\bx |\bt) / \pr (\bx)$, which is equivalent to $\pr(\bt  |\bx) / \pr (\bt)$ using Bayes's theorem (Equation~\eqref{eqn:Bayes_theorem}). Once $r(\bt, \bx)$ is known, the posterior can be recovered through \ac{MCMC} by sampling the prior weighted by the ratio, $r(\bt, \bx)$ \citep[e.g.,][]{Hermans2019, Miller2021, Bhardwaj2023}.
\end{itemize}

For the following study, we choose an \ac{NPE} approach to directly learn the posterior conditional on our simulated data (avoiding the additional sampling step required for \ac{NLE} and \ac{NRE}) and take advantage of the corresponding implementation in the open-source Python package {\tt sbi} \citep{Tejero-Cantero2020}.\footnote{\url{https://github.com/sbi-dev/sbi}}


\begin{deluxetable*}{c|cccc|ccLc|RcRR}[t]
	\tablecaption{Information for the 22 Machine-learning Experiments Conducted for This Study. \label{tab:experiments}}
	\tabletypesize{\small}
	\tablenum{2}
	\tablehead{
		\colhead{No.} &
		\colhead{Res} &
		\colhead{Surveys} &
		\colhead{Frac ($\%$)} & 
		\colhead{Input} & 
		\colhead{Comp} &
		\colhead{BS} &
		\colhead{LR} &
		\colhead{CNN} & 
		\colhead{VM} & 
		\colhead{Epochs} & 
		\colhead{Time (s)} & 
		\colhead{TM}
	}
	\startdata
	\textbf{1}	& 32			&  PMPS, SMPS, HTRU	&	100		& std			& 10			& 8			&  0.0005			& baseline		& 3.65	& 38		& 9,373	& 3.64	\\
	\textbf{2}	& 32			&  PMPS, SMPS, HTRU	&	100		& std			& 10			& 8			&  0.0005			& \textbf{deep}	& 3.71	& 49		& 14,292	& 3.71	\\
	\textbf{3}	& \textbf{64}	&  PMPS, SMPS, HTRU	&	100		& std			& 10			& 8			&  0.0005			& baseline		& 3.55	& 55		& 78,837	& 3.54	\\
	\textbf{4}	& \textbf{64}	&  PMPS, SMPS, HTRU	&	100		& std			& 10			& 8			&  0.0005			& \textbf{deep}	& 3.64	& 89		& 128,119	& 3.64	\\
	\textbf{5}	& 32			&  PMPS, SMPS, HTRU	&	\textbf{75}	& std			& 10			& 8			&  0.0005			& baseline		& 3.74	& 71		& 13,232	& 3.78	\\
	\textbf{6}	& 32			&  PMPS, SMPS, HTRU	&	\textbf{50}	& std			& 10			& 8			&  0.0005			& baseline		& 3.56	& 58		& 7,000	& 3.55	\\
	\textbf{7}$\star$	& 32			&  PMPS, SMPS, HTRU	&	100		& \textbf{norm}	& 10			& 8			&  \textbf{0.01}		& baseline		& 3.47	& 30		& 7,445	& 3.73	\\
	\textbf{8}	& 32			&  PMPS, SMPS, HTRU	&	100		& \textbf{norm}	& 10			& 8			&  \textbf{0.001}	& baseline		& 9.66	& 54		& 13,015	& 9.60	\\
	\textbf{9}	& 32			&  PMPS, SMPS, HTRU	&	100		& std			& \textbf{8}	& 8			&  0.0005			& baseline		& 3.74	& 52		& 12,389	& 3.73	\\
	\textbf{10}	& 32			&  PMPS, SMPS, HTRU	&	100		& std			& \textbf{5}	& 8			&  0.0005			& baseline		& 3.83	& 118	& 27,973	& 3.86	\\
	\textbf{11}	& 32			&  PMPS, SMPS, HTRU	&	100		& std			& 10			& \textbf{16}	&  0.0005			& baseline		& 3.99	& 85		& 10,476	& 3.97	\\
	\textbf{12}	& 32			&  PMPS, SMPS, HTRU	&	100		& std			& 10			& \textbf{32}	&  0.0005			& baseline		& 4.11	& 79		& 5,346	& 4.06	\\
	\textbf{13}	& 32			&  PMPS, SMPS, HTRU	&	100		& std			& 10			& 8			&  \textbf{0.001}	& baseline		& 3.36	& 61		& 14,785	& 3.33	\\
	\textbf{14}	& 32			&  PMPS, SMPS, HTRU	&	100		& std			& 10			& 8			&  \textbf{0.0001}	& baseline		& 4.22	& 75		& 18,369	& 4.22	\\
	\textbf{15}	& 32			&  \textbf{HTRU}		&	100		& std			& 10			& 8			&  0.0005			& baseline		& 3.43	& 63 		& 15,568	& 3.42	\\
	\textbf{16}	& 32			&  \textbf{SMPS, HTRU}	&	100		& std			& 10			& 8			&  0.0005			& baseline		& 3.58	& 40		& 9,979	& 3.59	\\
	\textbf{17}	& 32			&  \textbf{PMPS, SMPS}	&	100		& std			& 10			& 8			&  0.0005			& baseline		& 3.41	& 69		& 16,937	& 3.41	\\
	\textbf{18}$\star$	& \textbf{64}	&  PMPS, SMPS, HTRU	&	\textbf{50}	& std			& 10			& 8			&  0.0005			& baseline		& 3.45	& 47		& 5,766	& 3.44	\\
	\textbf{19}	& 32			&  PMPS, SMPS, HTRU	&	100		& \textbf{norm}	& 10			& \textbf{32}	&  \textbf{0.001}	& baseline		& 10.05	& 44		& 2,864	& 10.20	\\
	\textbf{20}	& 32			&  PMPS, SMPS, HTRU	&	100		& \textbf{norm}	& 10			& \textbf{32}	&  \textbf{0.0001}	& baseline		& 10.31	& 90		& 5,815	& 10.49	\\\
	\textbf{21}	& 32			&  PMPS, SMPS, HTRU	&	100		& \textbf{norm}	& 10			& \textbf{16}	&  \textbf{0.001}	& baseline		& 9.82	& 77		& 9,901	& 9.98	\\
	\textbf{22}$\star$	& 32			&  PMPS, SMPS, HTRU	&	100		& \textbf{norm}	& 10			& \textbf{16}	&  \textbf{0.0001}	& baseline		& 10.45	& 124	& 15,603	& 10.55	\\
	\enddata
\tablecomments{The columns summarize the specific training data and hyperparameters, as well as the resulting metrics: the experiment number; the resolution for our $P$--$\dot{P}$ density maps; the different surveys and the fraction of the $291,600$ populations in the training set used for training; information on whether we standardized (std) or normalized (norm) the input; the number of Gaussian components in our \ac{MDN}; the batch size; the learning rate; the CNN architecture (we distinguish our baseline setup and a deeper network; see Sections~\ref{sec:architecture} and \ref{sec:experiments} for details); the best metric computed over the validation set; the number of training epochs; the time it took to train the network in seconds and the average metric computed over our $3,600$ test samples. In bold, we highlight those parameters that we have varied with respect to the baseline experiment $\#1$. Experiments with an asterisk ($\star$) are removed from the following analysis due to training irregularities.}
\end{deluxetable*}

\subsection{Deep-learning Setup}
\label{sec:architecture}

For \ac{NPE}, we approximate the posterior using a family of densities, $q_{\bp}$, characterized by the distribution parameters, $\bp$. For our \ac{SBI} pipeline, we then use a neural network, $F$, to learn these $\bp$ for our simulator output, $\bx$, by adjusting the network weights, $\boldsymbol{\phi}$. In particular, we aim to optimize the neural density estimator such that $q_{F(\bx, \boldsymbol{\phi})}(\bt) \approx \pr (\bt | \bx)$. This can be achieved by minimizing the Kullback--Leibler divergence, $D_{\rm KL}(\pr_1 || \pr_2)$, which is a measure of the difference between two probability distributions, $\pr_1$ and $\pr_2$ \citep{Kullback1951}. \citet{Papamakarios2016} showed that this is equivalent to minimizing the expectation value of the loss function
\begin{equation}
	\mathcal{L}(\boldsymbol{\phi}) = - \sum_{i=1}^N \log q_{F(\bx_i, \boldsymbol{\phi})} (\bt_i)
		\label{eqn:loss}
\end{equation}
over a training data set $\{\bt_i, \bx_i\}$ of size $N$, provided that $N$ is large and the density estimator is sufficiently flexible. In practice, we maximize the negative of $\mathcal{L}(\boldsymbol{\phi})$, i.e., the total log-posterior. A key advantage of the resulting posterior approximation is that the evaluation of $q_{F(\bx, \boldsymbol{\phi})}(\bt)$ corresponds to a simple forward pass through a neural network (without the need to simulate additional data), which is very fast. We will take advantage of this \textit{amortized nature} of the posterior to assess the quality of our inferences below. 

For our pulsar study, we have drawn the model parameters $\bt_i = \{\mu_{\log B}, \sigma_{\log B}, \mu_{\log P}, \sigma_{\log P}, a_{\rm late}\}$ from uniform priors as defined previously in Equation~\eqref{eqn:priors}. The corresponding output, $\bx_i$, of a single run through the simulator are the three $P$--$\dot{P}$ density maps (one for each survey) illustrated in the bottom row of Figure~\ref{fig:pop_simulated}. In the following, we stack these maps together to form a three-channel input for our neural network. Of the 360,000 synthetic simulations produced, we use 90\% for training and validation, reserving the remaining 10\% for testing purposes. The former data set is further split into 90\% for training (291,600 populations) and 10\% for validation (32,400 populations). We note that as each population is represented by three density maps, we train the following inference pipeline on roughly 875,000 images. Performance results for the unseen test samples quoted in the following are computed for 10\% of the full test set (3600 populations) for computational reasons. The full workflow is illustrated schematically in Figure~\ref{fig:sbi_architecture}. 

Due to the complexity of these data, we do not train a neural density estimator directly on the density maps. We instead first apply a \ac{CNN} to extract features from our images and embed the corresponding information in a lower-dimensional latent vector. We choose the following baseline architecture for our embedding network:
\begin{itemize}
    \item Two-dimensional convolution layer with kernel size $3 \times 3$, 3 input channels, 32 output channels, stride 1, padding 1.
    \item Two-dimensional Max pooling layer with size $2 \times 2$, stride 2, no padding.
    \item Two-dimensional convolution layer with kernel size $3 \times 3$, 32 input channels, 64 output channels, stride 1, padding 1.
    \item Two-dimensional Max pooling layer with size $2 \times 2$, stride 2, no padding.
    \item Fully connected linear layer with the flattened output from the second pooling layer as input and 32 output neurons encoding the latent representation.
\end{itemize}
After each convolution and the fully connected layer, we apply a \acf{ReLU} activation function. The weights for the \ac{CNN} are initialized using the Kaiming prescription \citep{Kaiming2015} to avoid exploding or vanishing gradients during the training process.

We subsequently pass the latent vector generated by the \acs{CNN} to a neural density estimator. We implement an \ac{MDN} and specifically opt for a Gaussian mixture model in five dimensions to approximate the posterior, $q_{F(\bx, \boldsymbol{\phi})}(\bt)$, for our five free magnetorotational parameters. This implies
\begin{equation}
	q_{F(\bx, \boldsymbol{\phi})}(\bt) = \sum_{c = 1}^{C} \alpha_c \, \mathcal{N} (\bt | \boldsymbol{\mu}_c, \boldsymbol{\Sigma}_c ),
		\label{eqn:posterior_gmm}
\end{equation}
where $C$ denotes the total number of Gaussian components used, $\alpha_{c}$ is the mixture weight, and $\mathcal{N}(\bt | \boldsymbol{\mu}_c,\boldsymbol{\Sigma}_c)$ the multivariate Gaussian distribution with mean vector $\boldsymbol{\mu}_c$ and covariance matrix $\boldsymbol{\Sigma}_c$ for the $c$th component.

For our \ac{MDN}, we follow {\tt sbi}'s default implementation and use the following:
\begin{itemize}
    \item Three fully connected layers with 32 neurons each.
    \item Four fully connected output layers that encode the Gaussian mixture weights, $\alpha_{c}$, means, $\boldsymbol{\mu}_c$, and diagonal and upper triangular components of the covariance matrices, $\boldsymbol{\Sigma}_c$. These contain $c$, $5 c$, $5 c$ and $10 c$ neurons, respectively.
\end{itemize}
We again apply the \ac{ReLU} activation function after each hidden layer, while weights are now initialized with PyTorch's default initialization \citep{Glorot2010}.

We subsequently train the entire pipeline using the gradient descent optimizer Adam \citep{Kingma2014}. At each epoch the network undergoes a series of optimization steps based on the information provided in the entire training data set before epoch-averaged training and validation metrics are computed based on the negative losses defined in Equation~\eqref{eqn:loss}, i.e., we maximize our metrics. Note that we also set an early stop of 20 to prevent overfitting, which implies that the training process is interrupted (and the weights of the best validation epoch recorded) once the validation metric has not improved for 20 epochs.

\begin{figure}[t]
	\centering
	\includegraphics[width=0.95\columnwidth]{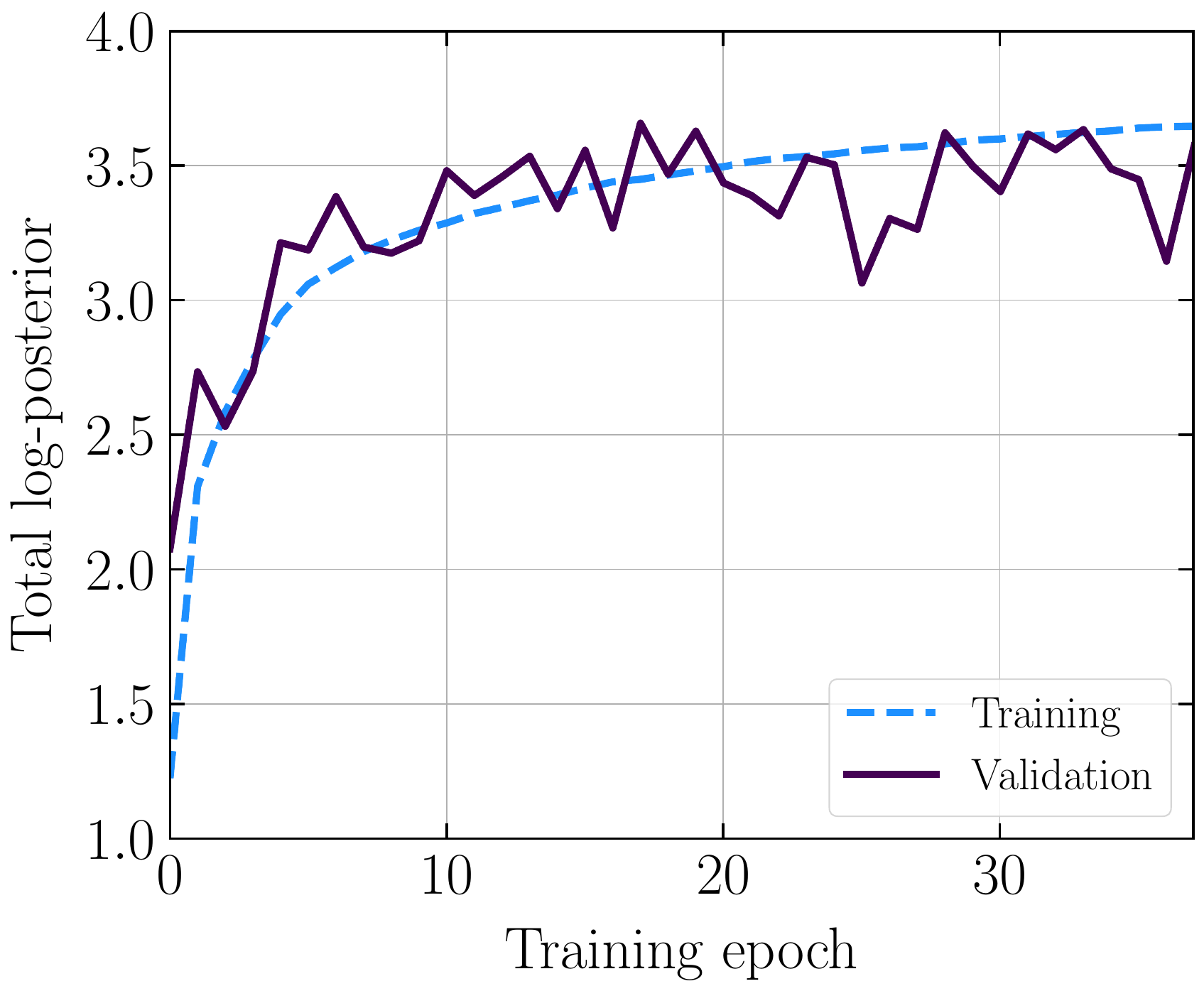}
	\caption{Training behavior for baseline experiment $\#1$. We show the training metric (light-blue dashed line) and the validation metric (purple solid line) as a function of the training epoch. We seek to maximize the total log-posterior, $\sum_{i=1}^N \log q_{F(\bx_i, \boldsymbol{\phi})} (\bt_i)$, over the training and validation data sets, respectively, as the network learns. Both metrics increase as expected, and the validation curve closely tracks the training curve, i.e., we see little overfitting. The best validation metric is reached at epoch $17$, and thus the early stop criterion halts the training after $37$ training epochs.}
	\label{fig:training_curve}
\end{figure}

\begin{deluxetable*}{ccc|DDDD}[t]
	\tablecaption{Magnetorotational Parameters for Three Random Test Samples and the Observed Pulsar Population. \label{tab:credible_intervals}}
	\tabletypesize{\small}
	\tablecolumns{6}
	\tablenum{3}
	\tablewidth{0pt}
	\tablehead{ \multicolumn3c{Parameters} &
		\multicolumn2c{Test Sample $1$} &
		\multicolumn2c{Test Sample $2$} &
		\multicolumn2c{Test Sample $3$} &
		\multicolumn2c{Observed Population}
	}
	\decimals
	\startdata
	\parbox[t]{0mm}{\multirow{5}{*}{\rotatebox[origin=c]{90}{Ground}}}	& \parbox[t]{2mm}{\multirow{5}{*}{\rotatebox[origin=c]{90}{truths, $\bt$}}}	& $\mu_{\log B}$	& $13.19$	& $13.86$	& $13.35$	& -	\\
	& 														& $\sigma_{\log B}$	& $0.96$	& $0.88$	& $0.24$	& -	\\
	& 														& $\mu_{\log P}$	& $-0.85$	& $-0.42$	& $-1.25$	& -	\\
	& 														& $\sigma_{\log P}$	& $0.51$	& $0.61$	& $0.60$	& -	\\
	&														& $a_{\rm late}$	& $-0.86$	& $-1.71$	& $-2.38$	& -	\\
	\hline
	\parbox[t]{0mm}{\multirow{5}{*}{\rotatebox[origin=c]{90}{$95\%$ \ac{CI}}}}	& \parbox[t]{2mm}{\multirow{5}{*}{\rotatebox[origin=c]{90}{experiment $\#1$}}}	& $\mu_{\log B}$	& $13.28^{+0.18}_{-0.18}$	& $13.73^{+0.15}_{-0.15}$	& $13.33^{+0.05}_{-0.04}$	& $13.07^{+0.07}_{-0.08}$	\\
	& 																	& $\sigma_{\log B}$	& $0.95^{+0.08}_{-0.08}$	& $0.79^{+0.07}_{-0.07}$	& $0.23^{+0.02}_{-0.02}$	& $0.43^{+0.03}_{-0.03}$	\\
	& 																	& $\mu_{\log P}$	& $-0.90^{+0.13}_{-0.13}$	& $-0.35^{+0.19}_{-0.18}$	& $-1.17^{+0.33}_{-0.34}$	& $-0.98^{+0.25}_{-0.29}$	\\
	& 																	& $\sigma_{\log P}$	& $0.49^{+0.10}_{-0.09}$	& $0.73^{+0.20}_{-0.15}$	& $0.73^{+0.25}_{-0.31}$	& $0.54^{+0.33}_{-0.25}$	\\
	&																	& $a_{\rm late}$	& $-0.83^{+0.06}_{-0.06}$	& $-1.88^{+0.35}_{-0.35}$	& $-2.47^{+0.43}_{-0.43}$	& $-1.77^{+0.35}_{-0.38}$	\\
	\hline
	\parbox[t]{0mm}{\multirow{5}{*}{\rotatebox[origin=c]{90}{$95\%$ \ac{CI}}}}	& \parbox[t]{2mm}{\multirow{5}{*}{\rotatebox[origin=c]{90}{ensemble}}}		& $\mu_{\log B}$	& $13.29^{+0.20}_{-0.20}$	& $13.74^{+0.19}_{-0.16}$	& $13.34^{+0.05}_{-0.05}$	& $13.10^{+0.08}_{-0.10}$	\\
	& 																	& $\sigma_{\log B}$	& $0.96^{+0.07}_{-0.08}$	& $0.78^{+0.09}_{-0.08}$	& $0.24^{+0.02}_{-0.02}$	& $0.45^{+0.05}_{-0.05}$	\\
	& 																	& $\mu_{\log P}$	& $-0.92^{+0.16}_{-0.15}$	& $-0.40^{+0.20}_{-0.27}$	& $-1.23^{+0.33}_{-0.34}$	& $-1.00^{+0.26}_{-0.21}$	\\
	& 																	& $\sigma_{\log P}$	& $0.49^{+0.10}_{-0.09}$	& $0.74^{+0.20}_{-0.17}$	& $0.67^{+0.30}_{-0.28}$	& $0.38^{+0.33}_{-0.18}$	\\
	&																	& $a_{\rm late}$	& $-0.84^{+0.06}_{-0.07}$	& $-1.76^{+0.39}_{-0.43}$	& $-2.34^{+0.43}_{-0.45}$	& $-1.80^{+0.65}_{-0.61}$	\\
	\enddata
\tablecomments{The first five rows show the ground truths, $\bt$, used to simulate the test populations. The second block gives medians and $95\%$ (CIs) obtained from inferences with the neural network from experiment $\#1$. The final block contains medians and $95\%$ \acp{CI} determined from the ensemble posterior combining $19$ experiments.}
\end{deluxetable*}


\subsection{Experiments}
\label{sec:experiments}

Table~\ref{tab:experiments} summarizes the 22 different experiments that we have conducted for this study to assess the performance of \ac{SBI} for pulsar population synthesis. For this purpose, we varied aspects of the training data, as well as the hyperparameters of our deep-learning pipeline. In particular, for the input we explored two different resolutions for the $P$--$\dot{P}$ maps, 32 and 64, assessed the network performance when all three density maps or only two/one are provided, and examined whether normalization or standardization during preprocessing leads to different results. We further studied the impact of using the full training data set or smaller subsets. Moreover, for the network we varied the number of Gaussian mixture components in our neural density estimator, the batch size, and the learning rate, and we explored two different \acp{CNN} for our embedding net. In addition to the baseline architecture described in Section~\ref{sec:architecture}, we also conducted two experiments with a deeper network composed of four convolutional blocks. Here the two convolutional layers introduced previously are followed by an additional layer with 32 and 64 input/output channels, respectively. Kernel size, stride, padding, subsequent pooling, and fully connected layers were kept as above. 

Due to the computational cost of each training experiment, a full grid search over all relevant configurations was beyond the scope of this work. We therefore opted to produce a representative set of experiments that provide sufficient information to study the variation of our inferred posteriors in Section~\ref{sec:results}. Finally, note that almost all of our optimizations are performed on a Tesla V100 SXM2 GPU with $\unit[32]{GB}$ memory. We only trained experiments $\#3$ and $\#4$, for which the full training data set with a resolution of 64 was too large to be optimized on the GPU, on a CPU with $\unit[32]{GB}$ RAM. In those two cases, training the network thus took markedly longer than for the other experiments (see below).


\section{Results}
\label{sec:results}

\begin{figure}[t]
	\centering
    	\includegraphics[width=1\columnwidth]{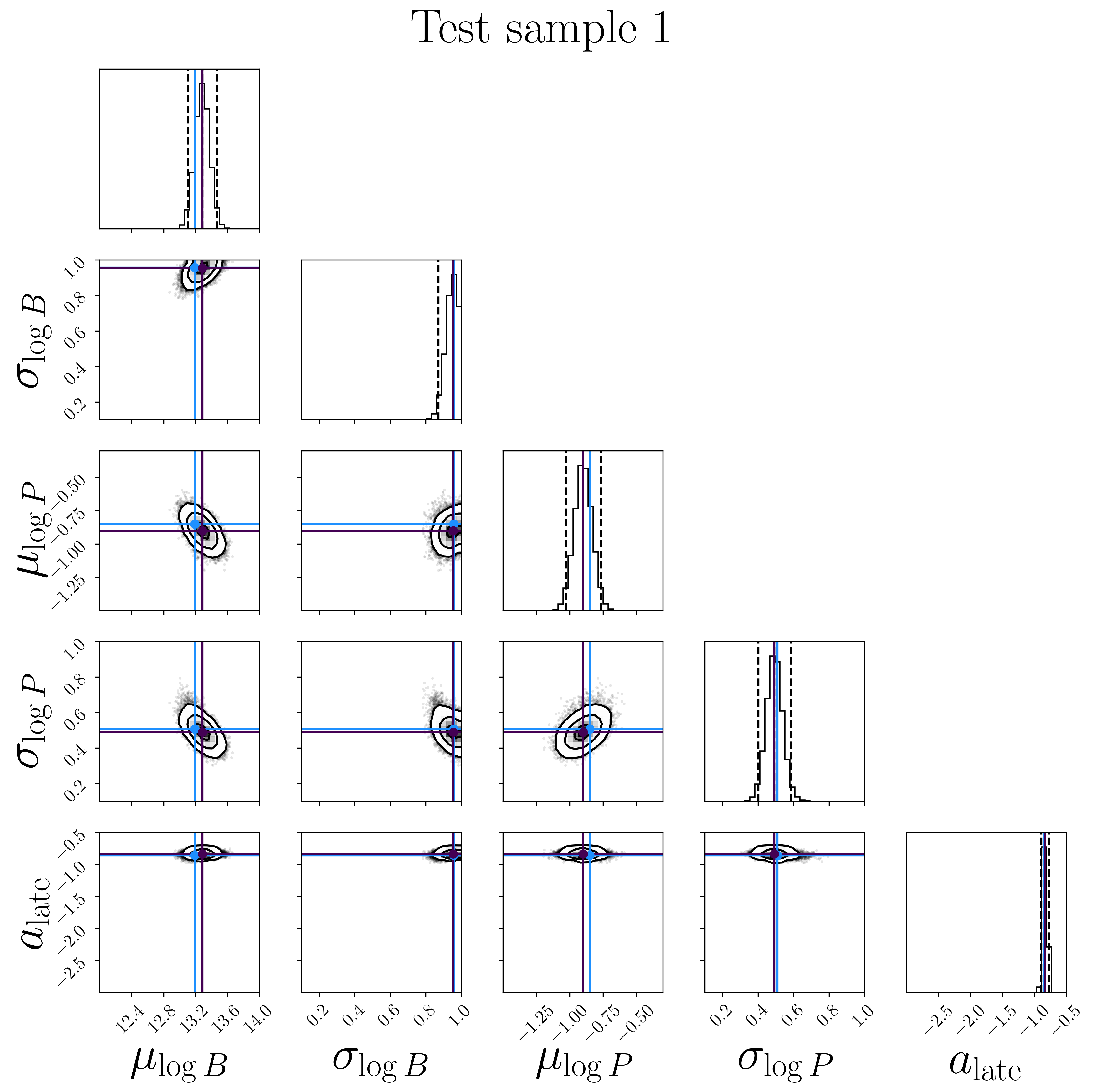}
	\caption{Benchmark inference for test simulation $1$ using the network from experiment $\#1$. The corner plot shows one- and two-dimensional marginal posterior distributions for the five magnetorotational parameters. We also show corresponding ground truths, $\bt$, in light blue, and the medians in purple. We observe that the posteriors cover the $\bt$ well. Corresponding $95\%$ \acp{CI} are summarized in Table~\ref{tab:credible_intervals}.}
	\label{fig:posterior_test12}
\end{figure}

\begin{figure}[t]
	\centering
	\includegraphics[width=1\columnwidth]{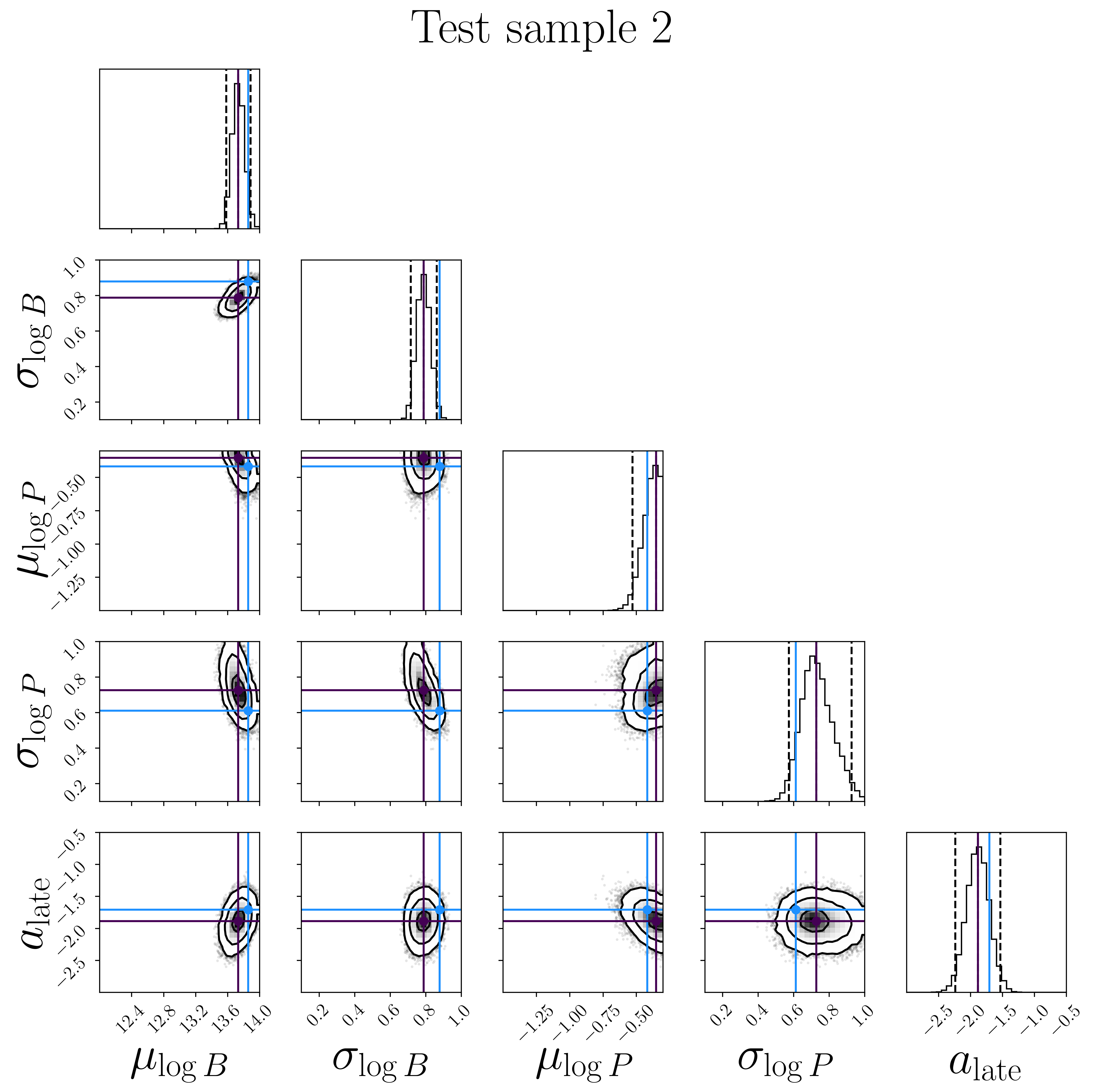}
	\vskip 0.2cm
	\includegraphics[width=1\columnwidth]{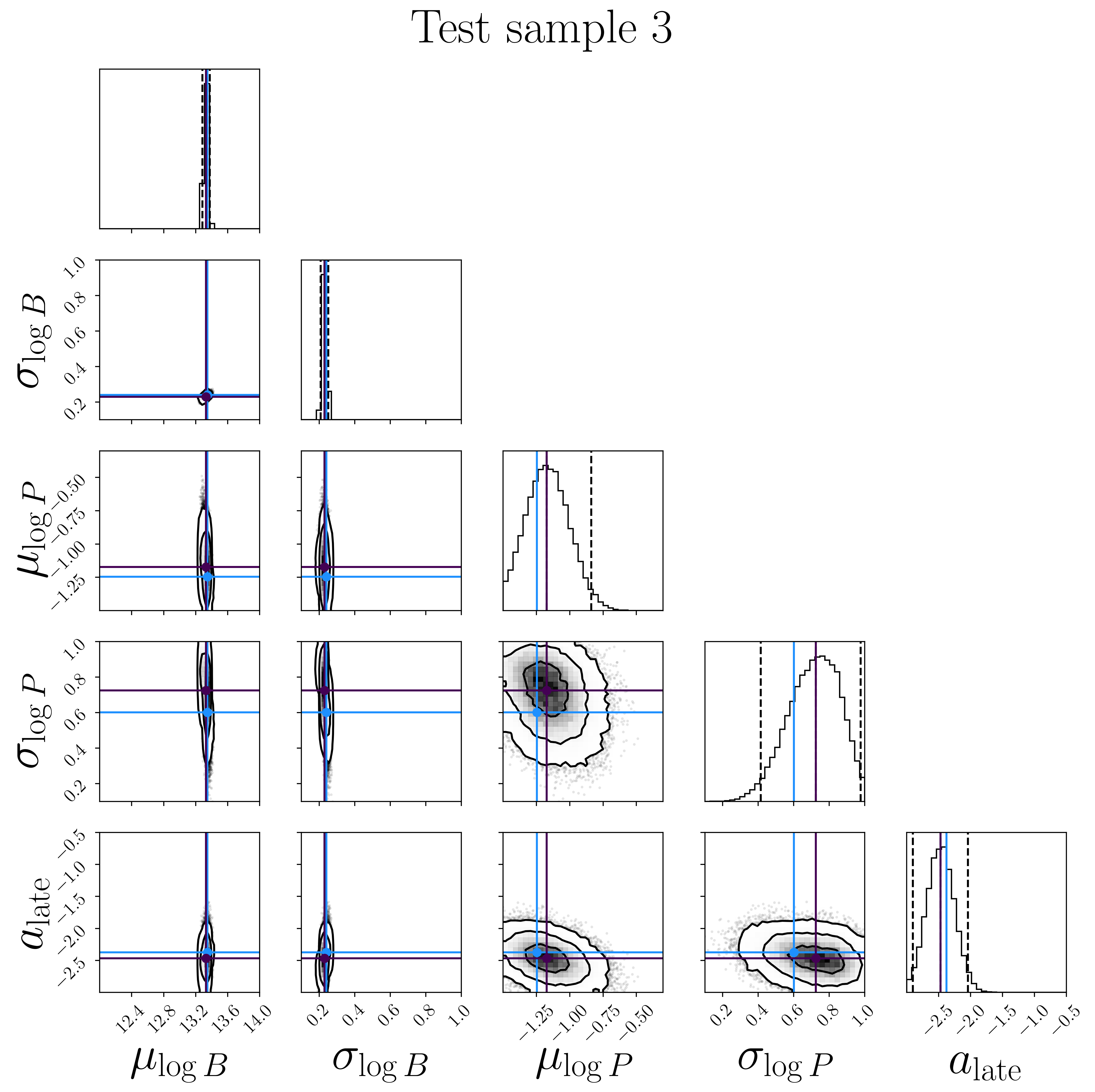}
	\caption{Same as Figure~\ref{fig:posterior_test12} but for test simulations $2$ and $3$.}
	\label{fig:posterior_test3}
\end{figure}


\subsection{Training}
\label{sec:training}

Several metrics for our experiments are summarized in the last four columns of Table~\ref{tab:experiments}. We observe that the optimization of our neural networks take $\sim \unit[1-8]{hr}$ on the GPU and on the order of a day on a CPU, completing $\sim 30- 124$ training epochs. In general, we find good training behavior, with the validation metric closely tracking the training metric and little or no overfitting. This is also evident in the network's generalization ability, illustrated by the average metrics computed over the unseen test set of 3600 simulations. The evolution of the training and validation metrics for experiment $\#1$ is shown in Figure~\ref{fig:training_curve} as an example. We remind the reader that we aim to maximize the total log-posterior. After visual inspection of all training curves, we remove experiment $\#7$ owing to irregularities in the training behavior and experiments $\#18$ and $\#22$ owing to a slight tendency to overfitting. Note that these shortcomings were not directly visible from the training metrics in Table~\ref{tab:experiments}. We also highlight that we find systematically larger training, validation, and test metrics in those experiments where our input density maps were normalized. In the following, however, we assess the quality of the corresponding posteriors and find that these do not result in better inferences. Beyond this difference, we cannot identify any significant variation in the metrics between the remaining configurations. Hence, we proceed with an analysis of all experiments apart from numbers $\#7, \#18$ and $\#22$.


\begin{figure*}
	\centering
	\includegraphics[width=0.92\textwidth]{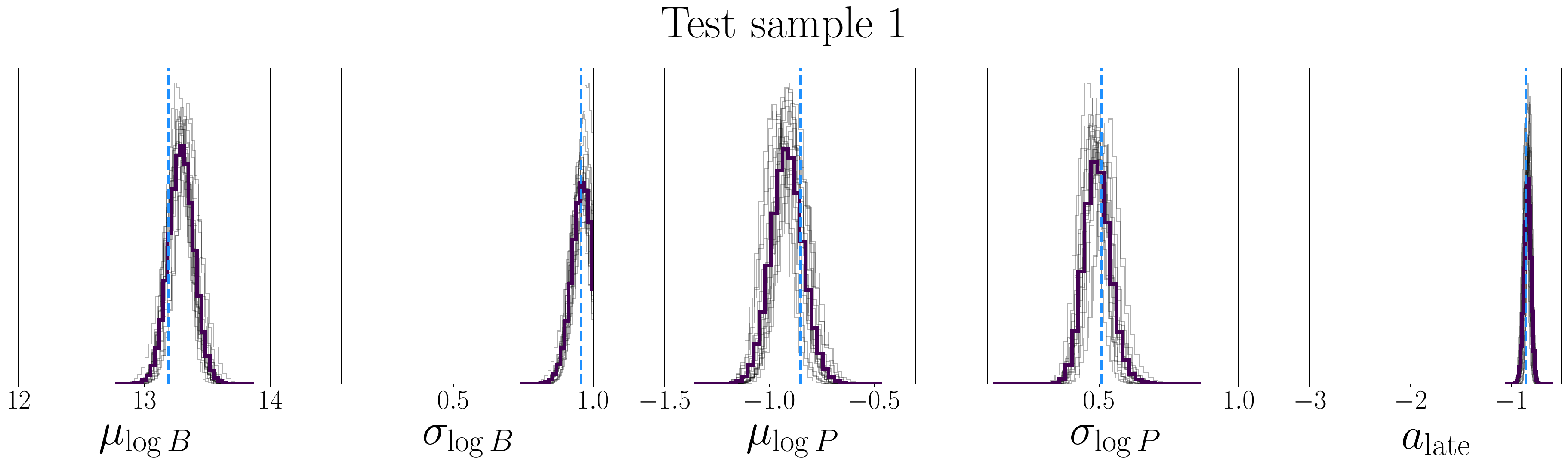}
	\vskip 0.2cm
	\includegraphics[width=0.92\textwidth]{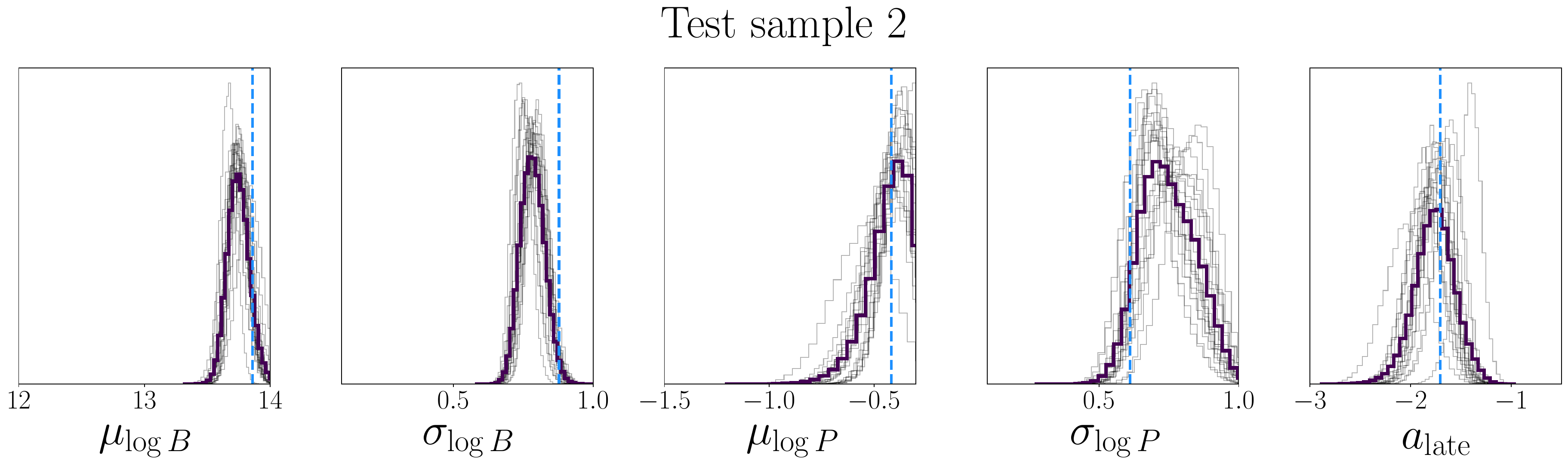}
	\vskip 0.2cm
	\includegraphics[width=0.92\textwidth]{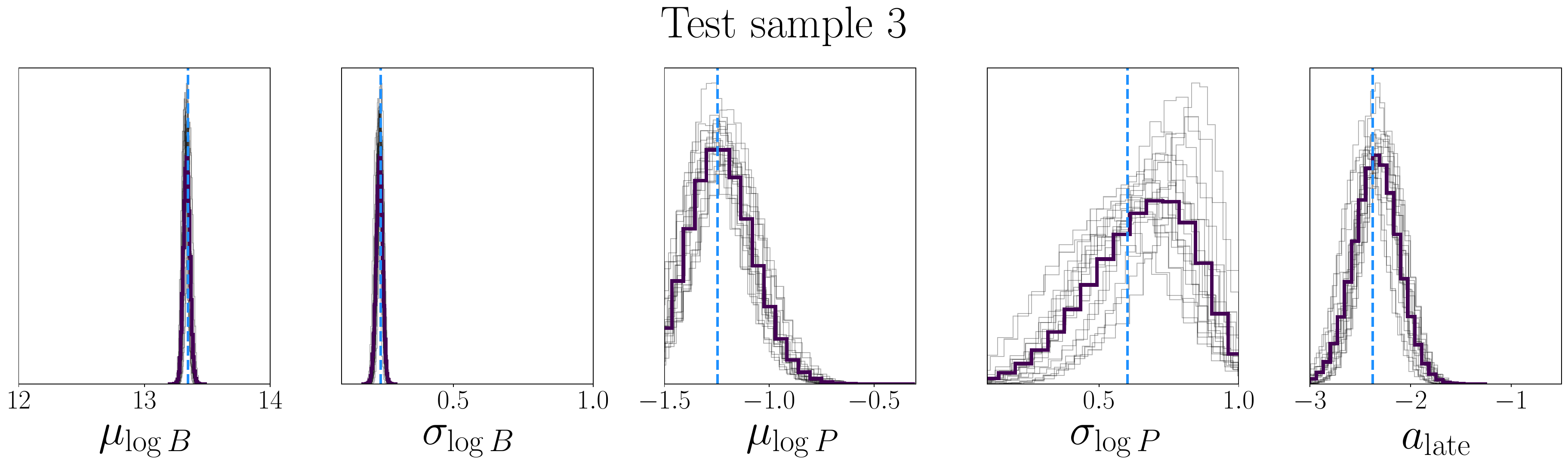}
	\caption{One-dimensional marginal posteriors for the five magnetorotational parameters for the three test simulations inferred using 19 different \ac{NPE} experiments shown in gray. The horizontal axes represent the parameters' prior ranges. The ground truths are shown as vertical light-blue dashed lines. We observe variation between the experiments, specifically for $\mu_{\log P}, \sigma_{\log P}$, and $a_{\rm late}$. We also plot the ensemble posteriors (purple) obtained as a weighted average of the individual posteriors.}
	\label{fig:posterior_comparison}
\end{figure*}

\subsection{Benchmark Inferences}
\label{sec:inferences_test}

As a first assessment of our approximated posteriors, we focus on inferring the five magnetorotational parameters, $\mu_{\log B}, \sigma_{\log B}, \mu_{\log P}, \sigma_{\log P}, a_{\rm late}$, for simulated populations where we know the input parameters, $\bt$. We specifically look at the three simulations, whose $P$--$\dot{P}$ diagrams were illustrated in the top row of Figure~\ref{fig:pop_simulated}. Corresponding ground truths, $\bt$, are summarized in the top five rows in Table~\ref{tab:credible_intervals}. In Figs.~\ref{fig:posterior_test12} and~\ref{fig:posterior_test3}, we show the resulting one- and two-dimensional marginal posterior distributions obtained by repeatedly sampling from the neural network optimized during experiment $\#1$. For all three cases, the posteriors are well-defined, significantly smaller than our prior ranges (Equation~\eqref{eqn:priors}) shown along the axes, and centered around the ground truths, $\bt$, highlighted in light blue. To quantify this, we calculate the $1\sigma, 2\sigma$, and $3\sigma$ credible regions, shown as contours in the two-dimensional posteriors. In the one-dimensional posterior panels, the corresponding $95\%$ credible intervals (\acp{CI}) are given as black dashed lines, while medians are illustrated as purple solid lines. Their numerical values are given in Table~\ref{tab:credible_intervals}. We observe that the ground truths, $\bt$, are typically contained within the $2\sigma$ credible regions, which we interpret as evidence that our \ac{NPE} approach is capable of producing reasonable posterior distributions. In general, the credible regions for the two parameters characterizing the initial magnetic field distribution are narrower than those for the initial period distribution and the late-time magnetic field decay. We confirm that this behavior is qualitatively similar for the remaining $P$--$\dot{P}$ simulations in our test set.

We next compare the inferences for our various training experiments. To visualize corresponding differences, we plot the one-dimensional marginalized posteriors for all $19$ experiments for the three test samples in gray in Figure~\ref{fig:posterior_comparison}. Ground truths, $\bt$, are shown as light-blue dashed lines. We observe that the widths of individual posterior approximations and their medians can vary somewhat between different test samples and magnetorotational parameters. Compared across the full test set, this behavior is again more dominant for the period and late-time magnetic field parameters than for the initial $B$-field properties. However, no individual \acp{NPE} stand out by exhibiting either particularly good or poor posteriors. Further note that we also do not see any differences for those experiments with normalized input maps that showed systematically better metrics than those experiments trained on standardized data. This highlights that training behavior alone does not provide sufficient information on the quality of the resulting inference.

In light of this, we also determine the combined posterior for all $19$ experiments. We calculate the corresponding \textit{ensemble posterior}, $\overline{q} (\bt)$, as the weighted average of the individual posteriors \citep{Hermans2021}:
\begin{equation}
	\overline{q}(\bt) = \sum_j^{19} w_j q_{F_j} (\bt),
\end{equation}
where $w_j$ represents the weight of the $j$th component. Giving equal importance to each experiment in the ensemble, we choose $w_j = 1/19$. The corresponding one-dimensional marginalized ensemble posteriors for $\mu_{\log B},$ $\sigma_{\log B}, \mu_{\log P}, \sigma_{\log P}$, and $a_{\rm late}$ for the three test simulations are illustrated as purple histograms in Figure~\ref{fig:posterior_comparison}. As expected, they fall within the individual posteriors. The corresponding $95\%$ \acp{CI} for the three test samples, which are typically comparable to or slightly wider than those calculated for experiment $\#1$ posteriors alone, are summarized in the bottom five rows of Table~\ref{tab:credible_intervals}.

\begin{figure}[t]
	\centering
	\includegraphics[width=0.95\columnwidth]{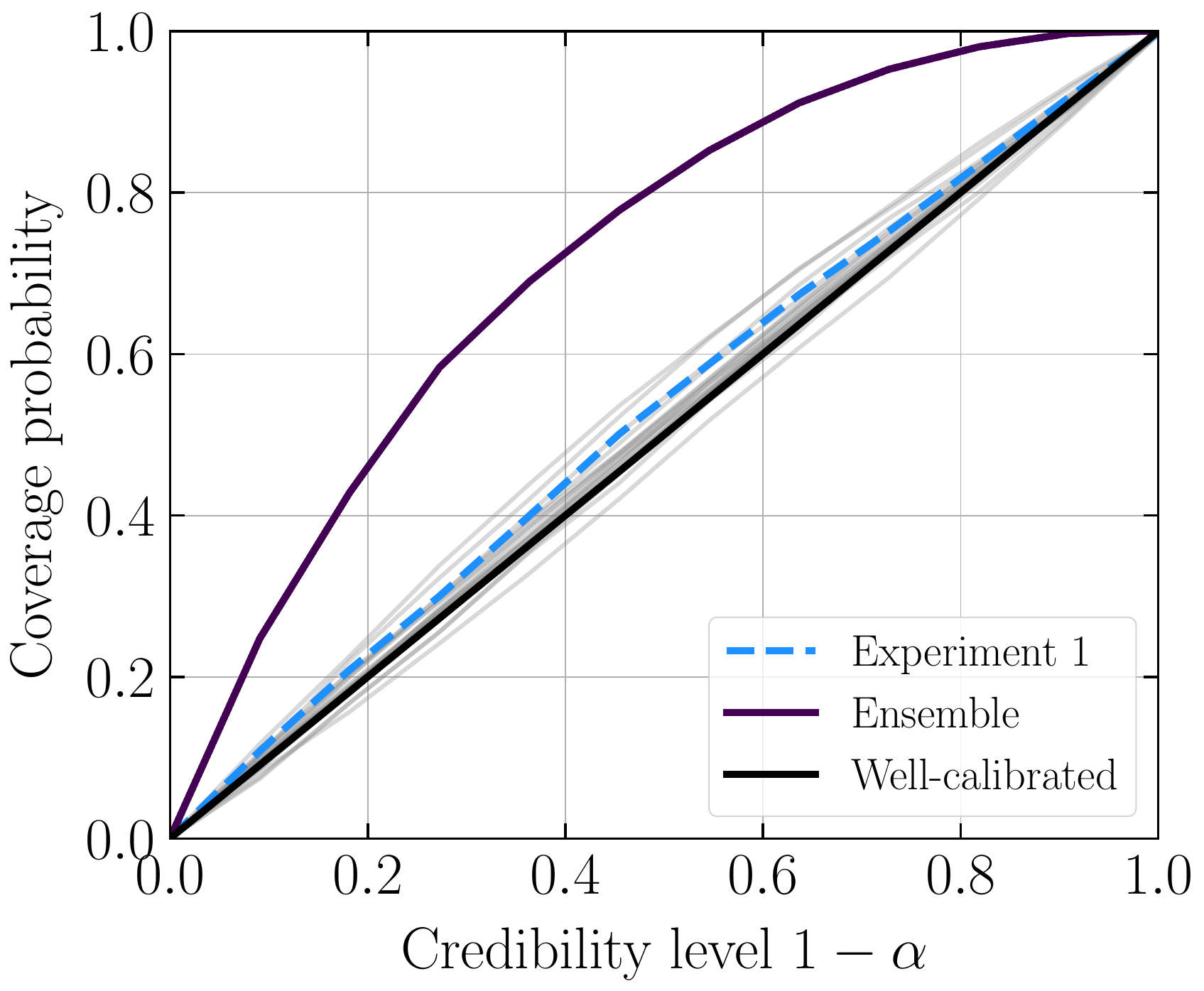}
	\caption{Coverage probability as a function of the credibility level, $1 - \alpha$, for our approximate posteriors calculated for $3600$ test simulations. We specifically highlight the coverage for experiment $\#1$ as a light-blue dashed line and that for the ensemble as a purple solid line. All remaining experiments are given in gray. For a well-calibrated posterior, the coverage follows the diagonal shown in black.}
	\label{fig:coverage}
\end{figure}

\begin{figure*}[t]
	\centering
	\includegraphics[width=0.92\textwidth]{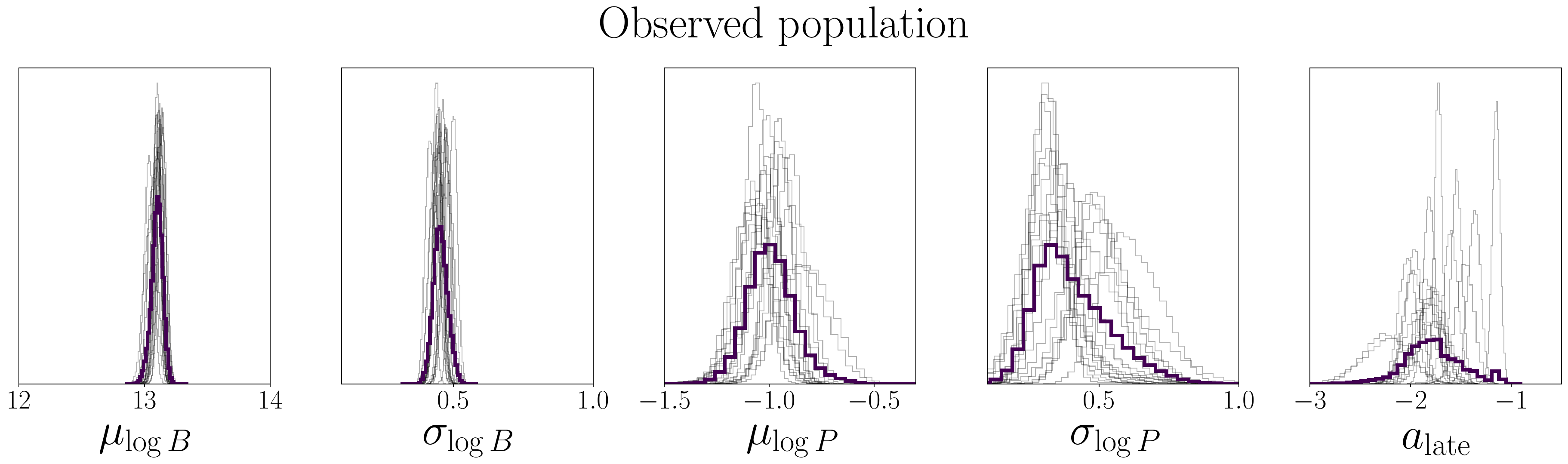}
	\caption{One-dimensional marginal posteriors for the five magnetorotational parameters for the observed pulsar population. We show inference results for 19 different \ac{NPE} experiments in gray and the ensemble posterior in purple.}
	\label{fig:posterior_comparison_atnf}
\end{figure*}


\subsection{Posterior Validation}
\label{sec:valdation}

To further assess whether posterior estimates are well calibrated, we determine their \textit{coverage}. As outlined in detail in Appendix~\ref{app:coverage}, the coverage probability measures the fraction of test samples for which (for a given credibility level $1 - \alpha$) the ground truths, $\bt$, fall within the corresponding $1 - \alpha$ region of their respective posteriors, $q_{F(\bx, \boldsymbol{\phi})} (\bt)$. For a well-calibrated posterior distribution and a sufficiently large number of test samples, this fraction should equal $1 - \alpha$. This implies that the coverage probability as a function of the credibility level is diagonal. In contrast, for a conservative posterior that is wider than the true posterior, we would recover a fraction larger than $1 - \alpha$. Conversely, for a narrower (overconfident) posterior, the corresponding fraction of test samples is less than $1 - \alpha$. In terms of the coverage, this corresponds to curves above and below the diagonal, respectively, and can therefore be used to assess the quality of approximate posteriors. 

We show the coverage probabilities for our different posterior estimates as a function of the credibility level, $1 - \alpha$, in Figure~\ref{fig:coverage}. We single out the coverage for the posterior from experiment $\#1$ (light-blue dashed line) and the ensemble posterior (purple solid line). All remaining experiments are shown in gray. We observe that the approximate posteriors for individual experiments closely follow the diagonal, exhibiting either slightly conservative or slightly overconfident behavior. As expected, the most conservative estimate is given by our ensemble posterior, which incorporates variations in the inference for $19$ different machine-learning configurations across all 3600 test samples. These results provide additional support that our neural posteriors are trustworthy and have indeed learned to accurately infer magnetorotational parameters from simulated $P$--$\dot{P}$ density maps.


\subsection{Inference on the Observed Population}
\label{sec:inference_observed}

Following the benchmark experiments and the coverage determination, we now turn our attention to the true pulsar populations observed with \ac{PMPS}, \ac{SMPS} and the low- and mid-latitude \ac{HTRU} survey. The corresponding $P$--$\dot{P}$ diagram was shown in the right panel of Figure~\ref{fig:pop_observed}. We represent these populations as three density maps, as outlined in Section~\ref{sec:sim_output} and subsequently feed them through our trained neural networks to infer the five parameters, $\mu_{\log B}, \sigma_{\log B}, \mu_{\log P}, \sigma_{\log P}$, and $a_{\rm late}$, assuming that our simulation framework provides a realistic description of the underlying physics.

\begin{figure*}
	\centering
	\includegraphics[width=0.8\textwidth]{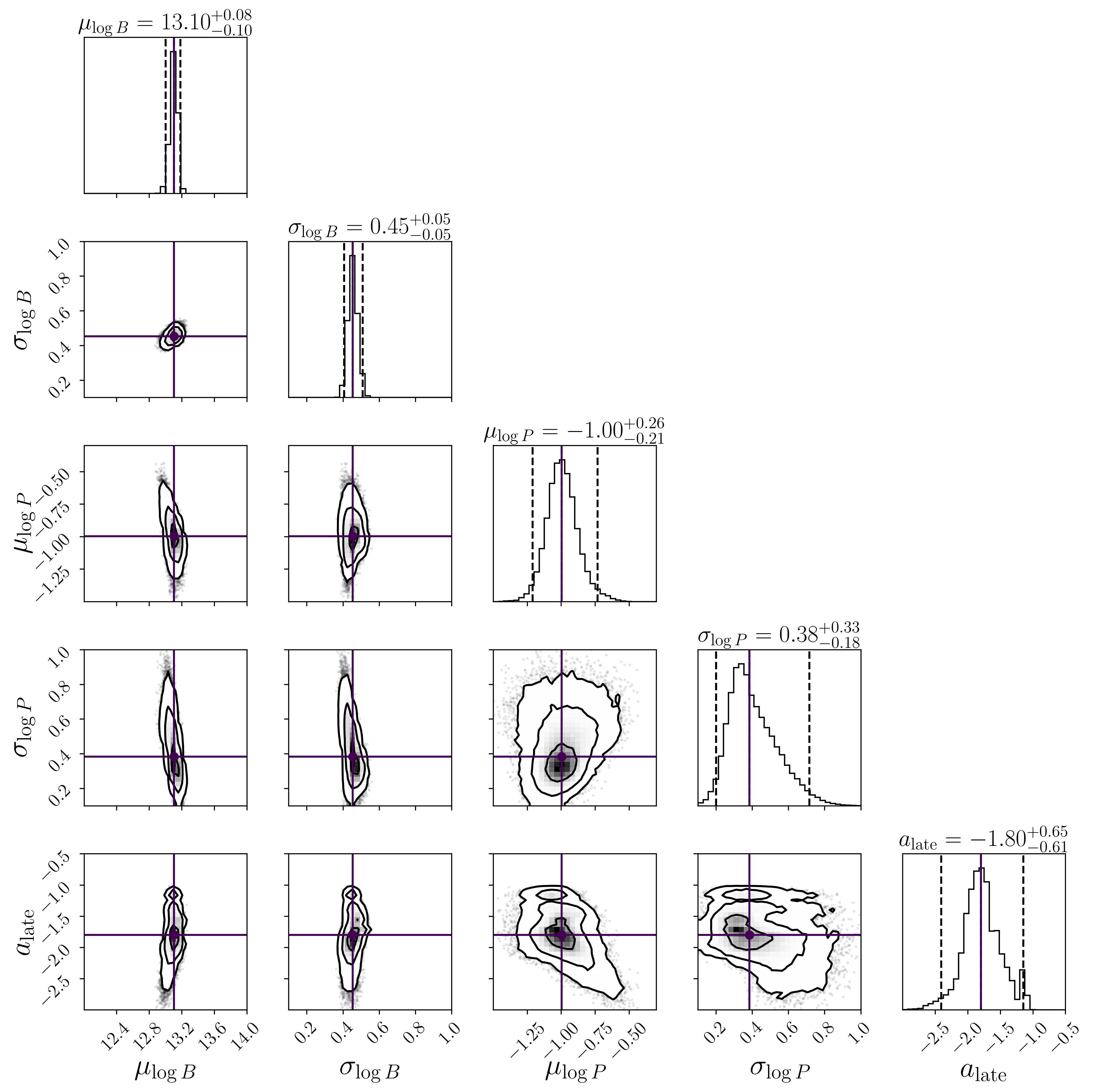}
	\caption{Inference results for the observed pulsar population using the ensemble posterior of $19$ different \acp{NPE}. The corner plot shows one- and two-dimensional marginal posterior distributions for the five magnetorotational parameters. We highlight the medians in purple. Corresponding values and $95\%$ \acp{CI} are summarized above the panels and in Table~\ref{tab:credible_intervals}.}
	\label{fig:posterior_ensemble_atnf}
\end{figure*}

We show the corresponding one-dimensional marginal posterior distributions for individual experiments (gray histograms) and the ensemble (purple histograms) in Figure~\ref{fig:posterior_comparison_atnf}. Additionally, a corner plot for the one- and two-dimensional ensemble posteriors is illustrated in Figure~\ref{fig:posterior_ensemble_atnf}. Corresponding medians (shown in purple in the corner plot) and $95\%$ \acp{CI} for experiment $\#1$ and the ensemble are also summarized in the last column of Table~\ref{tab:credible_intervals}.

The general trend (already observed for the simulated populations) that the initial magnetic field parameters, $\mu_{\log B}$ and $\sigma_{\log B}$, are much better constrained by our $\ac{NPE}$ framework than the remaining three values also holds for the observed population. As seen in the first two panels of Figure~\ref{fig:posterior_comparison_atnf}, all $19$ experiments recover narrow posteriors around similar medians. For the initial period distribution parameters, $\mu_{\log P}$ and $\sigma_{\log P}$ (see third and fourth panel, respectively), we obtain wider posteriors and a larger variety of median values between different experiments. These posteriors, however, cover similar regions within our prior ranges and are comparable to what we observed for the test samples. In contrast, the inferred posteriors for $a_{\rm late}$ (the final panel of Figure~\ref{fig:posterior_comparison_atnf}) exhibit different behavior from our benchmark experiments. In particular, posteriors vary significantly in width between different experiments, with those at the larger (smaller) end of the $a_{\rm late}$ range generally exhibiting narrower (larger) widths. Moreover, several distributions do not overlap at all. This is manifest as a relatively wide posterior in the ensemble that also shows a second peak, primarily driven by the rightmost individual posterior resulting from experiment $\#2$. Note that this configuration did not cause irregularities during the network optimization or unusual posteriors for our test samples. Therefore, we do not associate this behavior with the network itself. The corresponding bimodality is also visible in the final row of the corner plot in Figure~\ref{fig:posterior_ensemble_atnf}. We will discuss our interpretation of this below.

\begin{longrotatetable}
\begin{deluxetable}{c | c c c c c}
\tablecaption{Comparison between This Work and Several Population Synthesis Studies in the Literature. \label{tab:pop_syn_comparison}}
\tabletypesize{\footnotesize}
\tablecolumns{6}
\tablenum{4}
\tablehead{ 
& \colhead{\citet{Faucher2006}} & \colhead{\citet{Bates2014}} & \colhead{\citet{Gullon2014, Gullon2015}} & \colhead{\citet{Cieslar2020}} & \colhead{This Work}
}
\startdata
\hline
\multirow{2}{*}{$\boldsymbol{\mathcal{P}(r, \phi)}$} & spiral arms, & spiral arms, & spiral arms, & spiral arms, & $e$-density model \\
& $\mathcal{P}(r)$ & $\mathcal{P}(r)$ & $\mathcal{P}(r)$ & $\mathcal{P}(r)$ & \citet{Yao2017} \\
\hline
$\boldsymbol{\mathcal{P}(z)}$ & exponential & exponential & exponential & exponential & exponential \\
\hline
\textbf{Galactic} & \multirow{2}{*}{-} & \multirow{2}{*}{-} & \multirow{2}{*}{-} & \multirow{2}{*}{-} & \multirow{2}{*}{$T \approx \unit[250]{Myr}$} \\
\textbf{rotation} & & & & & \\
\hline
$\boldsymbol{\mathcal{P}(v_k)}$ & exponential & exponential, normal & exponential & Maxwell & Maxwell \\
\hline
$\boldsymbol{\mathcal{P}(B_0)}$ & lognormal & lognormal & lognormal & lognormal & lognormal \\
\hline
$\boldsymbol{\mathcal{P}(P_0)}$ & normal & normal, lognormal & normal & normal & lognormal \\
\hline
\multirow{3}{*}{$\boldsymbol{B(t)}$} & \multirow{3}{*}{-} & \multirow{3}{*}{-} & magnetothermal models & exponential & magnetothermal models \\
& & & \citet{Vigano2013} & decay & \citet{Vigano2021}, \\
& & & & & late-time power law \\
\hline
$\boldsymbol{P(t)}$ & vacuum dipole & vacuum dipole & plasma-filled dipole & vacuum dipole & plasma-filled dipole \\
\hline
$\boldsymbol{\chi(t)}$ & - & exponential & $P$-$\chi$ coupled & - & $P$-$\chi$ coupled \\
\hline
\multirow{2}{*}{\textbf{Beaming}} & \multirow{2}{*}{empirical} & empirical, & \multirow{2}{*}{empirical} & \multirow{2}{*}{empirical} & \multirow{2}{*}{geometrydependent} \\
& & geometrydependent & & & \\
\hline
\multirow{2}{*}{\textbf{Luminosity}} & pseudo, & pseudo, & pseudo, & pseudo, & intrinsic, \\
& $\propto |\dot{E}_{\rm rot}|^{\epsilon}$ & $\propto P^{\alpha} \dot{P}^{\beta}$ & $\propto |\dot{E}_{\rm rot}|^{\epsilon}$ & $\propto |\dot{E}_{\rm rot}|^{\epsilon}$ & $\propto |\dot{E}_{\rm rot}|^{\epsilon}$ \\
\hline
\multirow{2}{*}{\textbf{Surveys}} & \multirow{2}{*}{PMPS, SMPS} & \multirow{2}{*}{PMPS, SMPS} & PMPS, SMPS & \multirow{2}{*}{PMPS} & \multirow{2}{*}{PMPS, SMPS, HTRU} \\
& & & + X-ray pulsars (2015 study) & & \\
\hline
 \multirow{2}{*}{\textbf{Comparison}} & K-S test, & \multirow{2}{*}{K-S test} & annealing method, & \ac{MCMC} with & \multirow{2}{*}{\ac{SBI}} \\
 & by eye & & K-S test & Gaussian likelihood & \\ 
\enddata
\tablecomments{We compare the following ingredients, which are given as individual table rows: the distributions of sources in the Galactic plane, $\mathcal{P}(r, \phi)$, and along Galactic heights, $\mathcal{P}(z)$, in cylindrical galactocentric coordinates; the inclusion of Galactic rotation and, if present, the corresponding rotation period, $T$; the distribution of neutron star kick velocities, $\mathcal{P}(v_k)$; the distributions of initial dipolar magnetic field strengths and initial periods, i.e., $\mathcal{P}(B_0)$ and $\mathcal{P}(P_0)$, as well as the prescriptions for their evolution (denoted as $B(t)$ and $P(t)$, respectively); the treatment of the misalignment-angle evolution, $\chi(t)$; the description of the radio beaming, where pulsars that intercept our line of sight are either determined with an \textit{empirical} relation between the beaming fraction and the period obtained from polarization data \citep{Tauris1998} or with a \textit{geometry-dependent} approach that considers the radio beam aperture and the inclination angle, $\chi$. We further provide information on the luminosity (distinguishing between \textit{pseudo} and \textit{intrinsic} luminosities), the respective surveys used for comparison and, finally, the method used to contrast simulated and observed populations (where K-S denotes the Kolmogorov--Smirnov test).}
\end{deluxetable}
\end{longrotatetable}

\section{Discussion and conclusions}
\label{sec:conclusions}

In this study, we have successfully developed a new machine-learning pipeline that combines pulsar population synthesis with \ac{SBI} for the first time and tested the corresponding approach by inferring magnetorotational properties of neutron stars.


\subsection{Simulation Framework}
\label{sec:disc_simulator}

We first discussed our implementation of the forward model, i.e., the prescription for simulating the dynamical and magnetorotational properties of the Galactic population of isolated radio pulsars, modeling their radio emission and subsequently mimicking observational limitations for \ac{PMPS}, \ac{SMPS} and the low- and mid-latitude \ac{HTRU} survey. We followed earlier frameworks \citep[e.g.,][]{Faucher2006, Bates2014, Gullon2014, Gullon2015, Cieslar2020} but implemented several key differences, as compared in detail in Table~\ref{tab:pop_syn_comparison}. In particular, we sampled the birth positions of our pulsars from the Galactic electron distribution \citep{Yao2017} instead of following the typical approach of combining a spiral arm model with a radial pulsar distribution like that of \citet{Yusifov2004}. The latter is deduced for the observed, evolved pulsar sample and not the initial population. Moreover, we have included the (rigid) rotation of the Galaxy to treat the pulsar birth positions more consistently compared to earlier analyses. For the magnetic field evolution, we used a similar approach to that of \citet{Gullon2014, Gullon2015}, taking advantage of the newest two-dimensional magnetothermal simulations \citep{Vigano2021}, and solved for the coupled evolution of the spin period, $P$, and the misalignment angle, $\chi$, for a plasma-filled magnetosphere. To capture the field changes at late times, we developed a new physically motivated prescription in which the magnetic field, $B$, decays according to a power law captured by the index, $a_{\rm late}$. Together with the means, $\mu_{\log B}, \mu_{\log P}$, and standard deviations, $\sigma_{\log B}, \sigma_{\log P}$, which characterize the normally distributed logarithms of the initial periods and the initial fields, we hence obtained five parameters that control the neutron stars' magnetorotational evolution. 

To simulate the detection of our synthetic pulsars, we make the following changes compared to earlier studies: First, we do not model the pulsars' pseudoluminosity defined as $L_{\rm ps} \equiv S_{f, {\rm obs}} d^2$ (where $S_{f, {\rm obs}}$ is the detected flux at frequency, $f$, and $d$ is the pulsar distance), but instead assume that the intrinsic neutron star luminosity, $L_{\rm int}$, is proportional to the spin-down power, $\dot{E}_{\rm rot}$. In particular, we considered $L_{\rm int} \propto |\dot{E}_{\rm rot}|^{1/2}$ to determine the bolometric radio flux and subsequently propagate the corresponding pulsed emission toward Earth. We also used a geometry-based description to determine the pulsars that are beamed toward us, which earlier works typically treat in an empirical manner. In addition, we do not implement a pulsar death line to quench radio emission but instead let pulsars become undetectable naturally. Finally, we not only looked at \ac{PMPS} and \ac{SMPS} but also incorporated the \ac{HTRU} survey for the first time. Using the resulting simulation framework, we then produced 360,000 synthetic $P$--$\dot{P}$ diagrams, which we converted to one density map per survey in preparation for the neural networks. A total of $90\%$ of these simulations were used for training and validation, and the remaining $10\%$ were reserved for testing.


\subsection{Inference Procedure}
\label{sec:disc_inference}

The second part of this study is centered on the implementation of the \ac{SBI} approach, specifically focusing on \ac{NPE}, to learn a probabilistic association between our simulator output and the input parameters, $\bt = \{\mu_{\log B}, \sigma_{\log B}, \mu_{\log P}, \sigma_{\log P}, a_{\rm late}\}$. To do so, we first used a \ac{CNN} to extract features from our high-dimensional $P$--$\dot{P}$ maps and obtain a compressed representation, which was then transferred into a flexible neural density estimator. By taking advantage of the open-source Python package {\tt sbi} \citep{Tejero-Cantero2020},\footnote{\url{https://github.com/sbi-dev/sbi}} we specifically opted for a Gaussian mixture density model in five dimensions to approximate our posterior. To study the sensitivity of the \ac{NPE} results on the representation of our input data and the network hyperparameters, we conducted $22$ distinct experiments. An inspection of the corresponding training metrics led us to discard three experiments owing to irregular training behavior or overfitting. The remaining $19$ trained neural networks were analyzed further, and we found no significant differences in the resulting inferences when benchmarked on three random test simulations. The same was observed when validating the posteriors through a coverage calculation over the test set with $3600$ samples, highlighting that all $19$ posterior estimates are well calibrated. From this we concluded, in particular, that the training behavior is a poor identifier of subsequent inference quality, because normalization of input maps led to systematically better training, test, and validation metrics compared to standardizing the input but comparable inferences. Learning rate and batch size played a negligible role in both setups. 

We also point out that the use of smaller training data sets did not affect the inference quality either. While we expect that training sets of $\lesssim 10\%$ (i.e., 30,000 simulations) will eventually have an effect on this, databases of $50 \%$ (i.e., 150,000 simulations) are sufficient when inferring five parameters. For comparable studies, this would imply a significant reduction in simulation time, the most costly part of these analyses. Similar performances further justify optimizing our networks for density maps with a resolution of $32 \times 32$ bins instead of $64\times 64$ and the shallower baseline \ac{CNN} to speed up the training process. Additionally, we highlight that the use of different numbers of Gaussian mixture components also led to comparable optimization metrics and inference results. Extracting the corresponding mixture weights, $\alpha_c$, after the optimization, we find that across the entire test data set we only require two or three Gaussians to approximate our posteriors. However, we point out that training with a larger number of components was faster owing to fewer training epochs. Finally, note that the use of fewer surveys (i.e., one or two density maps only) did not change the inference results for our five magnetorotational parameters. Naively, one might think that complementary information on the pulsar population as, e.g., provided by \ac{SMPS}, which is sensitive to older stars at higher Galactic latitudes, would help the network learn better posteriors. However, we do not observe such behavior in our experiments. Although this might suggest that using single surveys in the future could be sufficient to constrain neutron star parameters through population synthesis, we caution that different surveys, in principle, provide additional information on the neutron star birth rate (see below) that was not supplied to our neural networks, i.e., we focused on the location and shape of the pulsar population in the $P$--$\dot{P}$ plane only.

Due to the variations in our inference results, and because we could not identify a single neural network as the best posterior estimator, we also determined the ensemble posterior through an equally weighted average of the individual experiments. The resulting posterior behaved as expected and showed more conservative behavior than the ensemble members. For the next section, we, will hence follow the recommendation by \citet{Hermans2021} and use our (most conservative) ensemble posterior to analyze the observed pulsar population.


\begin{figure}
	\centering
	\includegraphics[width=0.95\columnwidth]{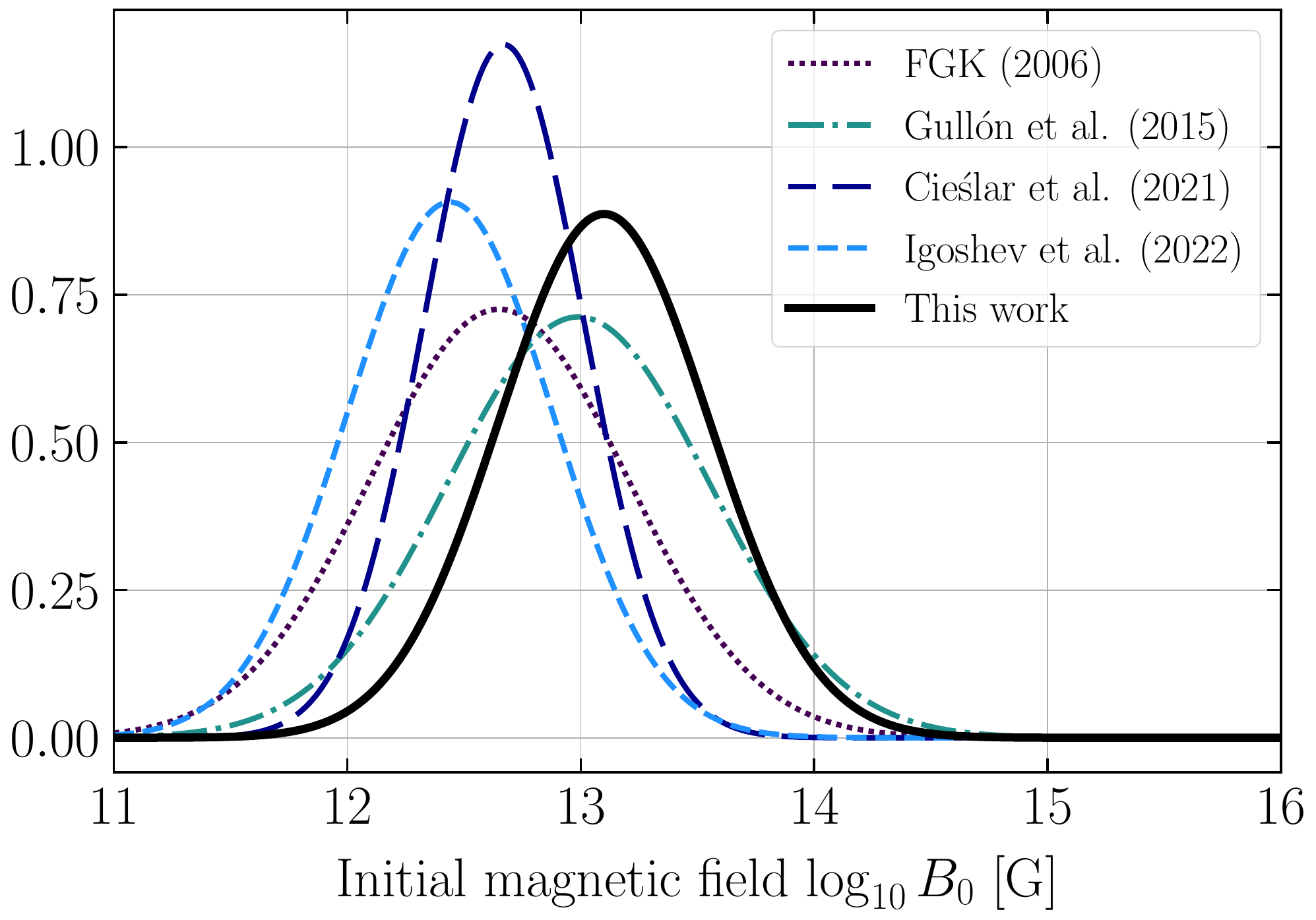}
	\vskip 0.3cm
	\includegraphics[width=0.95\columnwidth]{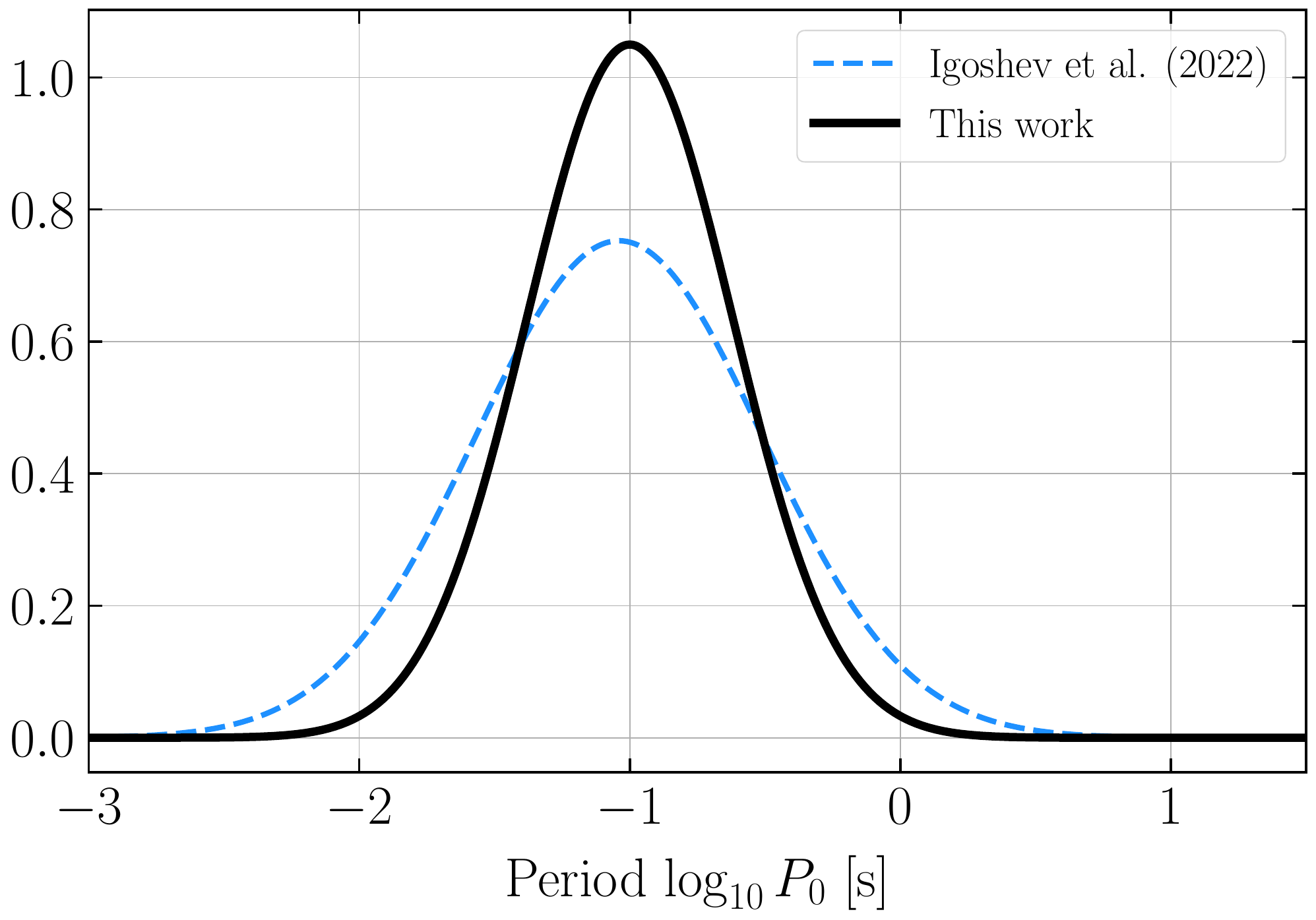}
	\caption{A comparison of initial magnetic field, $B_0$ (top), and period, $P_0$ (bottom), distributions for the radio pulsar population. The logarithms of $B_0$ and $P_0$ are normally distributed (see Eqns.~\eqref{eqn:B_pdf} and \eqref{eqn:P_pdf}) and characterized by means, $\mu_{\log B, P}$, and standard deviations, $\sigma_{\log B, P}$, respectively. That is, these distributions are normalized such that the total area under the curves equals 1 for logarithmic abscissa values. Corresponding numerical values are summarized in Table~\ref{tab:best_param}. The results of this work are illustrated as black solid lines. Additional studies are shown as detailed in the legends.}
	\label{fig:log_normal_distributions}
\end{figure}

\begin{deluxetable*}{c|DDDD}[t]
\tablecaption{Comparison between Best Parameters for the Lognormal Initial Magnetic Field and Initial Period Distributions in the Literature. \label{tab:best_param}}
\tabletypesize{\small}
\tablecolumns{5}
\tablenum{5}
\tablewidth{0pt}
\tablehead{
\colhead{References} &
\multicolumn2c{$\mu_{\log B}$} &
\multicolumn2c{$\sigma_{\log B}$} &
\multicolumn2c{$\mu_{\log P}$} &
\multicolumn2c{$\sigma_{\log P}$}
}
\decimals
\startdata
\citet{Faucher2006} & $12.65$ & $0.55$ & - & - \\
\citet{Gullon2015}  & $12.99$ & $0.56$ & - & - \\
\citet{Cieslar2020} & $12.67^{+0.01}_{-0.02}$ & $0.34^{+0.02}_{-0.01}$ & - & - \\
\citet{Igoshev2022} & $12.44$ & $0.44$ & $-1.04^{+0.15}_{-0.20}$ & $0.53^{+0.12}_{-0.08}$ \\
This work & $13.10^{+0.04}_{-0.05}$ & $0.45^{+0.03}_{-0.02}$ & $-1.00^{+0.11}_{-0.10}$ & $0.38^{+0.16}_{-0.10}$
\enddata
\tablecomments{We provide references and the four relevant parameters. Note that the first three studies use a different prescription for the initial period, which prevents a direct comparison with our study. For \citet{Gullon2015} and \citet{Cieslar2020}, we compare with \textit{model D} for the radio-pulsar population and the \textit{rotational model}, respectively. The corresponding distributions are illustrated in Figure~\ref{fig:log_normal_distributions}. Where available, we quote \acp{CI} at the $68\%$ level (including for this work), but note that these are difficult to compare owing to the difference in inference methods and underlying models and data.}
\end{deluxetable*}

\subsection{Inference results on the Observed Population}
\label{sec:disc_inference_obs}

Following the validation of our \ac{NPE} approach, we subsequently used the ensemble posterior estimator to infer the five magnetorotational parameters for the true population of isolated Galactic radio pulsars observed with our three surveys. In particular, we found the following best estimates at the $95\%$ credible level:
\begin{align}
	\mu_{\log B} &= 13.10^{+0.08}_{-0.10}, \nonumber \\
	\sigma_{\log B} &= 0.45^{+0.05}_{-0.05}, \nonumber \\
	\mu_{\log P} &= -1.00^{+0.26}_{-0.21}, \label{eqn:CI_best} \\
	\sigma_{\log P} &= 0.38^{+0.33}_{-0.18}, \nonumber \\
	a_{\rm late} &= -1.80^{+0.65}_{-0.61}. \nonumber
\end{align}
The corresponding corner plot was illustrated in Figure~\ref{fig:posterior_ensemble_atnf}, while we show the resulting distributions for the initial magnetic field and period as black solid lines in Figure~\ref{fig:log_normal_distributions}.

As noted during the benchmarking experiments, we generally obtain narrower posterior distributions for the initial magnetic field parameters when compared to the initial period parameters. Difficulties in constraining rotational birth properties are, however, not a shortcoming of our inference approach itself, as this was also noted by earlier population synthesis analyses \citep[e.g.,][]{Gullon2014, Gullon2015}. Instead, this has a physical reason that lies in the coupled evolution of the stars' misalignment angle, rotation period, and magnetic field. While the $B$-field initially stays constant (see Figure~\ref{fig:B_fields}), pulsars move from the top left in the $P$--$\dot{P}$ plane diagonally toward the bottom right, following lines of constant magnetic field (see, e.g., the right panel of Figure~\ref{fig:pop_observed}). As they do, stars with comparable field strengths but different initial periods evolve toward similar $P$ values. In addition, the misalignment angle evolution introduces further degeneracies because all $\chi$ decrease with time. However, as the field decays, spin-down and misalignment evolution slow down and pulsars begin to evolve almost vertically toward smaller $\dot{P}$ values. These processes depend further on $B_0$ and $P_0$ as stronger initial fields and smaller initial periods result in faster spin-down and faster evolution toward alignment. This is especially visible for test sample 3 (top right panel of Figure~\ref{fig:pop_simulated}), which is characterized by the smallest period mean, $\mu_{\log P}$, of all three test cases. The combined action of these effects is that stars born with different rotational properties attain similar $P$ at current times. This information loss on the initial period makes it harder to infer corresponding parameters. As expected, test simulation 3 thus shows the largest $95 \%$ \acp{CI} for $\mu_{\log P}$ and $\sigma_{\log P}$ out of our three test samples (third column in Table~\ref{tab:credible_intervals} and last row in Figure~\ref{fig:posterior_comparison}).


\subsection{Comparing Results with Earlier Works}
\label{sec:lit_comparison}

Contrasting the posterior medians from Equation~\eqref{eqn:CI_best} with the results of earlier population synthesis studies summarized in Table~\ref{tab:best_param} and Figure~\ref{fig:log_normal_distributions}, we first note that our $\mu_{\log B}$ estimate is roughly consistent with \citet{Gullon2014, Gullon2015} but somewhat larger than those of \citet{Faucher2006}, \citet{Cieslar2020} and \citet{Igoshev2022}. Moreover, while very close to \citet{Igoshev2022}, we obtain a smaller $\sigma_{\log B}$ than \citet{Gullon2014, Gullon2015} and \citet{Faucher2006} and a slightly larger estimate than \citet{Cieslar2020}. Although these works determine optimal parameter ranges different from ours (see Table~\ref{tab:pop_syn_comparison}), we expect the variation in the $B_0$ constraints to be mainly due to our more realistic prescription for the field and the coupled $P$-$\chi$ evolution. 

A direct comparison of our initial period parameters and earlier population synthesis literature is not possible because (following recent results by \citet{Igoshev2022}; see also \citet{Xu2023}) we considered the periods' logarithm and not the periods themselves to be normally distributed. However, we highlight that our inferred $\mu_{\log P}$ is comparable to that of \citet{Igoshev2022}, whereas our $\sigma_{\log P}$ is somewhat smaller (see bottom panel of Figure~\ref{fig:log_normal_distributions}). \citet{Igoshev2022} focused on a simplified analysis of $56$ young neutron stars in supernova remnants and looked at magnetorotational properties only. The authors were thus able to define an explicit likelihood function and perform statistical inference. Corresponding \acp{CI} given in Table~\ref{tab:best_param} are similar to ours, but we highlight that a systematic comparison is complicated owing to the distinct choices of underlying data and inference techniques. In this context, we also point out that although \citet{Cieslar2020} derive relatively narrow posteriors (see Table~\ref{tab:best_param}) for a range of pulsar properties using an \ac{MCMC} analysis, their underlying simulation framework is significantly reduced compared to ours invoking, e.g., (unrealistic) exponential field decay, vacuum magnetospheres, no coupling between periods and misalignment angles, and a simplified prescription for the beamed emission. In addition, they make an explicit assumption on the likelihood that might not accurately capture the complexity of the pulsar population synthesis even for their simplified model. We reiterate the robustness of our \ac{SBI} approach, which eliminates the need for an explicit expression for the likelihood and is therefore also suitable for more complex simulators like ours. Moreover, as outlined above, the use of a neural density estimator results in amortized posterior distributions that allow fast evaluation and sampling. We used this fact to determine the coverage and validate our posteriors, a procedure that is infeasible in \ac{MCMC} or nested sampling approaches owing to the time-consuming need for repeated sampling.

\begin{figure*}
	\centering
	\includegraphics[height=0.78\columnwidth]{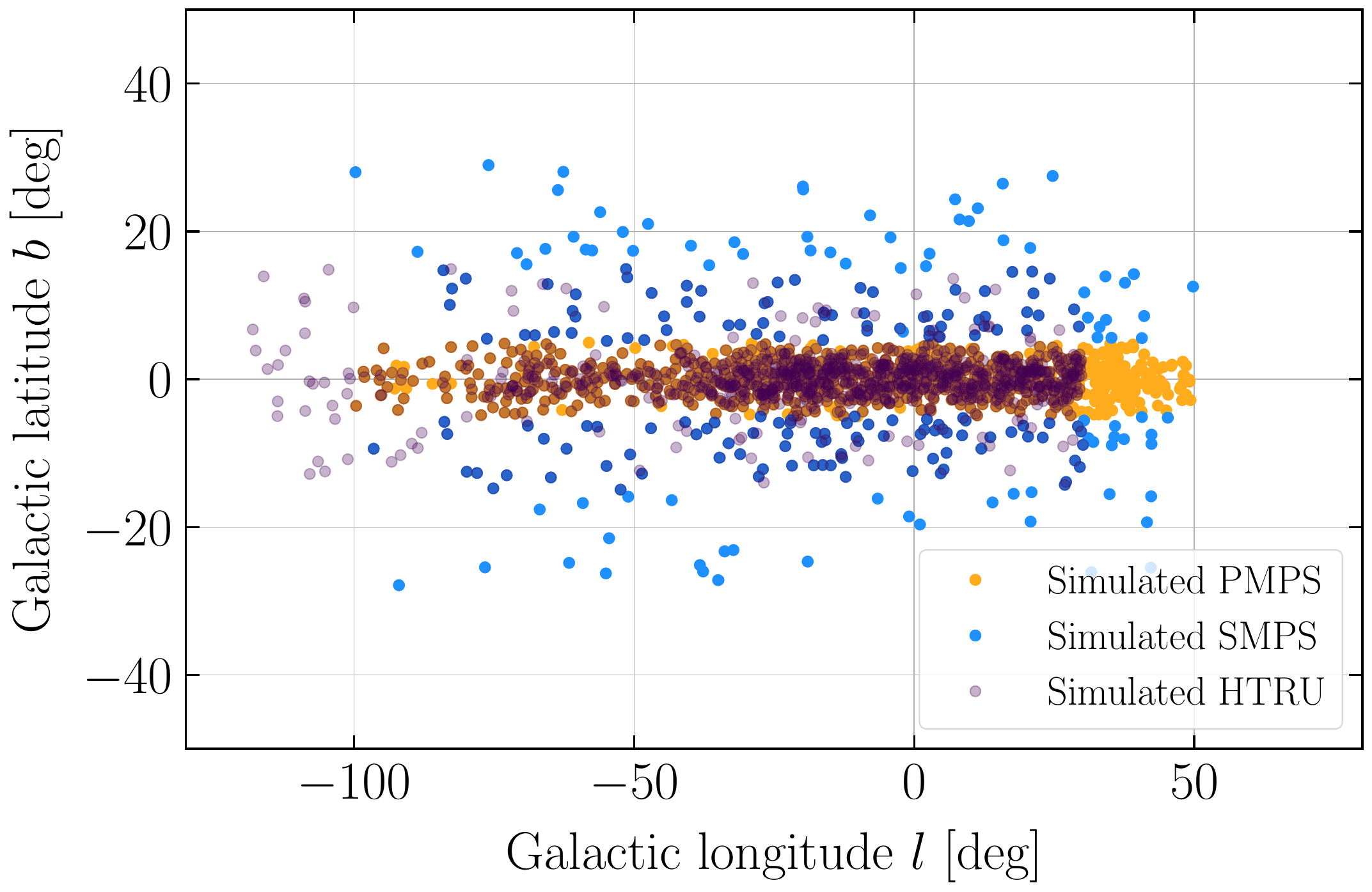}
	\hspace{0.2cm}
	\includegraphics[height=0.78\columnwidth]{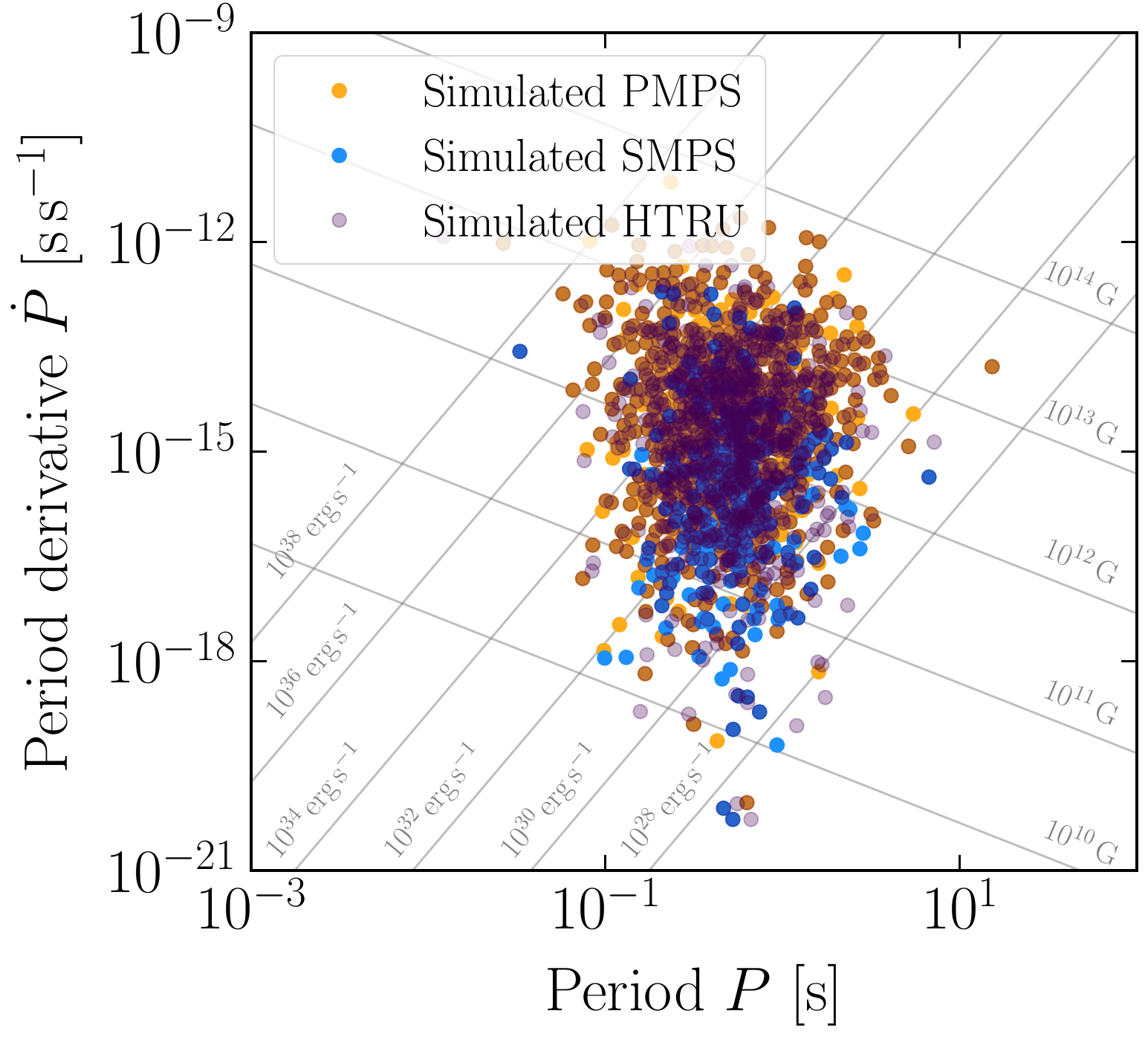}
	\caption{Simulated populations of isolated Galactic radio pulsars detected with \ac{PMPS}, \ac{SMPS} and the low- and mid-latitude \ac{HTRU} survey (highlighted in yellow, light blue, and purple, respectively) for the parameters inferred via \ac{SBI} from the observed radio pulsar population (see Equation~\eqref{eqn:CI_best}). The left panel shows the distribution of the simulated population in Galactic latitude, $b$, and longitude, $l$, while the right panel depicts the pulsars in the period, $P$, and period derivative, $\dot{P}$, plane. In the latter, we also give lines of constant spin-down power, $|\dot{E}_{\rm rot}|$, and constant dipolar surface magnetic field, $B$ (estimated via Equation~\eqref{eqn:P_ode} for an aligned rotator). Both plots directly compare to the true (observed) population shown in Figure~\ref{fig:pop_observed}.}
	\label{fig:pop_best_simulated}
\end{figure*}


\subsection{Late-time Magnetic Field Decay}
\label{sec:late-time_decay}

We now turn our attention to the parameter $a_{\rm late}$, the power-law index for the late-time magnetic field decay. We newly introduced $a_{\rm late}$ in pulsar population synthesis to account for the highly uncertain, core-dominated field evolution above $\unit[10^6]{yr}$ in a phenomenological way. While corresponding inferences were satisfactory for our benchmark experiments, we found that posteriors for $a_{\rm late}$ inferred from the observed population differed significantly between our $19$ experiments, resulting in systematically larger $95\%$ \acp{CI} for smaller $a_{\rm late}$ medians and vice versa (see rightmost panel of Figure~\ref{fig:posterior_comparison_atnf}). In addition, several posteriors did not overlap at all across our prior range, leading to a bimodality in the ensemble posterior. As we did not see anything similar for our synthetic simulations, we do not associate this behavior with the networks' performance or the \ac{SBI} approach itself. Instead, we hypothesize that this is due to shortcomings in our simulation framework. Put differently, our statistical inferences are only as good as the simulation model used to train our density estimator. Consequently, we see the complications in inferring $a_{\rm late}$ as an indication that our treatment of the late-time field evolution via a power law (albeit physically motivated by the behavior of known magnetic field evolution mechanisms) is insufficient to model the observed pulsar population. 

\begin{figure}[b]
	\centering
	\includegraphics[width=0.95\columnwidth]{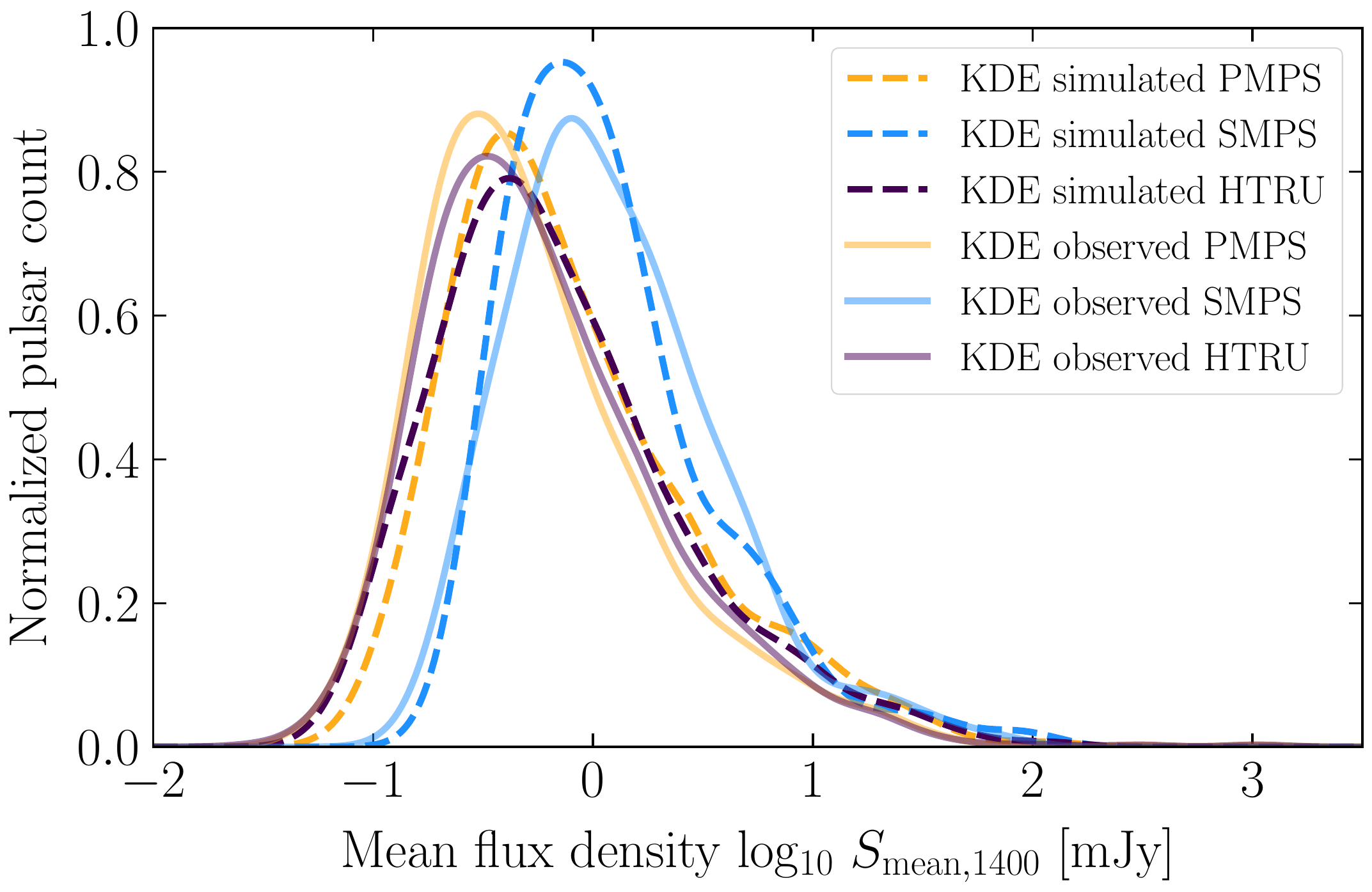}
    \caption{Distributions of mean radio flux densities, $S_{{\rm mean}, 1400}$, at \unit[1400]{MHz} for the populations of isolated Galactic radio pulsars in \ac{PMPS}, \ac{SMPS} and the low- and mid-latitude \ac{HTRU} survey (in yellow, light blue, and purple, respectively). To avoid overcrowding the plot, we omit the underlying histograms (see Figure~\ref{fig:flux_observed}) and only show the individual probability density functions obtained via \ac{KDE} using a Gaussian kernel. Estimates for the observed population are shown as solid lines, while one of our best-parameter simulations is shown with dashed lines. Data taken from the ATNF Pulsar Catalogue \citep[][\url{https://www.atnf.csiro.au/research/pulsar/psrcat/}, v1.69]{Manchester2005}.}
	\label{fig:flux_KDE_comp}
\end{figure}

Although further work is needed to better understand the late-time evolution of neutron star fields, we can assure ourselves that our current power-law prescription is not far from reality. To do so, we rerun our simulator with the best estimates summarized in Equation~\eqref{eqn:CI_best}. We show an example of the resulting population in Galactic longitude and latitude and in $P$ and $\dot{P}$ in Figure~\ref{fig:pop_best_simulated}. Both panels are analogous to the respective plots in Figure~\ref{fig:pop_observed}. Moreover, Figure~\ref{fig:flux_KDE_comp} shows a comparison between the estimated probability density functions for the radio flux density distributions for the observed populations (solid lines) and our best-parameter simulation (dashed lines).

While a detailed comparison between this simulated and the observed population and a study of implications for the neutron star birth rate are beyond the scope of this work, we will highlight a few main aspects. First, we note that the distributions look markedly similar, giving a reasonable level of confidence in our underlying simulation framework. This is particularly true for the Galactic longitude vs. latitude distribution and the mean radio flux densities. We attribute the small remaining differences in Figure~\ref{fig:flux_KDE_comp} primarily to uncertainties in the flux density measurements in the ATNF pulsar catalog discussed previously in Section~\ref{sec:obs_lims}, our choice of luminosity function (see Equation~\eqref{eqn:luminosity}) and systematics in the determination of pulsar survey sensitivities (see Table~\ref{tab:SurveyParam}). Finally, we do see a slight shift in the \ac{SMPS} population in the $P$--$\dot{P}$ diagram toward lower $\dot{P}$ values. This might again hint at missing physics at late times because \ac{SMPS} is sensitive to somewhat older pulsars compared to the other two surveys. 

\subsection{Neutron Star Birth Rate}
\label{sec:NS_birthrate}

We can further count the numbers of detected pulsars in all three synthetic surveys for our best-estimate simulation. Running our simulator 10 times to account for its stochastic nature, we obtain average pulsar counts of $1013$, $242$, and $1298$ for \ac{PMPS}, \ac{SMPS}, and the \ac{HTRU} survey, respectively. Comparing these to the true observed counts in Equation~\eqref{eqn:detected_objects}, we find an equivalent number of objects in \ac{PMPS} (within the sensitivity limits of our iterative approach of generating and detecting pulsars as summarized in Section~\ref{sec:sim_output}), while we overestimate the \ac{SMPS} population by $\sim 11\%$ and the \ac{HTRU} population by $\sim 27\%$ on average.

To understand these small discrepancies, we return to our earlier discussion of the neutron star birth rate in Section~\ref{sec:sim_output}. In particular, for our best estimates, we reach the observed target counts given in Equation~\eqref{eqn:detected_objects} for each survey for the following birth rates:
\begin{align}
&\text{\ac{PMPS}: $\sim 2.02 \pm 0.02$ neutron stars per century}, \nonumber \\
&\text{\ac{SMPS}: $\sim 1.84 \pm 0.03$ neutron stars per century}, \label{eqn:BR_estimated} \\
&\text{\ac{HTRU}: $\sim 1.66 \pm 0.02$ neutron stars per century}, \nonumber
\end{align}
where we quote means and standard errors for the 10 runs. These estimates are somewhat smaller than those obtained in earlier population synthesis studies \citep{Faucher2006, Gullon2014} and very close to the recent core-collapse supernova estimate from \citet{Rozwadowska2021} ($1.63 \pm 0.46$ per century). The differences in Equation~\eqref{eqn:BR_estimated} are sufficient to result in the slight overproduction of objects noted above. We remind that this is because we continue producing neutron stars until we hit the number of observed pulsars in all three surveys. In our specific case, \ac{PMPS} detections require a slightly larger birth rate than the other two surveys. As mentioned previously, the main reason for this is that we only expect the \textit{correct} physical model to produce the same birth rate across all surveys, again hinting that our simulator is missing some physics. Nonetheless, besides successfully constraining magnetorotational parameters for pulsar population synthesis using \ac{SBI} for the first time, we do recover birth rate results in Equation~\eqref{eqn:BR_estimated} that are very similar across all surveys.


\subsection{Future Directions}
\label{sec:future}

In light of the previous conclusions, we intend to further develop our current approach in a number of ways. 

On the simulation side, we will investigate additional luminosity prescriptions that go beyond our assumption, $L_{\rm int} \propto |\dot{E}|^{1/2}$, as this is another quantity that can significantly affect the pulsar distribution. Varying the exponent in our simulations, which was beyond the scope of this study owing to computational limitations, but using \ac{SBI} to constrain corresponding parameter ranges would be a first step in that direction. Moreover, while we followed \citet{Gullon2014, Gullon2015} and took a significant step forward in incorporating a realistic description of the neutron star magnetic field, we already noted above that further investigations into the field evolution of the neutron star core at late times will be important for future population synthesis frameworks. Finally, new pulsar surveys (in the radio band, as well as in other wavelengths) might hold the key to further constraining the neutron star population. While we did not see a significant improvement in our inferences using information from one, two, or three radio surveys, future studies will benefit from larger numbers of detected pulsars and accurate classification of telescope and detection biases. Furthermore, other wave bands, specifically X-rays or gamma rays, provide complementary information on the neutron star population. Our focus on realistic magnetic field evolution and the expansion of our approach to new three-dimensional magnetothermal simulations \citep[e.g.,][]{DeGrandis2021, Dehman2023} will be particularly crucial to determine realistic X-ray luminosities of the most strongly magnetized neutron stars. As highlighted by \citet{Gullon2015}, modeling these so-called magnetars and the isolated radio pulsar population consistently will be crucial to break degeneracies and constrain neutron star physics further.

The increase in simulator complexity associated with these improvements will not only result in more free parameters but also inevitably lead to larger computation times for our forward model. The approach taken here, i.e., simulating a large database for input parameter combinations that cover the entire space sufficiently, will become infeasible. To overcome these hurdles, we will also have to explore new \ac{SBI} approaches. Sequential methods \citep[e.g.,][]{Papamakarios2018, Deistler2022, Bhardwaj2023} that reduce the need for simulations by starting from a relatively small database and adaptively providing additional simulations (generated for those parts of the parameter space that are most useful for a neural density estimator to learn a posterior approximation) seem particularly suited to these tasks.


\section*{acknowledgments}
The authors thank Emilie Parent for useful exchanges on radio pulsar emission and detections, Clara Dehman for providing magnetic field evolution curves, and Jose Pons for insights on late-time magnetic field evolution. V.G., M.R. and C.P.A. also thank Jakob Macke and his group for valuable discussions on \ac{SBI}. The data production, processing, and analysis tools for this paper have been implemented and operated at the Port d'Informació Científica (PIC) data center. PIC is maintained through a collaboration of the Institut de F\'isica d'Altes Energies (IFAE) and the Centro de Investigaciones Energ\'eticas, Medioambientales y Tecnol\'ogicas (Ciemat). We particularly thank Christian Neissner and Carles Acosta for their support at PIC. V.G.  acknowledges support from a Juan de la Cierva Incorporaci\'on Fellowship. The authors are further supported by the ERC via the Consolidator grant ``MAGNESIA'' (No. 817661) and by the program Unidad de Excelencia Mar\'ia de Maeztu CEX2020-001058-M. We also acknowledge partial support from grant SGR2021-01269 (PI: Graber). M.R.'s and C.P.A.'s work has been carried out within the framework of the doctoral program in Physics at the Universitat Autonoma de Barcelona.


\vspace{5mm}

\software{Astropy \citep{astropy2013, astropy2018}, healpy \citep{Gorski2005, Zonca2019}, IPython \citep{PerezGranger2007}, JupyterLab, Matplotlib \citep{Hunter2007}, Numba \citep{Lam2015}, NumPy \citep{Oliphant2006, vanderWalt2011, Harris2020}, Pandas \citep{McKinney2010}, PyGEDM, PyTorch \citep{Paszke2019}, sbi \citep{Tejero-Cantero2020}, SciPy \citep{Jones2001, Virtanen2020}, Sphinx.}


\appendix

\section{magnetic field prescription}
\label{app:B-field}

As outlined in Section~\ref{sec:mr_evol}, a key ingredient for the magnetorotational evolution of radio pulsars is a realistic prescription for the evolution of the dipolar magnetic field strength, $B$, up to neutron star ages of $\unit[10^8]{yr}$. While earlier population synthesis studies have typically either neglected magnetic field decay entirely or relied on simplified descriptions invoking decaying exponentials or power laws, we choose a different approach and take advantage of recent progress in modeling the magnetothermal evolution of neutron star crusts. In particular, we use a set of five two-dimensional simulations \citep{Vigano2021} to fit the early-time magnetic field evolution which is driven by the combined action of the Hall effect and ohmic dissipation \citep[see, e.g.,][for details on these mechanisms]{Pons2019}. 

All five curves, shown as solid lines in Figure~\ref{fig:B_fields}, were simulated with realistic assumptions on relevant physics. In particular, the stellar structure and composition are based on the equation of state SLy4 \citep{Douchin2001} for a neutron star of mass $\unit[1.4]{M_{\odot}}$, resulting in a radius of $\unit[11.74]{km}$. The impurity parameter at the highest densities in the inner crust is set to $100$ \citep{Pons2013}, representing the presence of resistive nuclear pasta phases \citep[see, e.g.,][]{Chamel2008}, whereas the impurity profile for other crustal densities matches the results of \citet{Carreau2020} (see their Figure~5). Furthermore, the model for the neutron star envelope is taken from \citet{Potekhin2015}, while specific parameterization for the superfluid and superconducting energy gaps (SFB for the crustal neutrons, TToa for the core neutrons, and CCDKp for the core protons) were adopted from \citet{Ho2015}. 

What varies between the different simulations is the initial poloidal magnetic field strength, $B$, taking the values $10^{12}, 10^{13}, 10^{14}, 10^{15}, 5 \times 10^{15} \, {\rm G}$, respectively. This also implies different toroidal field strengths, which are typically a factor $10$ larger than the poloidal $B$ values. We observe in Figure~\ref{fig:B_fields} that those runs with larger magnetic fields decay faster. This is a direct result of the Hall effect, which depends on $B$ and acts to redistribute the magnetic field energy to smaller scales, where it subsequently decays owing to ohmic dissipation. For sources with $B \lesssim \unit[10^{12}]{G}$ and coupled thermal evolution, this Hall cascade does not take place and magnetic fields remain pretty much constant on timescales of the order of $\unit[10^6]{yr}$. 

Above this timescale, however, current magnetothermal simulations become unreliable because the implementation of relevant microphysics \citep{Potekhin2015} is unsuited to old neutron stars with temperatures $\lesssim \unit[10^6]{K}$. In addition, these simulations focus primarily on the crust and do not include a realistic treatment of the highly uncertain dynamics of the neutron star core, which should become relevant above $\sim \unit[10^6]{yr}$. As we require a prescription for the field above $\unit[10^6]{yr}$ for our population synthesis, we develop a simplified parameterization for the late-time magnetic field evolution that encodes the unknown evolution of the stellar core. As highlighted in Equation~\eqref{eqn:B_late}, we assume that field changes at late times can be captured by a power law characterized by the index, $a_{\rm late}$. This choice is physically motivated because several known magnetic field evolution mechanisms exhibit the same functional form. For example, Hall-like physics are encoded by $a_{\rm late} = -1$ \citep{Aguilera2008}, while ambipolar diffusion follows a power law with $a_{\rm late} = -0.5$ \citep{Goldreich1992}. 

To directly parameterize the behavior of the magnetic field across all relevant $B$ ranges and times $t$, we describe the field evolution with the following broken power laws:
\begin{align}
	B(t) &= B_0 \left(1 + \frac{t}{\tau_1} \right)^{a_1} \left(1 + \frac{t}{\tau_2} \right)^{a_2 - a_1} 
		\left(1 + \frac{t}{\tau_{\rm late}} \right)^{a_{\rm late} - a_2} \quad \text{for} \quad \tau_1 < \tau_2 < \tau_{\rm late}, 
			\\[1.4ex]
	B(t) &= B_0 \left(1 + \frac{t}{\tau_1} \right)^{a_1} \left(1 + \frac{t}{\tau_{\rm late}}\right)^{a_{\rm late} - a_1}
		\quad \text{for} \quad \tau_1 < \tau_{\rm late} < \tau_2, 
			\\[1.4ex]
	B(t) &= B_0\left(1 + \frac{t}{\tau_{\rm late}} \right)^{a_{\rm late}}
		\quad \text{for} \quad \tau_{\rm late} < \tau_1 < \tau_2. 
\end{align}
Here the two timescales $\tau_1 \equiv A_1  B_0^{b_1}$ and $\tau_2 \equiv A_2 B_0^{b_2}$ depend on the initial magnetic field, $B_0$, while $\tau_{\rm late}$ is a constant. The latter, together with the free parameters $A_{1,2}, b_{1,2}$ and the power-law indices $a_{1,2}$, can be adjusted to closely fit the numerical simulations. Measuring all three timescales in years and $B_0$ in gauss, we then choose $\tau_{\rm late} = \unit[2 \times 10^6]{yr}$, $A_{1} = \unit[10^{14}]{yr \, G}^{-b_1}$, $b_1 = -0.8$, $A_{2} =  \unit[6 \times 10^{8}]{yr \, G}^{-b_2}$, $b_{2} = -0.2$, $a_{1} = -0.13$, and $a_{2} = -3.0$. 

For particularly steep power-law indices, $a_{\rm late}$, the current prescription, in principle, allows the magnetic field to decay to unrealistically small values in contrast with observations of old millisecond pulsars \citep{Lorimer2008}. To prevent this, we assume that the magnetic field eventually settles at a constant value, $B_{\rm late}$, for very late times. In line with detected old neutron stars, we randomly sample the logarithm of $B_{\rm late}$ from a normal distribution with a mean $\mu_{\log B, {\rm final}} = 8.5$ and a standard deviation $\sigma_{\log B, {\rm final}} = 0.5$ as already outlined previously. The result of this magnetic field prescription for $a_{\rm late} = -3.0$ is shown as the dashed lines in Figure~\ref{fig:B_fields}.


\section{Coverage calculation}
\label{app:coverage}

To validate our neural posterior estimates, we follow \citet{Cook2006}, who demonstrated that for a well-calibrated posterior distribution the smallest volume that contains the ground truth, $\bt$, for a given sample in a test data set follows a uniform distribution. This, in turn, implies that the cumulative distribution function of these quantiles across the entire test set forms a diagonal line.
The graphical representation of this cumulative distribution function is commonly referred to as the \textit{coverage plot} (see Figure~\ref{fig:coverage}). Put differently, if we consider a credibility level $1-\alpha$, we expect the ground truth, $\bt$, to fall into this region for a fraction $1 - \alpha$ of test samples if the coverage is diagonal.

To calculate the corresponding coverage for our posteriors and assess how well they are calibrated, we take advantage of the amortized nature of our approximate posterior. In particular, for each of our $3600$ test samples, we have access to the ground truth, $\bt$, and the corresponding posterior approximation, $q_{F(\bx, \boldsymbol{\phi})}(\bt)$, where $F(\bx, \boldsymbol{\phi})$ represents a trained neural network. To determine the coverage, we need to calculate the quantiles for each $\bt$. In our case, where we infer five magnetorotational parameters and the posterior, $q_{F(\bx, \boldsymbol{\phi})}(\bt)$, is a five-dimensional probability density function (see Equation~\eqref{eqn:posterior_gmm}), we obtain corresponding quantiles by determining the so-called \acp{HDR}, i.e., those regions covering our sample space for a given probability $1-\alpha$ that have the smallest possible volume \citep{Hyndman1996}. To obtain these \acp{HDR} for each of our test samples, we first compute the total log-posterior at the ground truth, $\bt$, i.e., $\log q_{F(\bx, \boldsymbol{\phi})} (\bt)$. From each posterior, we subsequently draw samples, $\bt_s$, with $s\in \{1, \dots, S\}$, for which we also individually compute the log-posterior, i.e., $\log q_{F(\bx, \boldsymbol{\phi})} (\bt_s)$. The \ac{HDR} for a given test sample with ground truth, $\bt$, is now the percentage of samples, $\bt_s$, that satisfy the condition $\log q_{F(\bx, \boldsymbol{\phi})} (\bt_s) > \log q_{F(\bx, \boldsymbol{\phi})} (\bt)$. To compute the cumulative distribution function (coverage) across our test set, we repeat this process iteratively for all $3600$ test samples to determine, for a given credibility level $1 - \alpha$, the fraction of test samples where the \ac{HDR} is smaller than or equal to $1 - \alpha$. 

Deviations from the diagonal are present when posterior estimates are either too wide (conservative) or too narrow (overconfident). In the former case, ground truths would be enclosed within a given \ac{HDR} more often than expected for the true posterior, while in the latter scenario the opposite applies. The resulting coverage curves would, thus, lie above and below the diagonal, respectively, highlighting the benefit of the coverage plot in validating our posteriors.

Finally, note that for our ensemble approach we calculate the \ac{HDR} with the ensemble posterior, $\overline{q}(\bt)$, using the condition $\log \overline{q} (\bt_s) > \log \overline{q}(\bt)$. The remaining steps are identical to those outlined above.


\bibliography{bibliography}{}
\bibliographystyle{aasjournal}

\end{document}